\documentclass[letterpaper]{article} 
\usepackage{aaai25}  
\usepackage{times}  
\usepackage{helvet}  
\usepackage{courier}  
\usepackage[hyphens]{url}  
\usepackage{graphicx} 
\usepackage{natbib}  
\usepackage{caption} 
\usepackage{multirow}
\usepackage{url}
\usepackage{makecell}
\usepackage{ragged2e}

\usepackage[table,xcdraw]{xcolor}
\usepackage{algorithm}
\usepackage{algorithmic}
\usepackage{graphicx}
\usepackage{subfiles}
\usepackage{amssymb}
\usepackage{amsmath}
\usepackage{array}
\usepackage{booktabs} 
\usepackage{enumitem}
\usepackage{longtable}
\usepackage{listings}

 \usepackage{xspace}
\usepackage{newfloat}
\usepackage{environ}
\usepackage{tikz}

\definecolor{darkorange}{HTML}{724209}
\definecolor{lightorange}{HTML}{f7ce9d}
\definecolor{verylightorange}{HTML}{fbe6ce}
\definecolor{darkblue}{RGB}{31,78,120}
\definecolor{lightblue}{RGB}{221,235,247}
\definecolor{puce}{HTML}{513b41}
\definecolor{tifblue}{HTML}{c8f4f9}
\definecolor{lightbluegrey}{RGB}{230,240,255}
\definecolor{darkgrey}{RGB}{50,50,50}
\definecolor{darkgreen}{RGB}{27,64,74}
\definecolor{lightgreen}{RGB}{233,245,211}
\definecolor{brandblue}{HTML}{AFD2E9}
\definecolor{brandyellow}{HTML}{F9DB6D}
\definecolor{brandred}{HTML}{6e4047}

\DeclareFloatingEnvironment[placement=H]{frameenv} 

\newenvironment{myabstract}{
    \begin{frameenv} 
    \begin{mdframed}[roundcorner=10pt,backgroundcolor=lightorange,linecolor=darkgreen,%
                     linewidth=0pt,innertopmargin=10px,innerbottommargin=10px]
}{
    \end{mdframed}\end{frameenv}
}

\newenvironment{invsummary}{
    \begin{frameenv} 
    \begin{mdframed}[roundcorner=10pt,backgroundcolor=brandblue,%
                     linewidth=0pt,innertopmargin=10px,innerbottommargin=10px]
    \color{black}
}{
    \end{mdframed}\end{frameenv}
}
\definecolor{sweetred}{HTML}{E57373} 

\NewEnviron{todoenv}{
\par
\begin{tikzpicture}
\node[rectangle, minimum width=1\textwidth, fill=sweetred!20, rounded corners] (m) {\begin{minipage}{0.9\textwidth}\BODY\end{minipage}};
\draw[sweetred, dashed, rounded corners] (m.south west) rectangle (m.north east);
\end{tikzpicture}
}
\usepackage{mdframed}

\usepackage{newfloat}
\usepackage{listings}
\DeclareCaptionStyle{ruled}{labelfont=normalfont,labelsep=colon,strut=off} 
\floatstyle{ruled}
\newfloat{listing}{tb}{lst}{}
\floatname{listing}{Listing}
\newcommand{\il}{\texttt{iLLuMinaTE}\xspace}

\usepackage{comment}

\title{\il: An LLM-XAI Framework Leveraging Social Science Explanation Theories Towards Actionable Student Performance Feedback}

\author {
    Vinitra Swamy\thanks{These authors contributed equally to this work.}\textsuperscript{\rm 1},
    Davide Romano\footnotemark[1]\textsuperscript{\rm 1},
    Bhargav Srinivasa Desikan\textsuperscript{\rm 2},\\
    Oana-Maria Camburu\textsuperscript{\rm 3},
    Tanja Käser\textsuperscript{\rm 1}
}
\affiliations {
    \textsuperscript{\rm 1}EPFL, Switzerland  
    \textsuperscript{\rm 2}Institute for Public Policy Research, UK  
    \textsuperscript{\rm 3}University College London, UK\\
    vinitra.swamy@epfl.ch, davide.romano@epfl.ch, b.srinivasa-desikan@ippr.org, \\o.camburu@cs.ucl.ac.uk, tanja.kaeser@epfl.ch
}

\begin{document}

\maketitle

\begin{abstract}
Recent advances in eXplainable AI (XAI) for education have highlighted a critical challenge: ensuring that explanations for state-of-the-art models are understandable for non-technical users such as educators and students. In response, we introduce \il, a zero-shot, chain-of-prompts LLM-XAI pipeline inspired by \citet{miller2019explanation}'s cognitive model of explanation. \il is designed to deliver theory-driven, actionable feedback to students in online courses. \il navigates three main stages — causal connection, explanation selection, and explanation presentation — with variations drawing from eight social science theories (e.g. Abnormal Conditions, Pearl's Model of Explanation, Necessity and Robustness Selection, Contrastive Explanation). We extensively evaluate 21,915 natural language explanations of \il extracted from three LLMs (GPT-4o, Gemma2-9B, Llama3-70B), with three different underlying XAI methods (LIME, Counterfactuals, MC-LIME), across students from three diverse online courses. Our evaluation involves analyses of explanation alignment to the social science theory, understandability of the explanation, and a real-world user preference study with $114$ university students containing a novel actionability simulation. We find that students prefer \il explanations over traditional explainers $89.52\%$ of the time. Our work provides a robust, ready-to-use framework for effectively communicating hybrid XAI-driven insights in education, with significant generalization potential for other human-centric fields.
\end{abstract}

\section{Introduction}

\begin{figure*}[!ht]
    \centering
    \includegraphics[width=\linewidth]{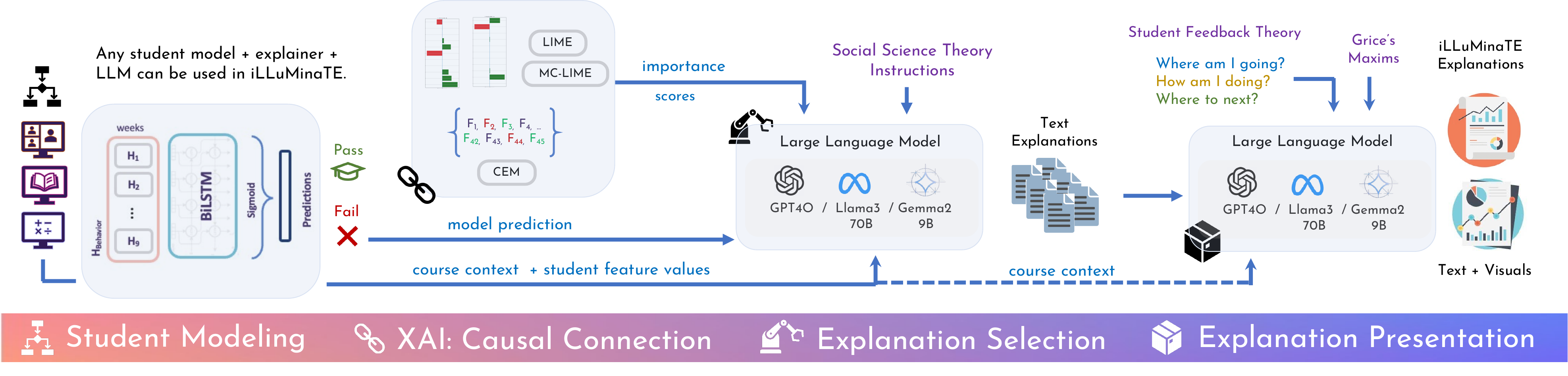}
    \caption{\textbf{\il involves four steps}: 1) modeling course data for student success prediction, 2) using XAI methods to extract feature importance scores, 3) selecting important aspects of the explanation through an LLM aligned with a given social science theory, 4) presenting the explanation with concise and actionable suggestions through an LLM.}
    \vspace{-4mm}
    \label{fig:pipeline}
\end{figure*}

Over the last decade, AI has seen widespread application in education, encompassing both learner-centric models — such as intelligent tutoring systems \cite{mousavinasab2021intelligent},  knowledge tracing \cite{piech2015deep}, and automated feedback systems \cite{jacobsen2023promises} — and teacher-centric models, including real-time classroom insights \cite{holstein2018classroom} and automated question generation \cite{hang2024mcqgen}. To adopt these solutions in real-world classrooms, model explainability is essential. \citet{nazaretsky2022teachers} underscore the importance of transparency for fostering educators' trust in AI-based educational technologies, while \citet{conati2018ai} stress the need for interpretable models in contexts where students see decision outcomes without understanding the underlying reasoning.

Recent literature on eXplainable AI (XAI) in education can be categorized into three main motivations: (1) allowing educational stakeholders to audit model mistakes \cite{khosravi2022explainable,pinto2024towards}, (2) building student and teacher trust in AI \cite{nazaretsky2024ai}, and (3) designing personalized interventions for students \cite{hur2022using, asadi2023ripple}. The most popular approaches in XAI for education are post-hoc explainers, like LIME \cite{hasib2022lime, scheers2021interactive} and SHAP \cite{baranyi2020interpretable, mu2020towards}. These explainers  treat the underlying model as a black-box, enabling explanations after model training.

Despite their rising popularity, XAI methods suffer from a major weakness: a lack of adequate understandability, especially for a non-technical audience. At a course level, STEM professors expressed difficulty in understanding explainer outputs, requesting ``more concrete and granular insights'' on the scale of individual students \cite{swamy2023trusting}. At the individual student level, \citet{hur2022using} designed XAI-based interventions, but found they were extensive to integrate and provided limited learning gains compared to expert feedback. With pervasive educational benchmarks like ASSISTments \cite{heffernan2014assistments} and MOOCRadar \cite{yu2023moocradar} having hundreds of features and temporal aspects, it becomes difficult for students and teachers to interpret feature importance. 

LLMs can be useful in making XAI more human-interpretable, especially towards building stakeholder trust in AI and designing personalized student interventions. A recent study \cite{kroeger2024large} suggests that LLMs can act as post-hoc explainers for complex models, finding that LLMs could identify relevant features when given examples of input data and model outputs. \citet{atanasova2022diagnostics} integrate explanation generation directly into the LLM’s training, optimizing over diagnostic properties like data consistency and confidence. However, in domains where explanations directly influence human decisions, the nature of LLMs as ``stochastic parrots'' can be accompanied by detrimental side effects \cite{bender2021dangers,sarkar2024large}. LLMs, while demonstrating potential in areas such as knowledge tracing or student synthesis, have not yet matured enough to act as student models that can be accompanied by inherent explanations \cite{neshaei2024towards,nguyen2023large}. We instead propose to use LLMs as \textit{communicators of explanations} \cite{zytek2024llms} to present XAI outputs in aligned text and visual formats that are \textit{actionable} for educational stakeholders. 

We therefore present \il, an in-context, chain-of-prompts, zero-shot LLM pipeline that is inspired by Miller's cognitive processes of explanation \cite{miller2019explanation}. \il follows three main stages: (1) causal connection, (2) explanation selection, and (3) explanation presentation. Our experiments range over eight prevalent social science theories of explanation \cite{hilton1990conversational,hilton1986knowledge, halpern2005causes,lombrozo2010causal,sep-scientific-explanation}, with three underlying explainers (LIME, Counterfactuals, MC-LIME), data from three online courses, evaluated using three LLMs (GPT-4o, Gemma2 9b, Llama3 70b) and a real-world user study with $114$ university students. Notably, we find that students preferred \il's explanations over baseline hybrid (text and visual) explanations from post-hoc methods $89.52\%$ of the time, and had a particular preference on actionability for abnormal, pearl, and contrastive explanations. With our study, we make the following main contributions:

\begin{enumerate}
    \item \textbf{\il, a chain-of-prompts framework in the education context} to extract theory-driven natural language explanations (NLE) for student feedback.
    \item \textbf{An LLM-XAI efficacy analysis} of $216$ variations of \il prompting strategies over explainers, LLMs, social science theories, and student populations.
    \item \textbf{A real-world evaluation of LLM-XAI preferences} conducted with $114$ university students. 
    \item \textbf{An XAI actionability study} simulating student performance gains based on actions they selected from generated explanations.
\end{enumerate}

We provide our modular implementation of \il publicly with adaptations for LangChain, Groq Cloud, and Replicate\footnote{\url{https://github.com/epfl-ml4ed/iLLuMinaTE}}. Our work provides a theory-driven methodology to communicate results of XAI to students, with broad generalization potential of explanation theory instruction prompts for other human-centric fields (e.g., healthcare, welfare, product recommendation).

\section{Methodology}
\label{sec:methods}
Our \il pipeline (see Fig. \ref{fig:pipeline}) consists of four stages towards communicating explanations in a human-understandable and actionable way through LLMs.

In the \textit{Student Modeling} phase, we extract behavioral features from raw clickstreams of student interactions and use BiLSTMs \cite{graves2005framewise} to predict student success following prior work \cite{asadi2023ripple,swamy2022meta}. We then employ post-hoc explainers to obtain feature importance scores, representing the \textit{XAI: Causal Connection} step. With the results from an explainer, the course context, and the student's feature values, we prompt an LLM using \textit{Explanation Selection} instructions specific to social science theories of explanations. We evaluate the obtained explanations using human expert and GPT-4o annotations. We then use \textit{Explanation Presentation} prompts to summarize the often verbose explanation selection reports into concise and actionable feedback for a student, taking into account theory on effective feedback \cite{hattie2007power}'s and maxims for communication \cite{grice1975logic}. We then evaluate the final explanations using expert annotations. We assess \textit{Student Preferences} and the actionability of \il explanations in comparison with the current state-of-the-art in XAI for education in a user study with $114$ students.

\subsection{Student Modeling}
\label{sec:student-modeling}
To create the models building the basis for the explanations, we use the same features and model architectures as prior research working with the same datasets \cite{swamy2022meta,galici2023not,swamy2023trusting,swamy2024multimodn}.

\vspace{1mm} \noindent \textbf{Data Collection}.
Our experiments are based on data collected from three MOOCs (Digital Signal Processing (DSP), Villes Africaines (VA), and Elements de Geomatique (Geo)) offered by a European University to a global student audience. The courses were organized into weekly modules including video lectures and quizzes and required students to complete graded assignments to earn course certificates. Students interacted with learning objects (videos, quizzes) associated with specific course weeks, enabling the creation of course-specific learning indicators. We represent a student's interations as a time series $I_s^c = {i_1, \ldots, i_K}$, where each interaction $i$ is a tuple $(t, a, o)$, including a \emph{timestamp} $t$, an \emph{action} $a$ (e.g., video play, pause; quiz submission), and a \emph{learning object} $o$. The \emph{binary success label} (pass-fail) for student $s$ in course $c$ is denoted as $y_{s,c}$. Data collection and analysis were approved by the university’s ethics review board (HREC 058-2020/10.09.2020).

\vspace{1mm} \noindent \textbf{Feature Extraction.} We use a broad set of $45$ behavioral features \(h\) derived from the student interactions, incorporating features from four feature sets shown to be predictive for student performance in MOOCs \cite{marras2021can}. \textbf{Regularity} features (3) capture consistent study habits \cite{boroujeni2016quantify}, \textbf{Engagement} features (13) measure course involvement \cite{chen2020utilizing}, \textbf{Control} features (22) analyze video usage \cite{lalle2020data}, and \textbf{Participation} features (7) track attendance in scheduled activities \cite{marras2021can}.

\vspace{1mm} \noindent \textbf{Modeling}. For a course \(c\) and student \(s\), our objective is to build a model that predicts \(y_{s,c}\) \textit{early}, using the features \(h_s\) from the first five weeks. Following prior work \cite{swamy2022meta}, we employ a BiLSTM for this task. We provide all reproducibility details in Appendix \ref{appendix:reproducibility}.

\vspace{-2mm}
\subsection{XAI: Causal Connection}
\label{sec:causal-connection}

We use three popular post-hoc explainers (LIME, CEM, and MC-LIME) to extract local, instance-specific explanations from the student models. We chose these three methods based on their popularity, but any in-hoc or interpretable-by-design model could be used.

\vspace{1mm} \noindent \textbf{LIME} (Local Interpretable Model-Agnostic Explanations) \cite{ribeiro2016why} provides interpretable explanations for individual predictions by approximating the complex model locally with an interpretable model. This process results in a set of feature weights indicating the positive or negative influence of each feature on the prediction.

\vspace{1mm} \noindent \textbf{CEM} (Contrastive Explanation Method) \cite{dhurandhar2018explanations} is a counterfactual method that identifies the features that need to be present (pertinent positives) or absent (pertinent negatives) for a model to maintain its prediction for a given instance. 

\vspace{1mm} \noindent \textbf{MC-LIME} (Minimal Counterfactual LIME) \cite{2022.EDM-short-papers.33} finds the minimal set of LIME features that, if changed, would alter the outcome. It focuses on features that increase the likelihood of an event (e.g., student dropout). MC-LIME applies changes to individual features, then pairs, and so on, until a change flips the prediction. This approach combines explanation sparseness with the advantages of counterfactuals and LIME.

\subsection{Explanation Selection}
\label{sec:explanation-selection}
Our \il pipeline generates explanations aligned to social science theories. Specifically, we have translated eight theories into prompts: \textit{Relevance Selection (RS), Abnormal Conditions (AC), Pearl's Model (Pearl's), Necessity and Robustness (NR), two contrastive explanations (BC, Con), Statistical Relevance (SR),} and \textit{Chain of Thought (CoT)}. Our prompt structure contains
 (1) descriptions of the model, features, XAI method, and course context, (2) explainer importance scores and relevant student feature values, and (3) social science theory instructions. We present the general prompt template and two theory-specific examples (RS, AC) in this section, all details are in Appendix \ref{appendix:exp-sel-prompt}.
 
%
\begin{figure}[t]
    \centering
    \includegraphics[width=0.93\linewidth]{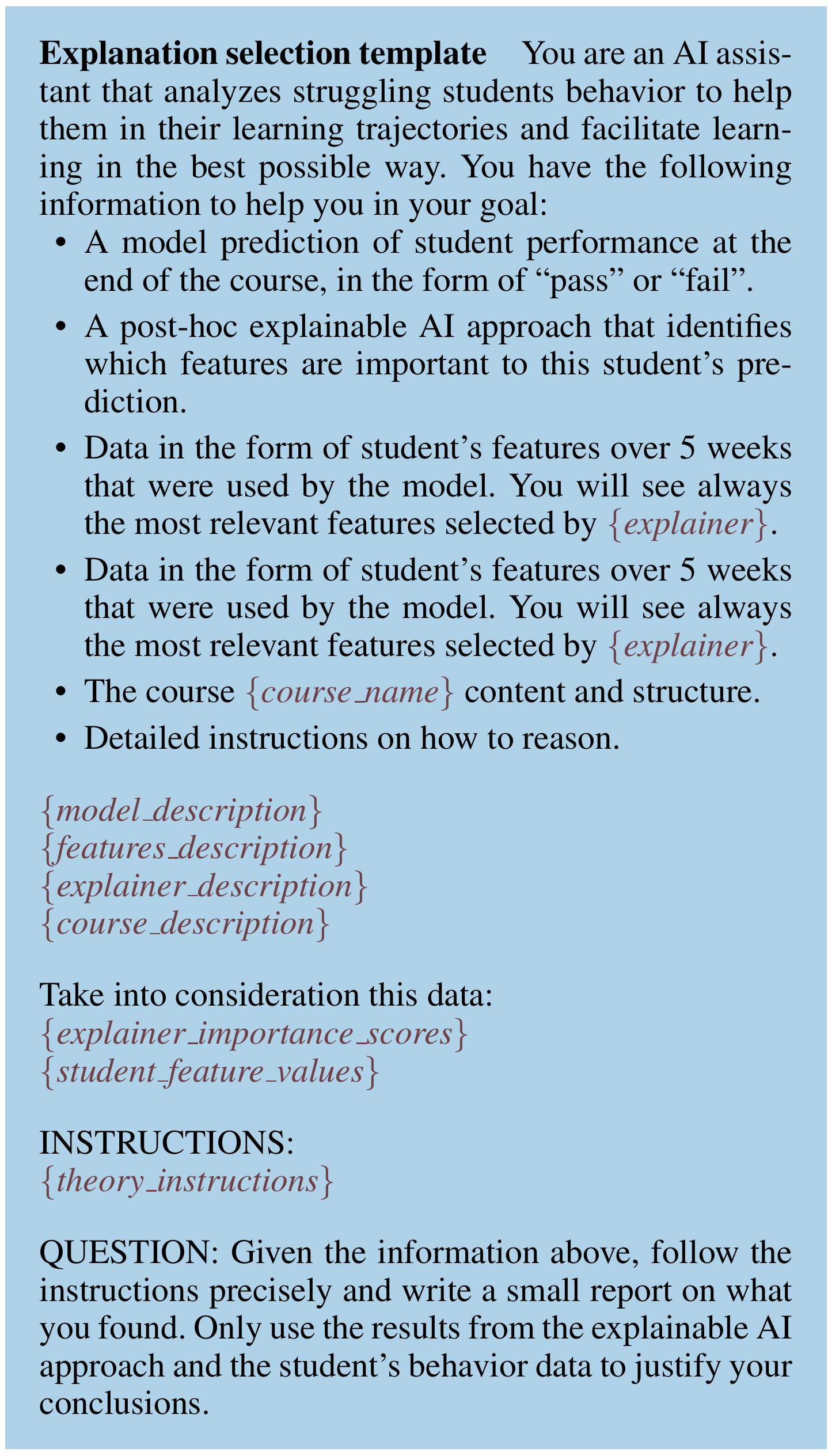}
    \vspace{-3mm}
\end{figure}

\vspace{1mm} \noindent \textbf{Relevance Selection.} Based on Hilton's conversational model of explanation \cite{hilton1990conversational}, relevance-based selection theory aims to ``resolve a puzzle in the explainee's mind'' by filling gaps in their knowledge. The theory emphasizes that shared knowledge between the explainer and the explainee are presuppositions of the explanations, and the other factors are the causes that should be explained. In short, the explainer should not explain any causes they think the explainee already knows.

\vspace{1mm} \noindent \textbf{Abnormal conditions.} This theory, based upon Hilton and Slugoski's abnormal conditions model \cite{hilton1986knowledge}, suggests that explanations often rely on unusual and temporally proximal events. People do not solely count on statistical likelihood but highlight uncommon factors contributing to an event to explain that event. During a conversation, the explainer is relying on the perceived common prior knowledge to identify potential causes that are considered abnormal, with greater weight given to temporally proximal events and factors that the explainee can control. This focus on controllable factors helps the explainee understand how to potentially avoid similar situations in the future.

\vspace{1mm} \noindent \textbf{Pearl's Model of explanation.} \citet{halpern2005causes} present a formal framework for selecting explanations based on epistemic relevance and structural causal models. The model distinguishes between \textit{exogenous variables}, whose values are determined by external factors, and \textit{endogenous variables}, whose values are influenced by relationships with other variables. 
Within a \textit{context} (a specific assignment of values to variables), the model defines an \textit{actual cause} as a minimal set of events that must occur for an event to happen. 

\vspace{1mm} \noindent \textbf{Necessity \& Robustness selection.} Two key criteria for selecting strong explanatory causes are necessity and robustness \cite{lipton1990contrastive}. \textit{Necessity} refers to whether a cause is essential for the effect to occur. \textit{Robustness} considers how generally a cause applies \cite{lombrozo2010causal}. This idea aligns with the concept of simplicity, where broader explanations with fewer specific requirements are favored.

\vspace{1mm} \noindent \textbf{Contrastive Explanation.} This theory suggests explanations are not simply cause and effect statements, but rather comparisons between what happened (the target event) and what could have happened (a counterfactual contrasting event) \cite{hilton1990conversational}. One possible way to make the identification of the counterfactual event (foil) successful is to ask the module to reformulate the question. This technique is known as Rephrase and Respond (RaR) \cite{deng2024rephrase}. 

\vspace{1mm} \noindent \textbf{Statistical Relevance.} This method is based on the SR model based on scientific causal reasoning \cite{sep-scientific-explanation}. The SR model explanations can be defined in simple terms as ``statistically relevant properties are explanatory and statistically irrelevant properties are not''. It follows this structure: \textit{‘‘Based on empirical data, factors A, B and C contribute to the probability of Y by the amount of X’’}.

\begin{figure}[t]
    \centering
    \includegraphics[width=0.93\linewidth]{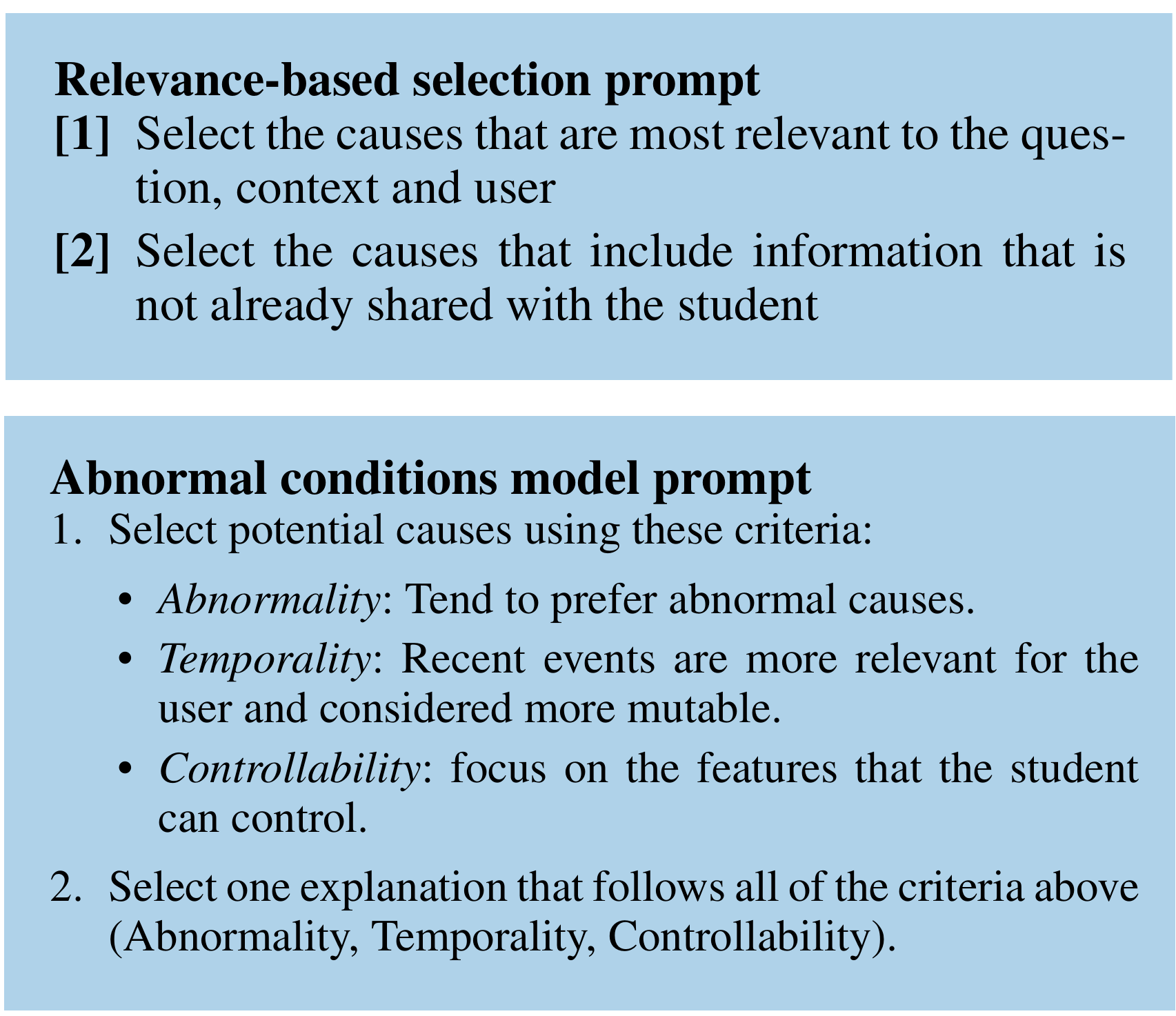}
    \vspace{-3mm}
\end{figure}

\vspace{1mm} \noindent \textbf{Chain-of-Thought (baseline).} Chain-of-Thought (CoT) prompting \cite{wei2023chainofthought} guides an LLM through sequential reasoning that mimics human thought processes.

\vspace{1mm} \noindent \textbf{Evaluation.} To assess whether \il responses align with post-hoc explanations, student feature values, and course context, we developed an annotation rubric based on decomposed questions as a basis for human and LLM annotation. Recent studies \cite{wang2023chatgpt} demonstrated that LLMs can match human annotators, especially when instructions are decomposed into simple criteria phrased as binary (``yes''/``no'') questions \cite{qin2024infobench}. We created four general decomposed questions applicable across all theories and additional theory-specific questions ($1-6$, depending on the theory). These were validated by a computational social scientist who was not involved in the prompt creation process. The general questions used items such as `Is the generated text correctly using the model's predicted outcome?' or `Is the generated text analysis based solely on the explainer results provided?'. An example of a theory-specific questions is: `Is the generated text selecting the causes that are most relevant to the user?' (\textit{RS}). The complete rubric is available in Appendix Table~\ref{appendix:tab-dqs}.

We instructed three experts to annotate the same block of $42$ responses (2 students, 3 explainers, 8 theories) across a set of up to eight decomposed questions for each setting, resulting in an inter-rater agreement of $\kappa=0.71\pm0.13$ (Cohen's Kappa). After that, each annotator proceeded with independently evaluating between $84$ to $105$ more responses, leading to a total of $315$ human-annotated responses and in total over $2,350$ annotations. We then proceeded with instructing GPT-4o to annotate the same $42$ responses using the same decomposed questions. We obtained an agreement of $94.68\% \pm 5.20$ (percentage of answers with agreement) between human and GPT-4o annotation. The inter-rater agreement per theory is in Appendix Table~\ref{tab:annot-agree-gpt}. Given the high agreement, we annotated all explanations by GPT-4o.

\subsection{Explanation Presentation}
\label{sec:explanation-presentation}
Inspired by Hilton's conversational model \cite{hilton1990conversational}, we employ a presentation prompt to refine and condense the output of the explanation selection prompt to make it relevant to the explainee's future actions, current context and prior knowledge. This prompt is formulated using Grice's maxims on communication \shortcite{grice1975logic} and learning science literature on best practices in communicating feedback to students \cite{shute2008focus,hattie2007power}. Grice's maxims provide a framework for understanding cooperative conversation, emphasizing providing the right amount of relevant and clear information to achieve a shared goal. Grice further divides this principle into four maxims: \textit{Quality} (truthfulness and evidence-based statements), \textit{Quantity} (providing enough but not excessive information), \textit{Relation} (relevance), and \textit{Manner} (clear and concise communication). \citet{hattie2007power} provide a prominent educational feedback framework with three steps: \textit{``Where am I going?''}, \textit{``How am I doing?''\footnote{We adapted 
 \citet{hattie2007power}'s question \textit{``How am I going?''} to \textit{``How am I doing?''} for ease of understanding.}}, and \textit{``Where to next?''} These steps require the feedback provider to clearly state the learning goal, provide a summary of relevant student performance, and suggest concrete actions for improvement in the near future. The full explanation presentation prompt is included in the Appendix \ref{appendix:exp-pres-prompt}.

\vspace{1mm} \noindent \textbf{Evaluation.} Similar to the explanation selection phase, we evaluated the final explanations using a rubric of decomposed `yes'/`no' questions (the detailed set of questions is illustrated in Appendix Table~\ref{appendix:tab-dqs}). Given the high agreement between GPT-4o and human annotations in the explanation selection phase, we used GPT-4o as an expert. Additionally, we conducted a readability evaluation using the following metrics: \textit{Flesch-Kincaid Grade Level} \cite{flesch1948new}, \textit{Gunning Fog Index} \cite{gunning1952technique}, \textit{SMOG index} \cite{mc1969smog}, and LanguageTool \textit{Grammar Issues} \cite{mozgovoy2011dependency}. These metrics evaluate the comprehensibility of the text in terms of sentence length, vocabulary complexity, grammatical correctness, and estimated years of schooling needed for understanding.

\begin{table*}[t]
\small
\centering
\resizebox{1.0\textwidth}{!}{
\begin{tabular}{@{}clllllll@{}}
\toprule
\textbf{Explainer} & \multicolumn{1}{c}{\textbf{Theory}} & \multicolumn{1}{c}{\textbf{Overall}} & \multicolumn{1}{c}{\textbf{\begin{tabular}[c]{@{}c@{}}Q1: Using \\ provided data\\ extensively\end{tabular}}} & \multicolumn{1}{c}{\textbf{\begin{tabular}[c]{@{}c@{}}Q2: Analysis\\ based solely\\ on explainer\end{tabular}}} & \multicolumn{1}{c}{\textbf{\begin{tabular}[c]{@{}c@{}}Q3: Correctly\\ using model's\\ prediction\end{tabular}}} & \multicolumn{1}{c}{\textbf{\begin{tabular}[c]{@{}c@{}}Q4: Using\\ course content\\ and structure\end{tabular}}} & \multicolumn{1}{c}{\textbf{\begin{tabular}[c]{@{}c@{}}Theory-\\ Specific Qs\end{tabular}}} \\ \midrule
 & \textit{RaR + Contrastive} & {\color[HTML]{036400} \textbf{0.985 ± 0.069}} & {\color[HTML]{036400} 0.99 ± 0.099} & {\color[HTML]{036400} 0.992 ± 0.087} & {\color[HTML]{036400} 0.998 ± 0.047} & {\color[HTML]{036400} 0.967 ± 0.178} & {\color[HTML]{036400} 0.984 ± 0.085} \\
 & \textit{Abnormal Conditions} & {\color[HTML]{036400} \textbf{0.958 ± 0.141}} & {\color[HTML]{036400} 0.997 ± 0.057} & {\color[HTML]{036400} 0.997 ± 0.057} & {\color[HTML]{036400} 0.998 ± 0.047} & 0.865 ± 0.342 & {\color[HTML]{036400} 0.952 ± 0.156} \\
 & \textit{Relevance Selection} & \textbf{0.949 ± 0.155} & {\color[HTML]{036400} 0.998 ± 0.047} & {\color[HTML]{036400} 0.999 ± 0.033} & {\color[HTML]{036400} 0.997 ± 0.057} & 0.898 ± 0.303 & 0.925 ± 0.195 \\
 & \textit{Necessity Robustness} & \textbf{0.949 ± 0.154} & {\color[HTML]{036400} 0.996 ± 0.066} & {\color[HTML]{036400} 0.997 ± 0.057} & {\color[HTML]{036400} 0.998 ± 0.047} & {\color[HTML]{9A0000} 0.676 ± 0.468} & {\color[HTML]{036400} 0.991 ± 0.047} \\
 & \textit{Pearl Explanation} & \textbf{0.948 ± 0.099} & {\color[HTML]{036400} 0.979 ± 0.142} & {\color[HTML]{036400} 0.983 ± 0.131} & {\color[HTML]{036400} 0.998 ± 0.047} & 0.869 ± 0.337 & 0.939 ± 0.071 \\
 & \textit{Statistical Relevance} & \textbf{0.884 ± 0.184} & {\color[HTML]{036400} 0.988 ± 0.109} & {\color[HTML]{036400} 0.993 ± 0.081} & {\color[HTML]{036400} 0.996 ± 0.066} & {\color[HTML]{9A0000} 0.453 ± 0.498} & {\color[HTML]{036400} 0.991 ± 0.001} \\
 & \textit{(Base) Contrastive} & {\color[HTML]{000000} \textbf{0.852 ± 0.194}} & {\color[HTML]{036400} 0.968 ± 0.175} & {\color[HTML]{9A0000} 0.647 ± 0.478} & {\color[HTML]{036400} 1.0 ± 0.0} & {\color[HTML]{9A0000} 0.681 ± 0.466} & 0.879 ± 0.179 \\
\multirow{-8}{*}{\textbf{CEM}} & \textit{Chain of Thought} & \textbf{0.872 ± 0.188} & {\color[HTML]{036400} 0.979 ± 0.142} & {\color[HTML]{036400} 0.984 ± 0.127} & {\color[HTML]{036400} 0.997 ± 0.057} & {\color[HTML]{9A0000} 0.585 ± 0.493} & 0.817 ± 0.001 \\ \midrule
 & \textit{RaR + Contrastive} & {\color[HTML]{036400} \textbf{0.975 ± 0.098}} & {\color[HTML]{036400} 0.974 ± 0.16} & 0.985 ± 0.123 & {\color[HTML]{036400} 1.0 ± 0.0} & 0.946 ± 0.227 & {\color[HTML]{036400} 0.975 ± 0.111} \\
 & \textit{Relevance Selection} & {\color[HTML]{036400} \textbf{0.96 ± 0.112}} & {\color[HTML]{036400} 0.987 ± 0.114} & 0.991 ± 0.093 & {\color[HTML]{036400} 0.998 ± 0.047} & 0.918 ± 0.274 & 0.946 ± 0.132 \\
 & \textit{Abnormal Conditions} & {\color[HTML]{036400} \textbf{0.951 ± 0.131}} & {\color[HTML]{036400} 0.99 ± 0.099} & 0.993 ± 0.081 & {\color[HTML]{036400} 0.995 ± 0.074} & 0.893 ± 0.309 & 0.938 ± 0.156 \\
 & \textit{Necessity Robustness} & \textbf{0.929 ± 0.147} & {\color[HTML]{036400} 0.982 ± 0.135} & 0.984 ± 0.127 & {\color[HTML]{036400} 0.998 ± 0.047} & {\color[HTML]{9A0000} 0.659 ± 0.474} & {\color[HTML]{036400} 0.959 ± 0.103} \\
 & \textit{Pearl Explanation} & \textbf{0.896 ± 0.106} & 0.904 ± 0.294 & 0.922 ± 0.269 & {\color[HTML]{036400} 0.989 ± 0.104} & 0.836 ± 0.371 & 0.88 ± 0.114 \\
 & \textit{Statistical Relevance} & \textbf{0.859 ± 0.144} & 0.919 ± 0.272 & 0.964 ± 0.186 & {\color[HTML]{036400} 0.984 ± 0.127} & {\color[HTML]{9A0000} 0.476 ± 0.5} & {\color[HTML]{036400} 0.953 ± 0.001} \\
 & \textit{(Base) Contrastive} & \textbf{0.822 ± 0.195} & 0.97 ± 0.172 & {\color[HTML]{9A0000} 0.45 ± 0.498} & {\color[HTML]{036400} 0.998 ± 0.047} & {\color[HTML]{9A0000} 0.624 ± 0.485} & 0.884 ± 0.183 \\
\multirow{-8}{*}{\textbf{LIME}} & \textit{Chain of Thought} & \textbf{0.82 ± 0.188} & 0.945 ± 0.229 & 0.956 ± 0.204 & {\color[HTML]{036400} 0.997 ± 0.057} & {\color[HTML]{9A0000} 0.526 ± 0.5} & {\color[HTML]{9A0000} 0.675 ± 0.001} \\ \midrule
 & \textit{RaR + Contrastive} & {\color[HTML]{036400} \textbf{0.985 ± 0.063}} & {\color[HTML]{036400} 0.987 ± 0.114} & {\color[HTML]{036400} 0.989 ± 0.104} & {\color[HTML]{036400} 0.999 ± 0.033} & {\color[HTML]{036400} 0.968 ± 0.175} & {\color[HTML]{036400} 0.983 ± 0.073} \\
 & \textit{Relevance Selection} & {\color[HTML]{036400} \textbf{0.963 ± 0.132}} & {\color[HTML]{036400} 0.996 ± 0.066} & {\color[HTML]{036400} 0.997 ± 0.057} & {\color[HTML]{036400} 1.0 ± 0.0} & 0.915 ± 0.279 & 0.949 ± 0.153 \\
 & \textit{Abnormal Conditions} & {\color[HTML]{036400} \textbf{0.96 ± 0.12}} & {\color[HTML]{036400} 0.991 ± 0.093} & {\color[HTML]{036400} 0.991 ± 0.093} & {\color[HTML]{036400} 0.993 ± 0.081} & 0.879 ± 0.326 & {\color[HTML]{036400} 0.957 ± 0.135} \\
 & \textit{Necessity Robustness} & \textbf{0.932 ± 0.162} & {\color[HTML]{036400} 0.99 ± 0.099} & {\color[HTML]{036400} 0.99 ± 0.099} & {\color[HTML]{036400} 1.0 ± 0.0} & {\color[HTML]{9A0000} 0.609 ± 0.488} & {\color[HTML]{036400} 0.978 ± 0.083} \\
 & \textit{Pearl Explanation} & \textbf{0.919 ± 0.108} & {\color[HTML]{000000} 0.942 ± 0.233} & {\color[HTML]{036400} 0.951 ± 0.216} & {\color[HTML]{036400} 0.992 ± 0.087} & 0.806 ± 0.395 & 0.914 ± 0.104 \\
 & \textit{Statistical Relevance} & \textbf{0.876 ± 0.171} & {\color[HTML]{036400} 0.97 ± 0.172} & {\color[HTML]{036400} 0.99 ± 0.099} & {\color[HTML]{036400} 0.993 ± 0.081} & {\color[HTML]{9A0000} 0.446 ± 0.497} & {\color[HTML]{036400} 0.978 ± 0.001} \\
 & \textit{(Base) Contrastive} & \textbf{0.856 ± 0.19} & {\color[HTML]{036400} 0.97 ± 0.172} & {\color[HTML]{036400} 0.615 ± 0.487} & {\color[HTML]{036400} 0.996 ± 0.066} & {\color[HTML]{9A0000} 0.688 ± 0.497} & {\color[HTML]{000000} 0.895 ± 0.187} \\
\multirow{-8}{*}{\textbf{MC-LIME}} & \textit{Chain of Thought} & \textbf{0.858 ± 0.199} & {\color[HTML]{036400} 0.973 ± 0.163} & {\color[HTML]{036400} 0.982 ± 0.135} & {\color[HTML]{036400} 0.999 ± 0.033} & {\color[HTML]{9A0000} 0.565 ± 0.496} & 0.77 ± 0.001 \\ \bottomrule
\end{tabular}}
\caption{\textbf{Alignment of GPT-4o, Gemma2 9b, and Llama3 70B generated explanations with theory}. Average (± std) of ``Yes'' answers, displayed separately for the first four general decomposed questions as well as averaged over the theory-specific questions and overall. Annotated by experts and GPT-4o. Scores over $95\%$ (less than $65\%$) are highlighted in green (red).}
\vspace{-4mm}
\label{tab:gpt-human-comparison}
\end{table*}

\subsection{Student Preferences (User Study)}
\label{sec:student-preferences}
To evaluate students' explanation preferences as well as the actionability of \il explanations, we conducted a user study comparing our explanations \textbf{(text and visual)} to post-hoc baselines. We recruited $114$ students on \texttt{Prolific}, (see Appendix \ref{appendix:study-background} for detailed information about the participants' demographics). Students were told that the explanations related to their own performance in three different online courses they were enrolled in.

For each course, we presented participants with eight explanations on their predicted success or failure in that course: four explanations at a time, with three randomly ordered \il variations and one baseline approach. We elected to use the six \il instructions with the highest instruction-following accuracy from the human-expert evaluation for this experiment, and an equal mixture of passing and failing behavior at different model confidence levels. Each explanation was provided as a brief text accompanied by a graph illustrating the features and concepts used by the model. GPT-4o created all \il visuals based on the first two responses (full prompt in Appendix \ref{appendix:exp-viz-prompt}). LIME visuals were used from the package, and CEM and MC-LIME visuals were expert-created and iterated upon with six pilot participants. Examples of the study format are included in our repository and Appendix \ref{appendix:study-materials}. Participants were asked to choose their preferred explanation for each set of comparisons and explain their choice (open-ended question). They were then asked to compare the explanations based on five criteria \cite{frej2024finding}:

\begin{enumerate}
\small
    \item \textbf{Usefulness}: This explanation is useful to understand the prediction based on my learning behavior.
    \item \textbf{Trustworthiness}: This explanation lets me judge if I should trust the suggestions.
    \item \textbf{Actionability}: This explanation helps me make a decision on how to improve my learning behavior.
    \item \textbf{Completeness}: This explanation has sufficient detail to understand why the prediction was made based on my learning behavior.
    \item \textbf{Conciseness}: Every detail of this explanation is necessary.
\end{enumerate}

Finally, participants were asked to choose one of ten suggested actions for the next week, based on their preferred explanation. These actions were aligned with behavioral features from the model, allowing us to simulate the impact if the student acted according to them. We trained a BiLSTM to predict student success on six weeks of student data, and conducted inference on simulated students increasing the relevant features by 25\% percentile. We also asked students which weeks of material they would focus on based on the explanation (between one to three weeks). 

\vspace{-1mm}
\section{Results}
We evaluated \il explanations by assessing the LLM's instruction-following abilities during explanation selection (Exp 1) and using readability metrics and an automated analysis to measure explanation understandability at the presentation stage (Exp 2). We then analyzed student preferences in the user study (Exp 3) and simulated whether actions derived from the explanations improved student performance (Exp 4).

\vspace{1mm} \noindent \textbf{Experimental Protocol.} We optimized the BiLSTM models (one for each course) using a train-validation-test split of 80:10:10 and including a hyperparameter search. We achieve balanced accuracies of $90.8$ (DSP), $80.3$ (Afr), and $76.8$ (Geo) respectively. These results for early student performance prediction at five weeks are in line with prior work \cite{swamy2022meta}. Explainers (LIME, CEM) were extracted with the same settings as per related work to ensure a fair comparison \cite{swamy2022evaluating, swamy2023trusting}.

\subsection{Exp 1: \il is aligned with social science theories of explanation} 
\label{sec:exp1}
In the first analysis, we evaluated how well the generated explanations aligned with the instructions. We selected $105$ representative students per course, distributed across six behavioral dimensions (regularity, effort, consistency, proactivity, control, and assessment) \cite{mejia2022identifying}. Table~\ref{tab:gpt-human-comparison} shows the average (with standard deviation) scores for each general decomposed question (see Section~\ref{sec:explanation-selection} and Appendix~\ref{appendix:decomposed-qs}), for the theory-specific questions, and overall. The general decomposed questions involve whether the response is using the provided data extensively (Q1), whether the analysis in the response is based solely on the results from the explainer (Q2), whether the response is correctly using the model's prediction (Q3), and whether the response is using the course content and structure. The scores represent the ratio of ‘Yes’ answers averaged across students and courses, with overall scores exceeding $0.82$ for all explainers and theories. There were no significant differences between explainers or theories, as indicated by overlapping $95\%$ CIs.
There were generally also no differences in scores between  questions. Only Q4 (``Using course content and structure'') had lower average scores than the other questions, suggesting that generating explanations fully incorporating the course content is a challenge. Notably, inter-rater agreement for Q4 was also lower than the average ($\kappa=0.52$ vs. $\kappa=0.71$ overall), indicating lower reliability of annotations for this question.

In a second analysis, we compared different LLMs' abilities to generate explanations according to instructed theories. Table~\ref{tab:exp1-llm-comparison} (column \textit{Explanation Selection}) shows the average scores for each model and explainer for the same representative students. Scores reflect the number of `Yes' answers to the decomposed questions described in Section~\ref{sec:explanation-presentation}, with annotation done automatically using GPT-4o. GPT-4o \il explanations scored the highest, closely followed by Gemma2 9b, and Llama3 70b. However, all of the $95\%$ CI overlap. Again, we found no differences between the different explainers. An experiment on the generalizability of the results to the flipped classroom context (smaller dataset, different domain) is included in Appendix \ref{appendix:generalization}.

\subsection{Exp 2: \il explanations are understandable} 
We assessed the presentation of \il explanations using both LLM annotation and readability metrics. Table~\ref{tab:exp1-llm-comparison} (column \textit{Explanation Presentation}) shows scores per LLM and explainer, averaged over $300$ representative students (see Section~\ref{sec:exp1}). Scores represent the ratio of ``Yes'' answers to the set of decomposed questions described in Section~\ref{sec:explanation-presentation}, as annotated by GPT-4o. For this stage, Llama3 70b achieved the highest scores, followed closely by GPT-4o and Gemma2 9b. This finding is in contrast to the explanation selection stage, where GPT-4o reached the highest score, suggesting that depending on the use case (whether selection or presentation is more important), it is possible to use much smaller and open source models for the task at hand.

\begin{table}[t]
  \centering
    \small
  \begin{tabular}{@{}llll@{}}
  \toprule
  \textbf{Model} & \textbf{Explainer} & \textbf{\begin{tabular}[c]{@{}l@{}}Explanation \\ Selection\end{tabular}} & \textbf{\begin{tabular}[c]{@{}l@{}}Explanation \\ Presentation\end{tabular}} \\ \midrule
  \multirow{3}{*}{\textbf{\begin{tabular}[c]{@{}l@{}}Gemma2\\ 9b\end{tabular}}} & \textit{CEM} & 0.941 ± 0.202 & 0.791 ± 0.212 \\
   & \textit{LIME} & 0.937 ± 0.217 & 0.801 ± 0.201 \\
   & \textit{MC-LIME} & 0.939 ± 0.208 & 0.767 ± 0.165 \\ \midrule
  \multirow{3}{*}{\textbf{GPT-4o}} & \textit{CEM} & 0.961 ± 0.148 & 0.992 ± 0.070 \\
   & \textit{LIME} & 0.954 ± 0.165 & 0.992 ± 0.072 \\
   & \textit{MC-LIME} & 0.965 ± 0.144 & 0.992 ± 0.072 \\ \midrule
  \multirow{3}{*}{\textbf{\begin{tabular}[c]{@{}l@{}}Llama3\\ 70b\end{tabular}}} & \textit{CEM} & 0.897 ± 0.222 & 0.998 ± 0.037 \\
   & \textit{LIME} & 0.848 ± 0.302 & 0.997 ± 0.042 \\
   & \textit{MC-LIME} & 0.881 ± 0.258 & 0.998 ± 0.037 \\ \bottomrule
  \end{tabular}
\caption{\textbf{Explanation quality by LLM}. Degree of instruction-following for explanations generated by GPT-4o, Gemma2 9b, and Llama3 70b. Average (with standard deviation) of ``Yes'' answers, annotated by GPT-4o.}
\vspace{-6mm}
 \label{tab:exp1-llm-comparison}
\end{table}

\begin{figure}[t]
    \centering
    \includegraphics[width=\linewidth, trim={0 10 0 0},clip]{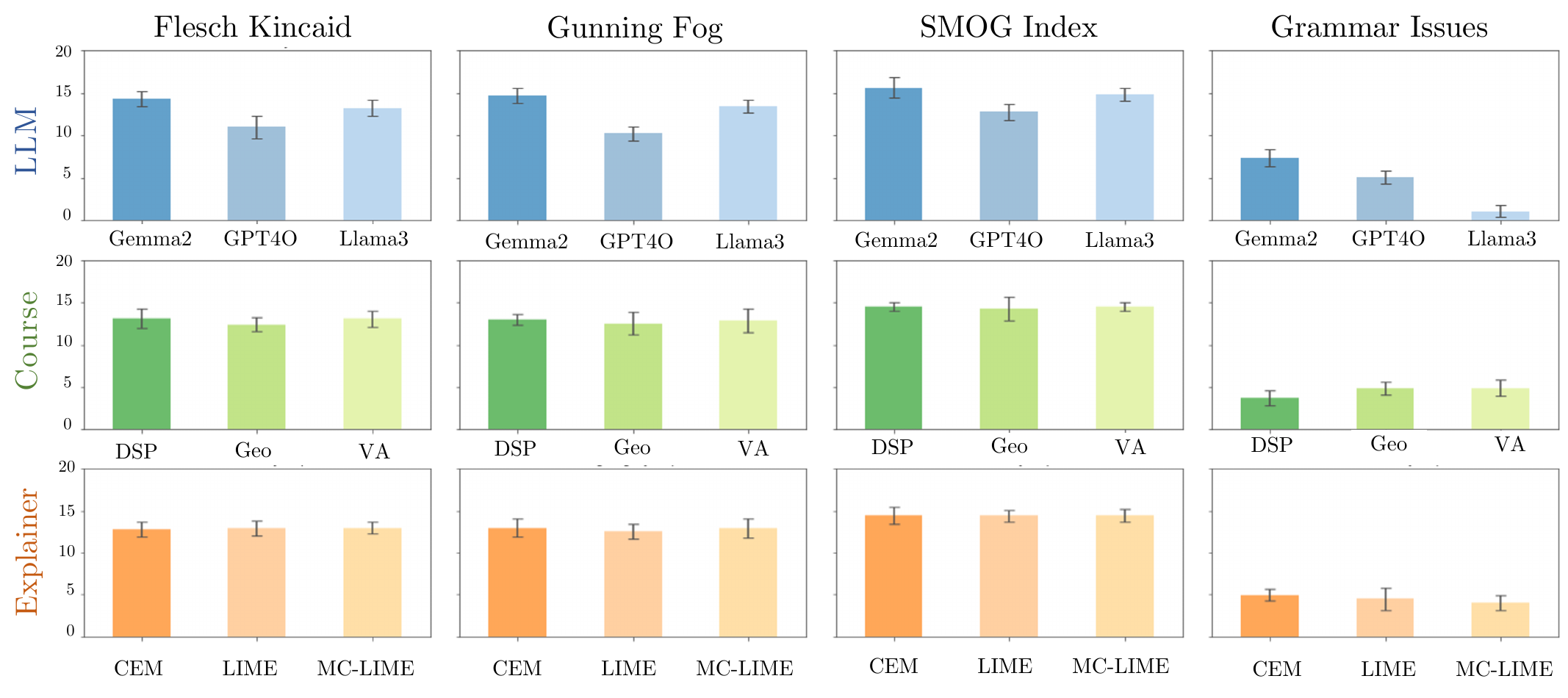}
    \caption{\textbf{Readability metrics.} Fleach Kincaid, Gunning Fog, SMOG Index, Grammar Issues across LLM (blue, top), course (green, middle), and explainer (orange, bottom). Lower scores are better.}
    \vspace{-3.5mm}
    \label{fig:eq2-readability}
\end{figure}

Figure~\ref{fig:eq2-readability} illustrates the readability scores (Flesch Kincaid, SMOG Index, Gunning Fog, Grammar Issues) of the final \il explanations after the explanation presentation prompt. Lower scores are better for all metrics. GPT-4o reached the best readability performance of the three LLMs (Fig~\ref{fig:eq2-readability}, top), with no overlap in $95\%$ confidence intervals. In contrast, Llama3 70b committed the least grammatical errors. We found no differences in readabilty or grammar between explanations for the different courses Fig~\ref{fig:eq2-readability}, middle), demonstrating that our approach is generalizable to diverse educational contexts. Similarly, all explainers achieved similar scores across all four metrics Fig~\ref{fig:eq2-readability}, bottom), showing that choice of source explainer does not have a strong impact on explanation understandability.

\begin{figure}[t]
    \centering
    \includegraphics[width=\linewidth,  trim={0 15 0 0},clip]{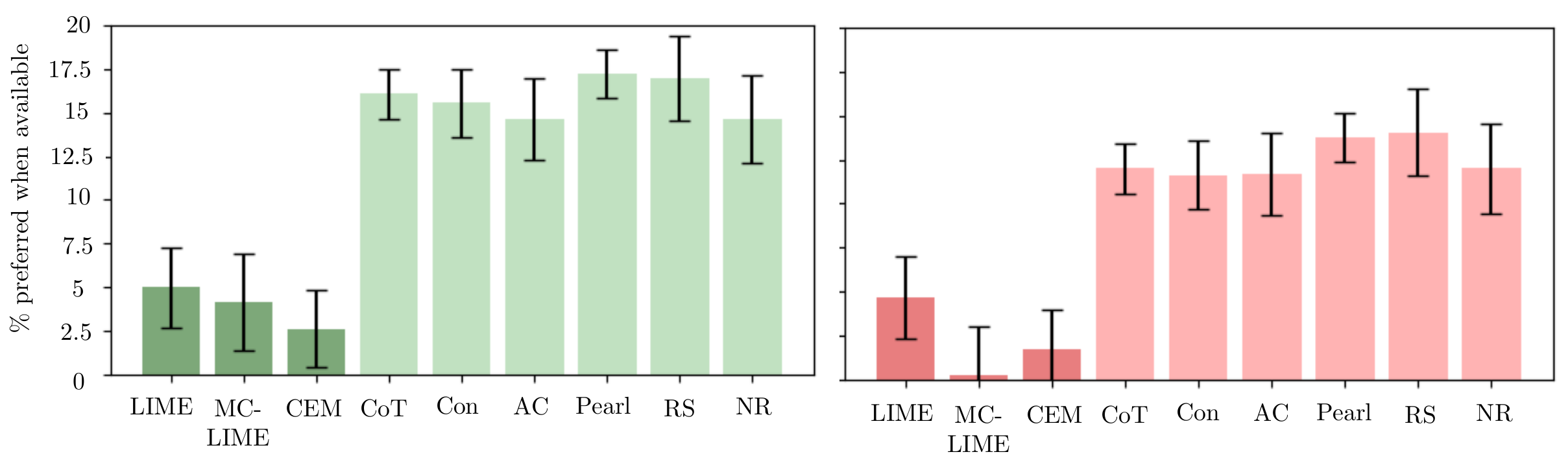}
    \caption{\textbf{Student preference of presented explanations for passing (left) and failing (right) student predictions.} Percentage of times a student chose each method when it was available. Higher scores are better.}
    \vspace{-3.5mm}
    \label{fig:exp3-studentpref}
\end{figure}

\begin{figure}[t!]
    \centering
    \includegraphics[width=\linewidth, trim={0 10 0 0},clip]{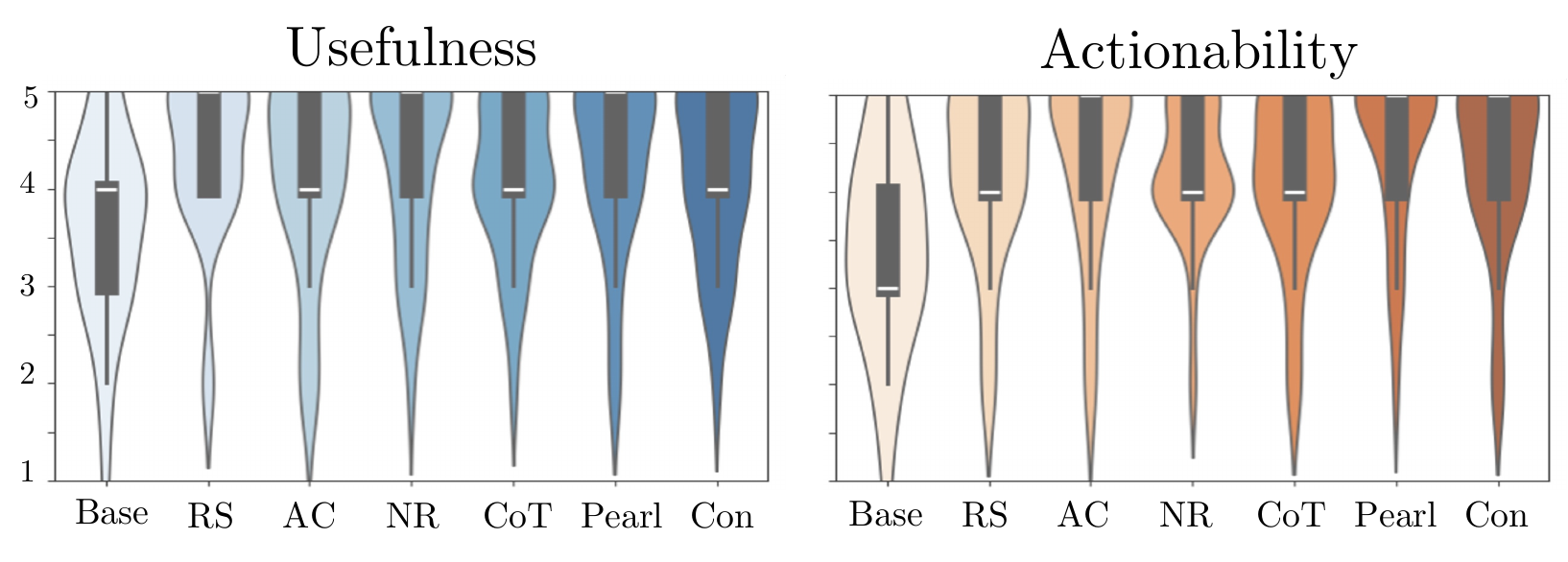}
    \caption{\textbf{Comparison of student preferences across two dimensions.} Distribution of Likert scores for \textit{Usefulness} and \textit{Actionability}, averaged over all participants and explainers. }
    \vspace{-5mm}
    \label{fig:exp3-criteria}
\end{figure}

\subsection{Exp 3: Students prefer \il} 
For each explanation in the user study, participants were asked to indicate their preferred explanation variation between one base explanation (LIME, CEM, MC-LIME) and three \il explanations. Figure~\ref{fig:exp3-studentpref} illustrates the percentage of times each method was preferred,
separately for either passing or failing student performance. Scores were averaged over all participants and explainers. Students overwhelmingly favored \il explanations over base ones. A Kruskal-Wallis test confirmed this preference, showing a significant difference between the base and \il explanations ($H=176.38, p<.0001$).

We evaluated students' responses to five Likert-scale questions on the usefulness, trustworthiness, actionability, completeness, and conciseness of explanations. Figure~\ref{fig:exp3-criteria} shows the Likert score distribution ($1$ - $5$) for two criteria per method; full results are in Appendix \ref{appendix:study-results}. Theory-based explanations received consistently high scores with no significant differences between theories. Base explanations were rated substantially lower (e.g., Usefulness${_\text{contrastive}}=4.18$, Usefulness${_\text{BASE}}=3.59$), highlighting the superior usefulness of \il explanations.

\vspace{-1mm}
\subsection{Exp 4: \il explanations can be effective at improving student performance} 
\label{sec:actionability}
Participants were asked to choose an action for the next week based on their preferred explanation. Over all explainers and theories, students most frequently selected actions to improve regularity of learning ($200$ responses) and attempt more problems ($147$ responses), while the least chosen action was to speed up quiz solving ($10$ responses). Participants also chose which weeks to focus on in the course, most commonly choosing weeks $6$ and $7$ ($329$, $248$ responses), which correspond to the weeks directly after the intervention. These were followed by review in weeks $5$ ($222$ responses) and $4$ ($198$ responses). This observation indicates that timing and proximity to the intervention influenced their choices. We also conducted a simulation experiment, applying participants' actions chosen for week 6 to student behavior in that week. For participants preferring \il responses, average performance improved significantly, independent of the underlying explainer: $13.5\%$ for LIME, $14.2\%$ for CEM, and $20.7\%$ for MC-LIME (full results in Appendix \ref{appendix:actionability}). MC-LIME’s effectiveness may stem from its minimal counterfactual approach, which greedily searches for the smallest set of features that cause the prediction to flip, making it suited for a single- or few-action intervention. Across different theories, both Contrastive and Necessity Robustness explanations result in the most actionable interventions with 28.2\% and 24.9\% average performance improvement respectively (Appendix Fig. \ref{fig:actionability}).

\section{Summary and Outlook}
In this work, we addressed the critical need for human-understandable explanations of complex models in education. We introduced \il, a theory-driven framework leveraging LLMs for generating NLEs through a chain of prompts, consisting of causal connection, explanation selection, and explanation presentation.

We tested our framework on $315$ students ($105$ per course) with all combinations of three post-hoc explainers, three LLMs, and eight prompting strategies, resulting in $21,915$ generated NLEs. We evaluated the instruction-following abilities using GPT-4o and human expert annotation, with both decomposed questions and readability metrics.
In a user study with $114$ university students, we found that students significantly preferred \il explanations over LIME, CEM, and MC-LIME and were able to derive actions that could improve their performance. 

Several challenges remain, including explainer variability and the difficulty of evaluating NLEs independently of prior beliefs and knowledge. In future work, we aim to explore interactivity of explanation dialogue with students \cite{slack2023explaining} and longitudinal LLM-XAI feedback effects. Our study highlights the shared potential of leveraging LLMs, eXplainable AI, and social science theories together towards scalable, personalized student support.

\section{Acknowledgments}
We kindly thank the Swiss State Secretariat for Education, Research and Innovation (SERI) for supporting this project. Oana-Maria Camburu was supported by a Leverhulme Early Career Fellowship. 

\bibliography{main}

\begin{thebibliography}{73}
\providecommand{\natexlab}[1]{#1}

\bibitem[{Abadi et~al.(2016)Abadi, Barham, Chen, Chen, Davis, Dean, Devin, Ghemawat, Irving, Isard et~al.}]{abadi2016tensorflow}
Abadi, M.; Barham, P.; Chen, J.; Chen, Z.; Davis, A.; Dean, J.; Devin, M.; Ghemawat, S.; Irving, G.; Isard, M.; et~al. 2016.
\newblock $\{$TensorFlow$\}$: A System for $\{$Large-Scale$\}$ Machine Learning.
\newblock In \emph{OSDI}.

\bibitem[{Asadi et~al.(2023)Asadi, Swamy, Frej, Vignoud, Marras, and Käser}]{asadi2023ripple}
Asadi, M.; Swamy, V.; Frej, J.; Vignoud, J.; Marras, M.; and Käser, T. 2023.
\newblock RIPPLE: Concept-Based Interpretation for Raw Time Series Models in Education.
\newblock \emph{AAAI}.

\bibitem[{Atanasova et~al.(2022)Atanasova, Simonsen, Lioma, and Augenstein}]{atanasova2022diagnostics}
Atanasova, P.; Simonsen, J.~G.; Lioma, C.; and Augenstein, I. 2022.
\newblock Diagnostics-guided explanation generation.
\newblock \emph{AAAI}.

\bibitem[{Baranyi, Nagy, and Molontay(2020)}]{baranyi2020interpretable}
Baranyi, M.; Nagy, M.; and Molontay, R. 2020.
\newblock Interpretable deep learning for university dropout prediction.
\newblock \emph{ITE}.

\bibitem[{Bender et~al.(2021)Bender, Gebru, McMillan-Major, and Shmitchell}]{bender2021dangers}
Bender, E.~M.; Gebru, T.; McMillan-Major, A.; and Shmitchell, S. 2021.
\newblock On the dangers of stochastic parrots: Can language models be too big?
\newblock In \emph{FAACT}.

\bibitem[{Boroujeni et~al.(2016)Boroujeni, Sharma, Kidzi{\'n}ski, Lucignano, and Dillenbourg}]{boroujeni2016quantify}
Boroujeni, M.~S.; Sharma, K.; Kidzi{\'n}ski, {\L}.; Lucignano, L.; and Dillenbourg, P. 2016.
\newblock How to quantify student’s regularity?
\newblock In \emph{ECTEL}.

\bibitem[{Cambria et~al.(2023)Cambria, Malandri, Mercorio, Mezzanzanica, and Nobani}]{cambria2023survey}
Cambria, E.; Malandri, L.; Mercorio, F.; Mezzanzanica, M.; and Nobani, N. 2023.
\newblock A survey on XAI and natural language explanations.
\newblock \emph{Information Processing \& Management}.

\bibitem[{Chen and Cui(2020)}]{chen2020utilizing}
Chen, F.; and Cui, Y. 2020.
\newblock Utilizing Student Time Series Behaviour in Learning Management Systems for Early Prediction of Course Performance.
\newblock \emph{JLA}.

\bibitem[{Cohausz(2022)}]{2022.EDM-short-papers.33}
Cohausz, L. 2022.
\newblock Towards Real Interpretability of Student Success Prediction Combining Methods of {XAI} and Social Science.
\newblock In \emph{EDM}. Durham, United Kingdom.

\bibitem[{Conati, Porayska-Pomsta, and Mavrikis(2018)}]{conati2018ai}
Conati, C.; Porayska-Pomsta, K.; and Mavrikis, M. 2018.
\newblock AI in Education needs interpretable machine learning: Lessons from Open Learner Modelling.
\newblock \emph{ICML WHI}.

\bibitem[{Danks(2022)}]{danks2022governance}
Danks, D. 2022.
\newblock Governance via explainability.
\newblock In \emph{The Oxford Handbook of AI Governance}.

\bibitem[{Deng et~al.(2024)Deng, Zhang, Chen, and Gu}]{deng2024rephrase}
Deng, Y.; Zhang, W.; Chen, Z.; and Gu, Q. 2024.
\newblock Rephrase and Respond: Let Large Language Models Ask Better Questions for Themselves.
\newblock \emph{arXiv}.

\bibitem[{Dhurandhar et~al.(2018)Dhurandhar, Chen, Luss, Tu, Ting, Shanmugam, and Das}]{dhurandhar2018explanations}
Dhurandhar, A.; Chen, P.-Y.; Luss, R.; Tu, C.-C.; Ting, P.; Shanmugam, K.; and Das, P. 2018.
\newblock Explanations based on the Missing: Towards Contrastive Explanations with Pertinent Negatives.
\newblock \emph{NeurIPS}.

\bibitem[{Flesch(1948)}]{flesch1948new}
Flesch, R. 1948.
\newblock A new readability yardstick.
\newblock \emph{Journal of App. Psychology}.

\bibitem[{Frej et~al.(2024)Frej, Shah, Knezevic, Nazaretsky, and K{\"a}ser}]{frej2024finding}
Frej, J.; Shah, N.; Knezevic, M.; Nazaretsky, T.; and K{\"a}ser, T. 2024.
\newblock Finding Paths for Explainable MOOC Recommendation: A Learner Perspective.
\newblock In \emph{LAK}.

\bibitem[{Galici et~al.(2023)Galici, K\"aser, Fenu, and Marras}]{galici2023not}
Galici, R.; K\"aser, T.; Fenu, G.; and Marras, M. 2023.
\newblock Do not trust a model because it is confident: Uncovering and characterizing unknown unknowns to student success predictors in online-based learning.
\newblock In \emph{LAK}.

\bibitem[{Graves and Schmidhuber(2005)}]{graves2005framewise}
Graves, A.; and Schmidhuber, J. 2005.
\newblock Framewise phoneme classification with bidirectional LSTM networks.
\newblock In \emph{IEEE IJCNN}.

\bibitem[{Grice(1975)}]{grice1975logic}
Grice, H.~P. 1975.
\newblock Logic and conversation.
\newblock \emph{Speech acts}.

\bibitem[{Gunning(1952)}]{gunning1952technique}
Gunning, R. 1952.
\newblock The technique of clear writing.

\bibitem[{Halpern and Pearl(2005)}]{halpern2005causes}
Halpern, J.~Y.; and Pearl, J. 2005.
\newblock Causes and explanations: A structural-model approach. Part II: Explanations.
\newblock \emph{BJPS}.

\bibitem[{Hang, Tan, and Yu(2024)}]{hang2024mcqgen}
Hang, C.~N.; Tan, C.~W.; and Yu, P.-D. 2024.
\newblock MCQGen: A Large Language Model-Driven MCQ Generator for Personalized Learning.
\newblock \emph{IEEE Access}.

\bibitem[{Hasib et~al.(2022)Hasib, Rahman, Hasnat, and Alam}]{hasib2022lime}
Hasib, K.~M.; Rahman, F.; Hasnat, R.; and Alam, M. G.~R. 2022.
\newblock A Machine Learning and Explainable \text{AI}Approach for Predicting Secondary School Student Performance.
\newblock In \emph{CCC}.

\bibitem[{Hattie and Timperley(2007)}]{hattie2007power}
Hattie, J.; and Timperley, H. 2007.
\newblock The power of feedback.
\newblock \emph{Review of educational research}.

\bibitem[{Heffernan and Heffernan(2014)}]{heffernan2014assistments}
Heffernan, N.~T.; and Heffernan, C.~L. 2014.
\newblock The ASSISTments ecosystem: Building a platform that brings scientists and teachers together for minimally invasive research on human learning and teaching.
\newblock \emph{IJAIED}.

\bibitem[{Hesslow(1988)}]{hesslow1988problem}
Hesslow, G. 1988.
\newblock The problem of causal selection.
\newblock \emph{Contemporary science and natural explanation: Commonsense conceptions of causality}.

\bibitem[{Hilton(1990)}]{hilton1990conversational}
Hilton, D.~J. 1990.
\newblock Conversational processes and causal explanation.
\newblock \emph{Psychological Bulletin}.

\bibitem[{Hilton and Slugoski(1986)}]{hilton1986knowledge}
Hilton, D.~J.; and Slugoski, B.~R. 1986.
\newblock Knowledge-based causal attribution: The abnormal conditions focus model.
\newblock \emph{Psychological review}.

\bibitem[{Holstein et~al.(2018)Holstein, Hong, Tegene, McLaren, and Aleven}]{holstein2018classroom}
Holstein, K.; Hong, G.; Tegene, M.; McLaren, B.~M.; and Aleven, V. 2018.
\newblock The classroom as a dashboard: Co-designing wearable cognitive augmentation for K-12 teachers.
\newblock In \emph{LAK}.

\bibitem[{Hume(2016)}]{hume2016enquiry}
Hume, D. 2016.
\newblock An enquiry concerning human understanding.
\newblock In \emph{Seven masterpieces of philosophy}. Routledge.

\bibitem[{Hur et~al.(2022)Hur, Lee, Bhat, and Bosch}]{hur2022using}
Hur, P.; Lee, H.; Bhat, S.; and Bosch, N. 2022.
\newblock Using Machine Learning Explainability Methods to Personalize Interventions for Students.
\newblock \emph{EDM}.

\bibitem[{Jacobsen and Weber(2023)}]{jacobsen2023promises}
Jacobsen, L.~J.; and Weber, K.~E. 2023.
\newblock The promises and pitfalls of ChatGPT as a feedback provider in higher education: An exploratory study of prompt engineering and the quality of AI-driven feedback.
\newblock \emph{OSF Preprints}.

\bibitem[{Josephson and Josephson(1996)}]{josephson1996abductive}
Josephson, J.~R.; and Josephson, S.~G. 1996.
\newblock \emph{Abductive inference: Computation, philosophy, technology}.
\newblock Cambridge University Press.

\bibitem[{Khosravi et~al.(2022)Khosravi, Shum, Chen, Conati, Tsai, Kay, Knight, Martinez-Maldonado, Sadiq, and Ga{\v{s}}evi{\'c}}]{khosravi2022explainable}
Khosravi, H.; Shum, S.~B.; Chen, G.; Conati, C.; Tsai, Y.-S.; Kay, J.; Knight, S.; Martinez-Maldonado, R.; Sadiq, S.; and Ga{\v{s}}evi{\'c}, D. 2022.
\newblock Explainable artificial intelligence in education.
\newblock \emph{CEAI}.

\bibitem[{Koh et~al.(2020)Koh, Nguyen, Tang, Mussmann, Pierson, Kim, and Liang}]{koh2020concept}
Koh, P.~W.; Nguyen, T.; Tang, Y.~S.; Mussmann, S.; Pierson, E.; Kim, B.; and Liang, P. 2020.
\newblock Concept bottleneck models.
\newblock In \emph{ICML}.

\bibitem[{Kroeger et~al.(2024)Kroeger, Ley, Krishna, Agarwal, and Lakkaraju}]{kroeger2024large}
Kroeger, N.; Ley, D.; Krishna, S.; Agarwal, C.; and Lakkaraju, H. 2024.
\newblock Are Large Language Models Post Hoc Explainers?
\newblock \emph{ArXiv}.

\bibitem[{Lall{\'e} and Conati(2020)}]{lalle2020data}
Lall{\'e}, S.; and Conati, C. 2020.
\newblock A data-driven student model to provide adaptive support during video watching across \text{MOOCs}.
\newblock In \emph{AIED}.

\bibitem[{Lipton(1990)}]{lipton1990contrastive}
Lipton, P. 1990.
\newblock Contrastive explanation.
\newblock \emph{Royal Institute of Philosophy Supplements}.

\bibitem[{Lombrozo(2006)}]{lombrozo2006structure}
Lombrozo, T. 2006.
\newblock The structure and function of explanations.
\newblock \emph{Trends in cognitive sciences}.

\bibitem[{Lombrozo(2010)}]{lombrozo2010causal}
Lombrozo, T. 2010.
\newblock Causal--explanatory pluralism: How intentions, functions, and mechanisms influence causal ascriptions.
\newblock \emph{Cognitive psychology}.

\bibitem[{Malle(2006)}]{malle2006mind}
Malle, B.~F. 2006.
\newblock \emph{How the mind explains behavior: Folk explanations, meaning, and social interaction}.
\newblock MIT press.

\bibitem[{Marras, Vignoud, and K\"aser(2021)}]{marras2021can}
Marras, M.; Vignoud, J. T.~T.; and K\"aser, T. 2021.
\newblock Can Feature Predictive Power Generalize? Benchmarking Early Predictors of Student Success across Flipped and Online Courses.
\newblock In \emph{EDM}.

\bibitem[{Mc~Laughlin(1969)}]{mc1969smog}
Mc~Laughlin, G.~H. 1969.
\newblock SMOG grading-a new readability formula.
\newblock \emph{Journal of reading}.

\bibitem[{Mejia-Domenzain et~al.(2022)Mejia-Domenzain, Marras, Giang, and K{\"a}ser}]{mejia2022identifying}
Mejia-Domenzain, P.; Marras, M.; Giang, C.; and K{\"a}ser, T. 2022.
\newblock Identifying and comparing multi-dimensional student profiles across flipped classrooms.
\newblock In \emph{AIED}.

\bibitem[{Miller(2019)}]{miller2019explanation}
Miller, T. 2019.
\newblock Explanation in artificial intelligence: Insights from the social sciences.
\newblock \emph{Artificial Intelligence}.

\bibitem[{Molnar(2020)}]{molnar2020interpretable}
Molnar, C. 2020.
\newblock Interpretable machine learning.

\bibitem[{Mousavinasab et~al.(2021)Mousavinasab, Zarifsanaiey, R.~Niakan~Kalhori, Rakhshan, Keikha, and Ghazi~Saeedi}]{mousavinasab2021intelligent}
Mousavinasab, E.; Zarifsanaiey, N.; R.~Niakan~Kalhori, S.; Rakhshan, M.; Keikha, L.; and Ghazi~Saeedi, M. 2021.
\newblock Intelligent tutoring systems: a systematic review of characteristics, applications, and evaluation methods.
\newblock \emph{ILE}.

\bibitem[{Mozgovoy(2011)}]{mozgovoy2011dependency}
Mozgovoy, M. 2011.
\newblock Dependency-based rules for grammar checking with LanguageTool.
\newblock In \emph{FedCSIS}.

\bibitem[{Mu, Jetten, and Brunskill(2020)}]{mu2020towards}
Mu, T.; Jetten, A.; and Brunskill, E. 2020.
\newblock Towards Suggesting Actionable Interventions for Wheel-Spinning Students.
\newblock \emph{EDM}.

\bibitem[{Nazaretsky et~al.(2022)Nazaretsky, Ariely, Cukurova, and Alexandron}]{nazaretsky2022teachers}
Nazaretsky, T.; Ariely, M.; Cukurova, M.; and Alexandron, G. 2022.
\newblock Teachers' trust in AI-powered educational technology and a professional development program to improve it.
\newblock \emph{BJET}.

\bibitem[{Nazaretsky et~al.(2024)Nazaretsky, Mejia-Domenzain, Swamy, Frej, and K{\"a}ser}]{nazaretsky2024ai}
Nazaretsky, T.; Mejia-Domenzain, P.; Swamy, V.; Frej, J.; and K{\"a}ser, T. 2024.
\newblock AI or Human? Evaluating Student Feedback Perceptions in Higher Education.
\newblock In \emph{ECTEL}.

\bibitem[{Neshaei et~al.(2024)Neshaei, Davis, Hazimeh, Lazarevski, Dillenbourg, and K{\"a}ser}]{neshaei2024towards}
Neshaei, S.~P.; Davis, R.~L.; Hazimeh, A.; Lazarevski, B.; Dillenbourg, P.; and K{\"a}ser, T. 2024.
\newblock Towards modeling learner performance with large language models.
\newblock \emph{AIED}.

\bibitem[{Nguyen, Tschiatschek, and Singla(2023)}]{nguyen2023large}
Nguyen, M.~H.; Tschiatschek, S.; and Singla, A. 2023.
\newblock Large Language Models for In-Context Student Modeling: Synthesizing Student's Behavior in Visual Programming from One-Shot Observation.
\newblock \emph{EDM}.

\bibitem[{Overton(2012)}]{overton2012explanation}
Overton, J.~A. 2012.
\newblock \emph{Explanation in Science}.
\newblock The University of Western Ontario (Canada).

\bibitem[{Peirce(1997)}]{peirce1997pragmatism}
Peirce, C.~S. 1997.
\newblock Pragmatism as a principle and method of right thinking.
\newblock \emph{The 1903 Harvard lectures on pragmatism}.

\bibitem[{Piech et~al.(2015)Piech, Bassen, Huang, Ganguli, Sahami, Guibas, and Sohl-Dickstein}]{piech2015deep}
Piech, C.; Bassen, J.; Huang, J.; Ganguli, S.; Sahami, M.; Guibas, L.~J.; and Sohl-Dickstein, J. 2015.
\newblock \text{Deep Knowledge Tracing}.
\newblock \emph{NeurIPS}, 28.

\bibitem[{Pinto and Paquette(2024)}]{pinto2024towards}
Pinto, J.~D.; and Paquette, L. 2024.
\newblock Towards a Unified Framework for Evaluating Explanations.
\newblock \emph{EDM HEXED}.

\bibitem[{Qin et~al.(2024)Qin, Song, Hu, Yao, Cho, Wang, Wu, Liu, Liu, and Yu}]{qin2024infobench}
Qin, Y.; Song, K.; Hu, Y.; Yao, W.; Cho, S.; Wang, X.; Wu, X.; Liu, F.; Liu, P.; and Yu, D. 2024.
\newblock InFoBench: Evaluating Instruction Following Ability in Large Language Models.
\newblock \emph{arXiv}.

\bibitem[{Ribeiro, Singh, and Guestrin(2016)}]{ribeiro2016why}
Ribeiro, M.~T.; Singh, S.; and Guestrin, C. 2016.
\newblock "Why Should I Trust You?": Explaining the Predictions of Any Classifier.
\newblock \emph{KDD}.

\bibitem[{Sarkar(2024)}]{sarkar2024large}
Sarkar, A. 2024.
\newblock Large Language Models Cannot Explain Themselves.
\newblock \emph{CHI HCXAI}.

\bibitem[{Scheers and De~Laet(2021)}]{scheers2021interactive}
Scheers, H.; and De~Laet, T. 2021.
\newblock Interactive and Explainable Advising Dashboard Opens the Black Box of Student Success Prediction.
\newblock In \emph{ECTEL}.

\bibitem[{Shute(2008)}]{shute2008focus}
Shute, V.~J. 2008.
\newblock Focus on formative feedback.
\newblock \emph{Review of educational research}.

\bibitem[{Slack et~al.(2023)Slack, Krishna, Lakkaraju, and Singh}]{slack2023explaining}
Slack, D.; Krishna, S.; Lakkaraju, H.; and Singh, S. 2023.
\newblock Explaining machine learning models with interactive natural language conversations using TalkToModel.
\newblock \emph{Nature Machine Intelligence}.

\bibitem[{Swamy et~al.(2023)Swamy, Du, Marras, and Käser}]{swamy2023trusting}
Swamy, V.; Du, S.; Marras, M.; and Käser, T. 2023.
\newblock Trusting the Explainers: Teacher Validation of Explainable Artificial Intelligence for Course Design.
\newblock \emph{LAK}.

\bibitem[{Swamy, Marras, and K{\"a}ser(2022)}]{swamy2022meta}
Swamy, V.; Marras, M.; and K{\"a}ser, T. 2022.
\newblock Meta transfer learning for early success prediction in MOOCs.
\newblock In \emph{L\@S}.

\bibitem[{Swamy et~al.(2024{\natexlab{a}})Swamy, Montariol, Blackwell, Frej, Jaggi, and K{\"a}ser}]{swamy2024interpretcc}
Swamy, V.; Montariol, S.; Blackwell, J.; Frej, J.~A.; Jaggi, M.; and K{\"a}ser, T. 2024{\natexlab{a}}.
\newblock InterpretCC: Intrinsic user-centric interpretability through global mixture of experts.
\newblock \emph{ArXiv}.

\bibitem[{Swamy et~al.(2022)Swamy, Radmehr, Krco, Marras, and Käser}]{swamy2022evaluating}
Swamy, V.; Radmehr, B.; Krco, N.; Marras, M.; and Käser, T. 2022.
\newblock Evaluating the Explainers: Black-Box Explainable Machine Learning for Student Success Prediction in MOOCs.
\newblock \emph{EDM}.

\bibitem[{Swamy et~al.(2024{\natexlab{b}})Swamy, Satayeva, Frej, Bossy, Vogels, Jaggi, K{\"a}ser, and Hartley}]{swamy2024multimodn}
Swamy, V.; Satayeva, M.; Frej, J.; Bossy, T.; Vogels, T.; Jaggi, M.; K{\"a}ser, T.; and Hartley, M.-A. 2024{\natexlab{b}}.
\newblock Multimodn—multimodal, multi-task, interpretable modular networks.
\newblock \emph{NeurIPS}.

\bibitem[{Van~Bouwel and Weber(2002)}]{van2002remote}
Van~Bouwel, J.; and Weber, E. 2002.
\newblock Remote causes, bad explanations?
\newblock \emph{Journal for the Theory of Social Behaviour}.

\bibitem[{Wang et~al.(2023)Wang, Liang, Meng, Sun, Shi, Li, Xu, Qu, and Zhou}]{wang2023chatgpt}
Wang, J.; Liang, Y.; Meng, F.; Sun, Z.; Shi, H.; Li, Z.; Xu, J.; Qu, J.; and Zhou, J. 2023.
\newblock Is chatgpt a good nlg evaluator? a preliminary study.
\newblock \emph{arXiv}.

\bibitem[{Wei et~al.(2023)Wei, Wang, Schuurmans, Bosma, Ichter, Xia, Chi, Le, and Zhou}]{wei2023chainofthought}
Wei, J.; Wang, X.; Schuurmans, D.; Bosma, M.; Ichter, B.; Xia, F.; Chi, E.; Le, Q.; and Zhou, D. 2023.
\newblock Chain-of-Thought Prompting Elicits Reasoning in Large Language Models.
\newblock \emph{arXiv}.

\bibitem[{Woodward and Ross(2021)}]{sep-scientific-explanation}
Woodward, J.; and Ross, L. 2021.
\newblock {Scientific Explanation}.
\newblock In \emph{The {Stanford} Encyclopedia of Philosophy}.

\bibitem[{Yu et~al.(2023)Yu, Lu, Zhong, Yao, Tu, Liao, Li, Li, Hou, Zheng et~al.}]{yu2023moocradar}
Yu, J.; Lu, M.; Zhong, Q.; Yao, Z.; Tu, S.; Liao, Z.; Li, X.; Li, M.; Hou, L.; Zheng, H.-T.; et~al. 2023.
\newblock Moocradar: A fine-grained and multi-aspect knowledge repository for improving cognitive student modeling in moocs.
\newblock In \emph{SIGIR}.

\bibitem[{Zytek, Pid{\`o}, and Veeramachaneni(2024)}]{zytek2024llms}
Zytek, A.; Pid{\`o}, S.; and Veeramachaneni, K. 2024.
\newblock LLMs for XAI: Future Directions for Explaining Explanations.
\newblock \emph{arXiv preprint arXiv:2405.06064}.

\end{thebibliography}
\vspace{14.5cm}
\onecolumn
\section{Reproducibility Details}
\label{appendix:reproducibility}

All code to replicate our student models and explainability results are provided in our repository: \texttt{https://github.com/epfl-ml4ed/iLLuMinaTE}. Any details not covered in the following reproducibility notes can also be found directly in the repository.

\vspace{1mm} \noindent \textbf{Student Model.} We input 45 student features in line with the exact extraction procedure from \cite{marras2021can, swamy2022meta} into a neural network architecture consisting of a \emph{Masking} layer (with a mask value of $-1$), followed by two Bidirectional LSTM (BiLSTM) layers with 64 and 32 units, respectively, and a loopback of $3$. The final layer is a \emph{Dense} layer with a Sigmoid activation function, which has a single output unit. This model predicts the probability that a student will fail the course. \citet{swamy2022meta} find that a BiLSTM architecture outperforms both traditional machine learning models (e.g., Support Vector Machines, Logistic Regression, Random Forest) and other deep learning models (e.g., Dense Fully-Connected Networks, RNNs, LSTMs, CNNs, and BiLSTMs) in the context of student performance prediction for online courses. They report a 26.8\% increase over traditional models (as measured by validation set for the DSP course). In our grid search for the best BiLSTM architecture, we tested various hidden size configurations, including \{32, 64, 128, 256, 32-32, 64-32, 128-32, 128-64, 128-128, 256-128, 64-64-32, 128-64-32, 64-64-64-32-32\}. The 64-32 configuration achieved the highest balanced accuracy for predicting outcomes in the \textit{DSP 1} course. We used the TensorFlow library \cite{abadi2016tensorflow} to train our models. Model training took approximately 15 minutes per model on a c2-standard-8 machine on Google Cloud (Linux Debian OS, 8vCPUs, 32 GB of RAM, 150 GB of disk). Each model was trained for 20 epochs, and the best performing model checkpoint was saved.

\vspace{1mm} \noindent \textbf{LIME.} The explanation highlights up to the 20 most influential features. The \texttt{mode} is set to \texttt{'classification'}, indicating that the function is tailored for the binary student performance prediction task. \texttt{discretize\_continuous} is \texttt{True}, meaning continuous features will be discretized to make the explanations more interpretable. The \texttt{num\_samples} parameter is set to 5000, specifying the number of perturbed samples LIME will generate to approximate the local decision boundary of the model. The \texttt{distance\_metric} is set to \texttt{'euclidean'}, which is used to measure the distance between the original instance and the perturbed samples. Lastly, the \texttt{sampling\_method} is set to \texttt{'gaussian'}, indicating that the perturbed samples will be drawn from a Gaussian distribution. The LIME setting is the exact same as used in \citet{swamy2022evaluating, swamy2023trusting}, but is expanded to select more features than the default (20 instead of 10) to not over-narrow the search space for MC-LIME.

\vspace{1mm} \noindent \textbf{MC-LIME.} The \texttt{init\_processing} function preprocesses the LIME results dataset by extracting feature names and reformatting the DataFrame. The \texttt{calculate\_std\_dev\_step\_sizes} function computes step sizes for feature modification based on a fraction of each feature's standard deviation, ensuring that changes applied to features during MC-LIME are proportionate. In the \texttt{get\_lime\_results} function, LIME important features and corresponding student-specific features are extracted, and step sizes are calculated to assess the sensitivity of model predictions. The \texttt{MC\_LIME} function iteratively modifies student features in small steps (guided by calculated step sizes) within specified bounds (0 to 1) to determine the minimal set of feature changes that could alter the model's prediction. The key hyperparameters in this procedure include the fraction of standard deviation used for step sizes, a prediction threshold (defaulting to 0.5), and the maximum group size for feature combinations (defaulting to 3), which controls the complexity and granularity of the search for influential features. The script ensures that feature modifications are meaningful and within realistic bounds, aiming to identify the minimal changes required to change a student's predicted outcome.

\vspace{1mm} \noindent \textbf{CEM.} The \texttt{pn\_all} function generates pertinent negative (PN) explanations for a set of instances using the Contrastive Explanation Method (CEM) applied to our underlying BiLSTM model. The function is configured with several key hyperparameters: \texttt{kappa} (set to 0.0) controls the minimum probability difference needed to minimize the first loss term, \texttt{beta} (0.1) weights the L1 loss term, and \texttt{gamma} (100) weights the optional auto-encoder loss term. The initial weight \texttt{c\_init} (1.0) influences the loss term that encourages a prediction different from the original class, with \texttt{c\_steps} (10) updates and \texttt{max\_iterations} (1000) iterations per value of \texttt{c}. The \texttt{feature\_range} is set based on the minimum and maximum values of the features, and gradient clipping is applied within the range \texttt{clip} (-1000.0, 1000.0). The learning rate \texttt{lr} is initialized to 0.01, and \texttt{no\_info\_val} is set to -1.0, representing a non-informative value for prediction. The function iterates over the specified instances, reshaping the features and generating explanations through CEM, with results stored in \texttt{changes}, \texttt{explanations}, and \texttt{final\_num\_instances}. Errors during processing are handled and logged, ensuring robustness in explanation generation. The CEM setting is the exact same as used in \citet{swamy2022evaluating, swamy2023trusting}.

\vspace{1mm} \noindent \textbf{LLMs.} We use three LLMs through these experiments: GPT-4o (gpt-4o-2024-05-13), Gemma2 9b (google/gemma-2-9b-it, July 2024), and Llama3 70B (meta/meta-llama-3-70b-instruct, April 18 2024 release). We use the instruct variations of each of these models, using the OpenAI API\footnote{https://platform.openai.com/} for GPT-4o and Groq Cloud\footnote{https://console.groq.com/} for Gemma2 and Replicate \footnote{https://replicate.com/} for Llama3. The visualizations for the user study (Sec. \ref{sec:student-preferences}, Appendix \ref{appendix:study-materials}) were generated through the ChatGPT user interface with underlying GPT-4o. It can also be generated through running LLM code snippets, but we chose to use the ChatGPT interface to remain closest to an actual teacher's workflow (using models as presented with a zero-shot chain-of-prompts, ensuring minimal post-processing steps).

\vspace{1mm} \noindent \textbf{Dataset.} 
Our objective is to predict student success during the initial weeks of three massive open online courses (MOOCs) by analyzing students' clickstream data. The log data collected from students includes detailed interactions with videos (such as play, pause, fast forward, and seek actions) as well as quiz-related events (such as submissions). To ensure the protection of student privacy, all student data is fully anonymized and the dataset is kept confidential, as required by the ethical guidelines outlined in HREC 058-2020/10.09.2020 and HREC 096-2020/09.04.2022. 

Formally, given the interactions $I_s$ generated by students $S$ up until course week $w$, we construct a matrix $H \subset \mathbb{R}^{|S| \times w \times f}$. This indicates that each feature in the feature set is computed for each student on a weekly basis, where $f \in \mathbb{N}$ represents the dimensionality of the feature set. We extract these features for each student $s$ and concatenate the feature sets to produce the final combined behavioral feature vector $h_s$ for each student. The complete matrix of features is defined as $H \in \mathbb{R}^{|S| \times w \times 42}$, with $H = [H_1 \cdot H_2 \cdot H_3 \cdot H_4]$ (where $\cdot$ represents concatenation). Due to the varying scales of the features, we apply min-max normalization to each feature in $H$, scaling them between 0 and 1, considering all students and weeks for that feature. The appendix table in \cite{swamy2022meta} lists all the features used in this dataset. Detailed statistics for each course can be found in Table \ref{tab:EDU_course_stats}. 

\begin{table}[h!]
\centering
\resizebox{\textwidth}{!}{%
\begin{tabular}{@{}llcccccc@{}}
\toprule
\textbf{Title} & \textbf{Identifier} & \textbf{Topic} & \textbf{Level} & \textbf{Language} & \textbf{No.\ Weeks} & \textbf{No.\ Students} & \textbf{Passing Rate [\%]}  \\ \midrule
Digital Signal Processing & DSP &  CS & MSc & English & 10 & 4,012 & 23.1 \\
Éléments de Géomatique & Geo & Math & BSc & French & 11 & 452 & 45.1 \\
Villes Africaines & VA &  SS & BSc & En/Fr & 12 & 5,643 & 9.9 \\
\bottomrule
\end{tabular}}
\vspace{2mm}
\caption{Course Details and Statistics.\\
Topic abbreviations: Math: Mathematics; NS: Natural Science; CS: Computer Science; SS: Social Science; Arch: Architecture; Bus: Economics and Business.}
\label{tab:EDU_course_stats}
\end{table}

\section{Prompting Framework}

In this Appendix, we present the explanation selection prompts (Sec. \ref{sec:explanation-selection}), explanation presentation prompts (Sec. \ref{sec:explanation-presentation}), and explanation visualization prompts (used for the hybrid explanation user study in Sec. \ref{sec:student-preferences}). We then discuss the development procedure of these prompts, and the exact frameworks and models used such that the results can be replicated.
\label{appendix:prompts}
\subsection{Explanation Selection Prompts}
\label{appendix:exp-sel-prompt}

The explanation selection template structure is highlighted in Section \ref{sec:explanation-selection}, and detailed again here. We provide an example of the model description, features description, explainer description (CEM), and course description (DSP 1). The details of the theory instructions are included below for each prompting strategy. 
\begin{invsummary}
\subsubsection*{Explanation selection template}
You are an AI assistant that analyzes struggling students behavior to help them in their learning trajectories and facilitate learning in the best possible way. 
You have the following information to help you in your goal:
\begin{itemize}
    \item A model prediction of student performance at the end of the course, in the form of “pass” or “fail”.
    \item A post-hoc explainable AI approach that identifies which features are important to this student’s prediction.
    \item Data in the form of student's features over 5 weeks that were used by the model. You will see always the most relevant features selected by \textcolor{brandred}{\textit{\{explainer\}}} \textcolor{blue}{CEM}.
    \item Data in the form of student's features over 5 weeks that were used by the model. You will see always the most relevant features selected by \textcolor{brandred}{\textit{\{explainer\}}} \textcolor{blue}{CEM}.
    \item The course \textcolor{brandred}{\textit{\{course\_name\}}} \textcolor{blue}{Digital Signal Processing 1} content and structure.
    \item Detailed instructions on how to reason.

\end{itemize}

\vspace{1em}

\textcolor{brandred}{\textit{\{model\_description\}}}
\textcolor{blue}{The model you are using is a recurrent neural network that is trained on predicting the student performance at the end of the course, in the form of “pass” or “fail”. The features of that the model are using are derived from student behavior:\\}

\end{invsummary}

\begin{invsummary}

\textcolor{brandred}{\textit{\{features\_description\}}} \\ \\
\textcolor{blue}{\textbf{delay\_lecture}: The average delay in viewing video lectures after they are released to students.}\\

\textcolor{blue}{\textbf{ratio\_clicks\_weekend\_day}: The ratio between the number of clicks on the weekend and the weekdays.}\\

\textcolor{blue}{\textbf{total\_clicks}: The number of clicks that a student has made overall.}\\

\textcolor{blue}{\textbf{total\_clicks\_problem}: The number of clicks that a student has made on problems this week.}\\

\textcolor{blue}{\textbf{total\_clicks\_video}: The number of clicks that a student has made on videos this week. The number of times a student clicked on a video (load, pause, play, forward).}\\

\textcolor{blue}{\textbf{total\_clicks\_weekday}: The number of clicks that a student has made on the weekdays.}\\

\textcolor{blue}{\textbf{total\_clicks\_weekend}: The number of clicks that a student has made on the weekends.}\\

\textcolor{blue}{\textbf{competency\_strength}: The extent to which a student passes a quiz getting the maximum grade with few attempts.}\\

\textcolor{blue}{\textbf{competency\_alignment}: The number of problems this week that the student has passed.}\\

\textcolor{blue}{\textbf{content\_alignment}: The number of videos this week that have been watched by the student.}\\

\textcolor{blue}{\textbf{competency\_anticipation}: The extent to which the student approaches a quiz provided in subsequent weeks.}\\

\textcolor{blue}{\textbf{content\_anticipation}: The number of videos covered by the student from those that are in subsequent weeks.}\\

\textcolor{blue}{\textbf{student\_speed}: The average time passed between two consecutive attempts for the same quiz.}\\

\textcolor{blue}{\textbf{student\_shape}: The extent to which the student receives the maximum quiz grade on the first attempt.}\\

\textcolor{blue}{\textbf{regularity\_periodicity\_m1}: The extent to which the hourly pattern of the user's activities repeats over days.}\\

\textcolor{blue}{\textbf{regularity\_peak\_dayhour}: The extent to which students' activities are centered around a particular hour of the day.}\\

\textcolor{blue}{\textbf{number\_sessions}: The number of unique online sessions the student has participated in.}\\

\textcolor{blue}{\textbf{time\_sessions\_mean}: The average of the student's time per session.}\\

\textcolor{blue}{\textbf{time\_sessions\_sum}: The sum of the student's time in sessions.}\\

\textcolor{blue}{\textbf{time\_sessions\_std}: The standard deviation of the student's time in sessions.}\\

\textcolor{blue}{\textbf{time\_between\_sessions\_std}: The standard deviation of the time between sessions of each user.}\\

\textcolor{blue}{\textbf{time\_in\_problem\_sum}: The total (cumulative) time that a student has spent on problem events.}\\

\textcolor{blue}{\textbf{time\_in\_video\_sum}: The total (cumulative) time that a student has spent on video events.}\\

\textcolor{blue}{\textbf{total\_clicks\_Video}: The number of clicks that a student has made on videos this week. The number of times a student clicked on a video (load, pause, play, forward).}\\

\textcolor{blue}{\textbf{total\_clicks\_Video.Load}: The number of times a student loaded a video.}\\

\textcolor{blue}{\textbf{frequency\_action\_Video.Load}: The frequency between every \texttt{Video.Load} action and the following action.}\\

\textcolor{blue}{\textbf{weekly\_prop\_watched\_mean}: The ratio of videos watched over the number of videos available.}\\

\textcolor{blue}{\textbf{weekly\_prop\_replayed\_mean}: The ratio of videos replayed over the number of videos available.}\\

\end{invsummary}

\begin{invsummary}

\textcolor{brandred}{\textit{\{explainer\_description\}}}\\ \\
\textcolor{blue}{We use CEM Counterfactuals as our explainable AI approach, which finds the smallest number of changes necessary to change a prediction from student failure to student success. The output is the minimal difference in the feature values that would change the prediction.\\}

\textcolor{brandred}{\textit{\{course\_description\}}}\\

\textcolor{blue}{The course the student is taking is Digital Signal Processing 1, which is a Master’s level course over 10 weeks under the topic of Electrical Engineering and Computer Science. This is the course content:\\}

\textcolor{blue}{\textbf{WEEK 1} \\
SKILLS: Digital Signals \\
TOPICS: Welcome to the DSP course, Introduction to signal processing \\}

\textcolor{blue}{\textbf{WEEK 2} \\
SKILLS: Digital Signals \\
TOPICS: Discrete time signals, The complex exponential, The Karplus-Strong Algorithm \\}

\textcolor{blue}{\textbf{WEEK 3} \\
SKILLS: Hilbert (Linear Alg.) \\
TOPICS: Motivation and Examples, From Euclid to Hilbert, Hilbert Space, properties and bases, Hilbert Space and approximation \\}

\textcolor{blue}{\textbf{WEEK 4} \\
SKILLS: DFT, DTFT DFS, FFT \\
TOPICS: Exploration via a change of basis, The Discrete Fourier Transform (DFT), DFT Examples, DFT, DFS, DTFT, DTFT formalism, Relationship between transforms, Sinusoidal modulation, FFT: history and algorithms \\}
 
\textcolor{blue}{\textbf{WEEK 5} \\
SKILLS: Ideal Filters, Filter Design \\
TOPICS: Linear Filters, Filtering by example, Filter stability, Frequency response, Ideal filters, Filter design - Part 1: Approximation of ideal filters, Realizable filters, Implementation of digital filters, Filter design - Part 2: Intuitive filters, Filter design - Part 3: Design from specs, Real-time processing, Dereverberation and echo cancelation \\}

\textcolor{blue}{\textbf{WEEK 6} \\
SKILLS: Modulation, Interpolation \& Sampling \\
TOPICS: Introduction to continuous-time paradigm, Interpolation, The space of bandlimited signals, Sampling and aliasing: Introduction, Sampling and aliasing, Discrete-time processing and continuous-time signals, Another example of sampled acquisition \\}

\textcolor{blue}{\textbf{WEEK 7} \\
SKILLS: Multirate \\
TOPICS: Stochastic signal processing, Quantization, A/D and D/A conversion \\}

\textcolor{blue}{\textbf{WEEK 8} \\
SKILLS: DFT, DTFT DFS, Ideal Filters \\
TOPICS: (Revisiting the topics of week 4 with additional context) Image processing, Image manipulations, Frequency analysis, Image filtering, Image compression, The JPEG compression algorithm \\}

\textcolor{blue}{\textbf{WEEK 9} \\
SKILLS: Modulation, Quantization \\
TOPICS: Digital communication systems, Controlling the bandwidth, Controlling the power, Modulation and demodulation, Receiver design, ADSL \\}

\textcolor{blue}{\textbf{WEEK 10} \\
SKILLS: Applications \\
TOPICS: The end, Startups and DSP, Acknowledgements, Conclusion video \\}
\end{invsummary}
    
\begin{invsummary}
\vspace{1em}

Take into consideration this data:\\

\textcolor{brandred}{\textit{\{explainer\_importance\_scores\}}}\\ 

\textcolor{blue}{\textbf{MODEL PREDICTION}: pass, with 74.538538\% of confidence.}\\

\textcolor{blue}{\textbf{FEATURE IMPORTANCES} \\ \\
These are the features found important by CEM: \\\\
number\_sessions\_InWeek5 - 0.466320 \\
regularity\_peak\_dayhour\_InWeek5 - 0.458857\\
number\_sessions\_InWeek4 - 0.390114\\
regularity\_peak\_dayhour\_InWeek4 - 0.386096\\
number\_sessions\_InWeek3 - 0.307134\\
frequency\_action\_Video.Load\_InWeek5 - 0.279141\\
number\_sessions\_InWeek2 - 0.271637\\
time\_sessions\_mean\_InWeek1 - 0.240877\\
time\_sessions\_mean\_InWeek2 - 0.237525\\
number\_sessions\_InWeek1 - 0.199099\\}

\textcolor{brandred}{\textit{\{student\_feature\_values\}}}\\

\textcolor{blue}{The relevant feature values found by \textcolor{brandred}{\{explainer\}} \textcolor{blue}{CEM} for the student are included below: \\}

\textcolor{blue}{'number\_sessions\_InWeek5': 0.0908182726364544, \\'regularity\_peak\_dayhour\_InWeek5': 0.0073937247167162, \\'number\_sessions\_InWeek4': 0.0, \\'regularity\_peak\_dayhour\_InWeek4': 0.0088844825597196,\\ 'number\_sessions\_InWeek3': 0.0, \\'frequency\_action\_Video.Load\_InWeek5': 0.0,\\ 'number\_sessions\_InWeek2': 0.0, \\'time\_sessions\_mean\_InWeek1': 0.0, \\'time\_sessions\_mean\_InWeek2': 0.0, \\'number\_sessions\_InWeek1': 0.0}

\vspace{1em}

INSTRUCTIONS:

\textcolor{brandred}{\textit{\{theory\_instructions\}}}

\textcolor{blue}{Use the social science theory ``Relevance-Based Selection" to derive key insights from the model prediction, explainer, and student feature values.}

\textcolor{blue}{\\
1. Select the causes that are most relevant to the question, context and user \\
2. Select the causes that include information that is not already shared with the student\\}
    
QUESTION: Given the information above, follow the instructions precisely and write a small report on what you found. Only use the results from the explainable AI approach and the student's behavior data to justify your conclusions.
\end{invsummary}

Other examples of \textcolor{brandred}{\{explainer\_instructions\}} include:\\

\textcolor{darkblue}{LIME: We use LIME as our explainable AI approach, which gives importance scores for 20 features that contributed the most to the prediction of the model. Positive scores positively contribute to the model's decision to reach the predicted outcome. Negative scores would push the prediction towards the opposite class. The magnitude of the score indicates the strength of the feature's contribution.\\}

\textcolor{darkblue}{MC\_LIME: We use Minimal Counterfactual LIME as our explainable AI approach, which finds the smallest number of changes necessary to change a prediction from student failure to student success. It uses LIME scores to select the features to change. The output are the sets of features with the new values that would change the prediction.\\}

We have included the social science instructions from each of the eight explanation theories (Relevance Selection, Abnormal Conditions, Pearl's Model of explanation, Necessity and Robustness selection, Base Contrastive Explanation, RaR + Contrastive Explanation, Statistical Relevance, and Chain-of-Thought).

\begin{invsummary}
\subsubsection*{Relevance-based selection prompt}
    \begin{enumerate}[leftmargin=*]
        \item Select the causes that are most relevant to the question, context and user
        \item Select the causes that include information that is not already shared with the student
    \end{enumerate}
\end{invsummary}
\vspace{-5mm}
\begin{invsummary}
\subsubsection*{Abnormal conditions model prompt}
    \begin{enumerate}[leftmargin=*]
        \item  Select potential causes using these criteria:\begin{itemize}
            \item \textit{Abnormality}: Tend to prefer abnormal causes.
            \item \textit{Temporality}: Recent events are more relevant for the user and considered more mutable.
            \item \textit{Controllability}: focus on the features that the student can control.
        \end{itemize}
        \item Select one explanation that follows all of the criteria above (Abnormality, Temporality, Controllability).
    \end{enumerate}
\end{invsummary}
\vspace{-5mm}

\begin{invsummary}
\subsubsection*{Pearl's model of explanation prompt}
    \begin{enumerate}[leftmargin=*]
        \item Identify variables that are determined by factors external to the model (\textit{exogenous variables}) and variables that are determined by relationships with other variables (\textit{endogenous variables}).
        \item Select a set of factors representing the explanation by following the criteria:\begin{itemize}
            \item The set of factors is minimal - there are no irrelevant causes in the set.
            \item A change in these factors' values would result in a reversed prediction.
        \end{itemize}
    \end{enumerate}
\end{invsummary}
\vspace{-5mm}

\begin{invsummary}
\subsubsection*{Necessity \& Robustness prompt}
    \begin{enumerate}[leftmargin=*]
        \item Define sets of factors that would explain the model’s output given the feature scores and the explainer scores.
        \item Select a few of these sets of factors that are the most necessary to cause the event.
        \item Robustness is defined as the extent to which the model’s output would still have occurred if conditions were somewhat different. Select an explanation that is the most robust.
    \end{enumerate}
\end{invsummary}
\vspace{-5mm}

\begin{invsummary}
\subsubsection*{Base Contrastive Explanation prompt}
    \begin{enumerate}[leftmargin=*]
        \item Reformulate the question as “Why P rather than Q?”
        \item Using the explainable AI importance scores, determine the causes that explain the difference between P and not-Q. Which causes are a cause of P, but are not present in the causal chain of Q?
    \end{enumerate}
\end{invsummary}
\vspace{-5mm}

\begin{invsummary}
\subsubsection*{RaR + Contrastive Explanation prompt}
    \begin{enumerate}[leftmargin=*]
        \item An explanation is always of the form “Why P rather than Q?”, in which P is the target event and Q is a counterfactual contrast case that did not occur. P is the fact (the event that did occur) and Q the foil (event that did not).
        \item  Select one foil based on two criterias: abnormality and relevance given the question, context, and user.
        \item Using the explainable AI importance scores, determine the causes that explains the difference between P and not-Q. Which causes are a cause of P, but are not present in the causal chain of Q?
    \end{enumerate}
\end{invsummary}
\vspace{-5mm}

\begin{invsummary}
\subsubsection*{SR prompt}
    \begin{enumerate}[leftmargin=*]
        \item 1. Select the most likely explanations for this model output. ``Based on the data, factors A, B and C contribute to the probability of Y''. 
        \item Complete this phrase with the corresponding factors of our data.
    \end{enumerate}
\end{invsummary}
\vspace{-5mm}

\begin{invsummary}
\subsubsection*{CoT prompt}
Let’s think step-by-step.
\end{invsummary}
\vspace{-5mm}

\subsubsection{\il Explanation Selection Examples}

We present an example for the same sample student from the DSP 1 course using the CEM explainer and the abnormal conditions explanation strategy with Gemma-2 9B, GPT-4o, and Llama-3 70B.

\begin{myabstract}
\textbf{Gemma-2 9B, DSP 1, CEM, Abnormal Conditions -- \il Selection Prompt}\\

\textbf{Analysis of Student Performance in Digital Signal Processing 1}\\

\textbf{Introduction:} This report analyzes the performance prediction of a student in the \textit{Digital Signal Processing 1} course using an explainable AI approach and student behavior data. The model predicts a "pass" with 99.74\% confidence.\\

\textbf{Applying the ``Abnormal Conditions" Theory:} The ``Abnormal Conditions" theory proposes that deviations from typical behavior patterns can indicate underlying issues. By applying this theory to the student's data, we aim to identify potential causes for their predicted success.\\

\textbf{Identifying Potential Causes:} The explainable AI approach (CEM) highlights several features as crucial for the prediction:

\begin{itemize}
    \item \textbf{Number of sessions in Week 5:} Significantly low (0.09) compared to the average.
    \item \textbf{Time sessions standard deviation in Week 5:} Zero, indicating a very consistent session length, which could be unusual.
    \item \textbf{Regularity peak day hour in Week 4:} Low (0.07) suggesting a lack of a strong peak time for activity.
    \item \textbf{Total clicks on videos in Week 5:} Low (0.04) compared to other features.\\
\end{itemize}

\textbf{Selected Explanation:} The most compelling explanation based on the "Abnormal Conditions" theory is the \textbf{low number of sessions in Week 5 (0.09)}. This deviates significantly from the average and suggests a potential issue with engagement during this crucial week.\\

\textbf{Justification:}

\begin{itemize}
    \item \textbf{Abnormality:} The low number of sessions is significantly below the expected range.
    \item \textbf{Temporality:} Week 5 is a recent period, making its impact more immediate and relevant to the current prediction.
    \item \textbf{Controllability:} The student can directly control the number of sessions they participate in.\\
\end{itemize}

\textbf{Conclusion:}
While the model predicts a high probability of success, the low number of sessions in Week 5 raises a concern. This could indicate a temporary dip in engagement or a potential learning barrier encountered during this week. Further investigation into the specific content and activities of Week 5 might reveal the underlying cause and allow for targeted interventions to ensure the student's continued success.

\end{myabstract}
\vspace{-5mm}
\begin{myabstract}
\textbf{GPT-4o, DSP 1, CEM, Abnormal Conditions -- \il Selection Prompt}\\

\textbf{Student Performance Analysis Report}\\

\textbf{Model Prediction:} The model predicts that the student will pass the course with a confidence of 99.74\%.\\

\end{myabstract}
\begin{myabstract}

\textbf{Key Insights from Explainable AI and Student Behavior Data:}

\begin{itemize}
    \item \textbf{Feature Importances (CEM):}
    \begin{enumerate}
        \item \textbf{number\_sessions\_InWeek5:} 1.000000
        \item \textbf{number\_sessions\_InWeek4:} 0.771101
        \item \textbf{time\_sessions\_std\_InWeek5:} 0.708540
        \item \textbf{regularity\_peak\_dayhour\_InWeek4:} 0.668573
        \item \textbf{total\_clicks\_Video\_InWeek5:} 0.647907
        \item \textbf{total\_clicks\_video\_InWeek5:} 0.598572
        \item \textbf{number\_sessions\_InWeek3:} 0.547558
        \item \textbf{regularity\_peak\_dayhour\_InWeek5:} 0.538825
        \item \textbf{total\_clicks\_Video.Load\_InWeek5:} 0.492897
        \item \textbf{time\_in\_video\_sum\_InWeek5:} 0.454281\\
    \end{enumerate}
    \item \textbf{Relevant Feature Values:}
    \begin{itemize}
        \item \textbf{number\_sessions\_InWeek5:} 0.090818
        \item \textbf{number\_sessions\_InWeek4:} 0.181637
        \item \textbf{time\_sessions\_std\_InWeek5:} 0.0
        \item \textbf{regularity\_peak\_dayhour\_InWeek4:} 0.068038
        \item \textbf{total\_clicks\_Video\_InWeek5:} 0.041874
        \item \textbf{total\_clicks\_video\_InWeek5:} 0.041874
        \item \textbf{number\_sessions\_InWeek3:} 0.166500
        \item \textbf{regularity\_peak\_dayhour\_InWeek5:} 0.102560
        \item \textbf{total\_clicks\_Video.Load\_InWeek5:} 0.041874
        \item \textbf{time\_in\_video\_sum\_InWeek5:} 0.017867\\
    \end{itemize}
\end{itemize}

\textbf{Analysis Using Abnormal Conditions Theory:}\\

\textbf{- Potential Causes of the Result:}
\begin{enumerate}
    \item \textbf{Abnormality:} 
    The student's number of sessions in Week 5 (0.090818) is significantly low compared to the importance placed on this feature. Additionally, the total clicks on videos in Week 5 (0.041874) are notably low.
    
    \item \textbf{Temporality:} 
    Recent weeks (Weeks 4 and 5) show more relevance for the student's current and future performance. Features from Week 5 are particularly critical as they are the most recent data points available.
    
    \item \textbf{Controllability:} 
    The student has control over their number of sessions and video clicks, making these features actionable for improving their performance.
\end{enumerate}

\textbf{- Explanation:} Based on the criteria of Abnormality, Temporality, and Controllability, the most significant insight is the student's low engagement in Week 5, particularly in terms of the number of sessions (0.090818) and total clicks on videos (0.041874), which are abnormally low.\\

\textbf{Conclusion:} The student's low engagement in Week 5 is a controllable and recent abnormality that significantly affects their predicted success. To improve their learning trajectory and maintain their predicted pass, the student should:
\begin{itemize}
    \item \textbf{Increase the number of sessions:} Aim to participate in more online sessions.
    \item \textbf{Increase video engagement:} Watch and interact with more video content.
\end{itemize}

By addressing these specific areas, the student can enhance their engagement, solidify their understanding of the course material, and increase their likelihood of success.

\end{myabstract}
\vspace{-8mm}
\begin{myabstract}
\textbf{Llama-3 70B, DSP 1, CEM, Abnormal Conditions -- \il Selection Prompt}\\

\textbf{Report:} Based on the model prediction, explainer, and student feature values, I applied the social science theory \textit{Abnormal Conditions} to derive key insights.\\

\textbf{Potential Causes of the Result:}\\

\begin{enumerate}
    \item \textbf{Abnormality:} The feature \texttt{time\_sessions\_std\_InWeek5} has a value of 0.0, which is an abnormal condition, indicating that the student's time spent in sessions during Week 5 has no standard deviation. This could be a cause of the model's prediction of "pass" with high confidence.
    \item \textbf{Temporality:} The features \texttt{number\_sessions\_InWeek5}, \texttt{total\_clicks\_Video\_InWeek5}, \texttt{total\_clicks\_video\_InWeek5}, and \texttt{time\_in\_video\_sum\_InWeek5} are all related to the student's behavior during Week 5, which is a recent event. These features are more relevant and mutable than earlier events.
    \item \textbf{Controllability:} The features \texttt{number\_sessions\_InWeek5}, \texttt{total\_clicks\_Video\_InWeek5}, and \texttt{total\_clicks\_video\_InWeek5} are all related to the student's engagement with the course content, which is within their control.\\
\end{enumerate}

\textbf{Selected Explanation:} Based on the criteria above, I select the explanation that the student's high engagement with the course content during Week 5, as indicated by the abnormal feature values, is a key cause of the model's prediction of "pass" with high confidence. Specifically, the student's high number of sessions (\texttt{number\_sessions\_InWeek5}) and clicks on videos (\texttt{total\_clicks\_Video\_InWeek5} and \texttt{total\_clicks\_video\_InWeek5}) during Week 5 suggest that they are actively participating in the course and making an effort to learn the material.\\

This explanation meets all the criteria: it is an abnormal condition (high engagement), it is a recent event (Week 5), and it is within the student's control (engagement with course content).

\end{myabstract}

Extended examples across different instruction strategies and base explainers can be found in our code repository.

\subsection{Explanation Presentation Prompts}
\label{appendix:exp-pres-prompt}

We use the framework of Hattie et al. \cite{hattie2007power} to present the explanation in the form of student feedback:

\begin{enumerate}
    \item Where Am I Going? - A brief description of the student’s performance and explicitly state the learning goal
    \item How Am I Doing? - A brief description of the explanation findings
    \item Where to Next? - Two recommended actions that the student can take that are specific to weeks of the course (make connections between previous weeks and upcoming weeks)
\end{enumerate}

We also include Grice's Maxims to guide the communication of the explanation \cite{grice1975logic}. Grice's maxims provide a framework for understanding cooperative conversation. They propose that communication is a collaborative effort, with participants adhering to the ``cooperative principle". This principle emphasizes providing the right amount of relevant and clear information to achieve a shared goal. Grice further divides this principle into four maxims: \begin{itemize}
    \item Quality (truthfulness and evidence-based statements)
    \item Quantity (providing enough but not excessive information)
    \item Relation (staying relevant)
    \item Manner (clear and concise communication)
\end{itemize}

Notably, Grice acknowledges that maxims can be strategically ignored to express meaning indirectly, as seen in irony or metaphors. We have expressed Grice's maxims in the prompt as follows:

\begin{enumerate}
    \item do not say things that you believe to be false
    \item do not say things for which you do not have sufficient evidence.
    \item do not add information that is not relevant
    \item only say what is relevant
    \item be orderly
\end{enumerate}

\begin{invsummary}
\subsubsection*{Explanation Presentation Structure}
Given this report, I want you to write a shorter version using the theory of feedback from Hattie et al.:\vspace{4mm}

\textbf{Where Am I Going? }- A brief description of the student's performance and explicitly state the learning goal\\
\textbf{How Am I Doing? }- A brief description of the explanation findings\\
\textbf{Where to Next? }- Two recommended actions that the student can take that are specific to weeks of the course (make connections between previous weeks and upcoming weeks)\\

The student who is going to interact with you is the same student that the data is about, so use the tone of a professor who is talking to a student in a comforting and personable manner. \\

Follow the instructions underneath in the INSTRUCTIONS section. \vspace{4mm}

\textbf{INSTRUCTIONS} In the explanation findings section (How am I going?) explicitly use the following structure:
\textcolor{brandred}{\textit{\{presentation\_instruction\}}}

\textcolor{brandred}{\textit{\{course\_description\}}} \\

5 weeks of the course have concluded.\\

Follow these rules:

\begin{itemize}
    \item do not include the output of the model or a prediction
    \item if you include a feature name, describe what it means
    \item try to be as concise as possible
    \item do not include the headers from Hattie et al. in the response, but keep the structure
    \item limit yourself to 200 words
    \item do not include a sign-off, simply include the main content
\end{itemize}

\vspace{4mm}
To communicate this intervention most effectively, use Grice's maxims of conversation.

\begin{itemize}
    \item do not say things that you believe to be false
    \item do not say things for which you do not have sufficient evidence.
    \item do not add information that is not relevant
    \item only say what is relevant
    \item be orderly
\end{itemize}

Your goal is to explain to the student what they should do to improve their performance in the course in the best way possible. Follow the instructions above.

\end{invsummary}

We use the following format template to guide the LLM's response into an easily post-processable format. 

\begin{invsummary}
\subsection*{Format Template}

\textcolor{brandred}{\textit{\{format\_instructions\}}}
Return a JSON file with a string with your feedback to the student. \\

\textcolor{brandred}{\textit{\{conversation\_template\}}}\\

Current conversation:\\
    \textcolor{brandred}{\textit{\{chat\_history\}}} \\
    \textcolor{brandred}{\textit{\{input\}}} \\

AI Assistant: 
\end{invsummary}

For the \textcolor{brandred}{\textit{\{presentation\_instruction\}}} portion of the prompt, we have the following instructions.

\begin{invsummary}
\subsubsection*{Relevance-Based Selection Presentation Prompt}
Relevant Causes: Say to the student which causes you selected as relevant based on the question, context, and your background. Be explicit.

New Information: Say to the student which information you thought they knew already, such as "Assuming that you know…" and highlight aspects that haven't been previously communicated to the student. Be clear and honest.

Finally: Say to the student that you focused on the most relevant causes that provide new insights. Specify which causes you selected and why they are important.
\end{invsummary}
\vspace{-5mm}
\begin{invsummary}
\subsubsection*{Abnormal Conditions Presentation Prompt}

Abnormal Causes: Tell the student which causes you selected as abnormal and explain why these are important based on the recent events and their uniqueness.

Recent Events: Highlight the recent events that are relevant to the result. Emphasize why these events are important for the student to consider.

Controllable Factors: Point out the factors that the student has control over. Explain how focusing on these factors can help them improve their outcomes. Be clear and direct.
\end{invsummary}
\vspace{-5mm}

\begin{invsummary}
\subsubsection*{Pearl's Explanation Presentation Prompt}
External Factors: Identify the external factors (exogenous variables) that influence the result. Explain why these factors are important and how they impact the outcome.

Internal Relationships: Point out the variables determined by their relationships with other variables (endogenous variables). Describe these relationships and their significance.

Focus on Improvement: Highlight any features that negatively impact the student’s success. Offer clear and actionable advice on how to improve these areas.
\end{invsummary}
\vspace{-5mm}

\begin{invsummary}
\subsubsection*{Necessity \& Robustness Presentation Prompt}

Necessary Causes: Identify the most necessary factors that explain the result. Explain why these factors are essential for understanding the outcome.

Robust Explanation: Describe the explanation that would still hold true even if conditions were slightly different. Highlight why this robust explanation is reliable and important.
\end{invsummary}
\vspace{-5mm}

\begin{invsummary}
\subsubsection*{Base Contrastive Presentation Prompt}

Target Event: Explain the event that did occur (P). Describe this event clearly and why it is significant.

Contrast Event: Identify the contrasting event that did not occur (Q). Explain why this event is relevant and why it was chosen as a contrast.

Key Differences: Highlight the main differences between the target event and the contrast event. Explain why these differences are important for understanding the result.
\end{invsummary}
\vspace{-5mm}

\begin{invsummary}
\subsubsection*{RaR + Contrastive Presentation Prompt}

Target Event (P): Describe the event that did occur (P). Explain why this event is important.

Contrast Event (Q): Describe the event that did not occur (Q). Explain why this event was chosen as a contrast.

Key Causes: Identify the causes that explain the difference between P and Q. Highlight which causes led to P and were not present in Q. Explain why these causes are significant for understanding the outcome.
\end{invsummary}
\vspace{-5mm}

\begin{invsummary}
\subsubsection*{Statistical Relevance Presentation Prompt}

Likely Causes: List the factors (A, B, and C) that most likely contribute to the result (Y) based on the feature scores. Explain why these factors are important.

Explanation: Provide a concise explanation of how these factors lead to the result. Be clear and straightforward, avoiding complex terms.
\end{invsummary}
\vspace{-5mm}

\begin{invsummary}
\subsubsection*{Chain of Thought Presentation Prompt}
Let's process the explanation step by step.
\end{invsummary}

\subsubsection{\il Presentation Examples}

We present examples from the presentation prompt for a sample student in DSP 1, with explainer CEM, and the abnormal conditions strategy across three language models.

\begin{myabstract}

\textbf{Gemma-2 9B, DSP 1, CEM, Abnormal Conditions -- \il Presentation Prompt}\\

You're doing great in the \textit{Digital Signal Processing} course! Let's take a look at your progress so far and see what we can do to help you continue on this successful path.\\

\textbf{How Am I Doing?}\\

It seems like you've been engaging well with the course material. However, I noticed that in Week 5, you had a lower number of sessions compared to other weeks. This is a bit unusual because Week 5 covers important concepts like ideal filters and filter design, which build upon the knowledge from earlier weeks. Remember, consistent engagement is key to mastering these complex topics.\\

\textbf{Where to Next?}\\

\begin{itemize}
    \item \textbf{Connect the Dots:} Take some time to review the material from Week 4, especially the concepts related to the Discrete Fourier Transform (DFT) and its applications. This will help solidify your understanding of the foundational principles needed for Week 5's content.
    \item \textbf{Active Learning:} In Week 6, you'll delve into modulation, interpolation, and sampling. To make the most of these concepts, try actively applying them to real-world examples. Think about how these techniques are used in everyday technologies like audio and video compression.
\end{itemize}
\end{myabstract}

\begin{myabstract}

\textbf{GPT-4o, DSP 1, CEM, Abnormal Conditions -- \il Presentation Prompt}\\

\textbf{Feedback on Your Course Progress}\\

\textbf{Where Am I Going?}\\
You're progressing well in \textit{Digital Signal Processing 1}. Your goal is to master the course content and successfully pass the course.\\

\textbf{How Am I Doing?}
\begin{itemize}
    \item \textbf{Abnormal Causes:} Your engagement in Week 5 is notably low, with very few online sessions and minimal interaction with video content. These are critical because they deviate from the norm and significantly impact your learning.
    \item \textbf{Recent Events:} Weeks 4 and 5 are crucial as they cover essential topics like Filter Design and FFT. Your low engagement during these weeks is concerning.
    \item \textbf{Controllable Factors:} You can control the number of online sessions you participate in and your interaction with video materials. Increasing these activities will enhance your understanding and performance.\\
\end{itemize}

\textbf{Where to Next?}\\
\begin{enumerate}
    \item \textbf{Week 6:} Focus on Modulation and Interpolation. Engage more in online sessions and interact with videos to catch up.
    \item \textbf{Week 7:} As you move into Multirate processing, ensure you maintain high engagement. Build on the foundation from previous weeks to better understand complex concepts.
\end{enumerate}

\end{myabstract}
\begin{myabstract}

\textbf{Llama-3 70B, DSP 1, CEM, Abnormal Conditions -- \il Presentation Prompt}\\

\textbf{Where Am I Going?}\\
You're doing great so far, and your goal is to master the skills in \textit{Digital Signal Processing 1}. You're currently on track to pass the course, but let's work together to ensure you continue to excel.

\textbf{How Am I Doing?}
\begin{itemize}
    \item \textbf{Abnormal Causes:} Your high engagement with the course content during Week 5 is unusual and important. This suggests you're actively participating in the course and making an effort to learn the material.
    \item \textbf{Recent Events:} Your behavior during Week 5, such as having a high number of sessions and clicks on videos, is crucial to your success. This recent engagement is a key factor in your predicted performance.
    \item \textbf{Controllable Factors:} You have control over your engagement with the course content. Focus on maintaining your current level of participation, and consider revisiting topics from previous weeks, like Week 4's DFT and DTFT, to reinforce your understanding.
\end{itemize}

\textbf{Where to Next?}
To continue improving, I recommend:
\begin{enumerate}
    \item Review and practice the skills learned in Week 4, specifically DFT and DTFT, to solidify your understanding before moving forward.
    \item As you enter Week 6, focus on applying the concepts of Modulation, Interpolation, and Sampling to real-world problems, and explore how they relate to previous topics.
\end{enumerate}

By following these recommendations, you'll be well on your way to achieving your goal of mastering \textit{Digital Signal Processing 1}.

\end{myabstract}

Extended examples across different instruction strategies and base explainers can be found in our code repository.

\subsection{Visualization Prompt}
\label{appendix:exp-viz-prompt}

We use this prompt with the ChatGPT-4o interface to directly extract the visualizations from the native chat environment. It is also possible to ask language models to generate the code and run it separately. We did not evaluate this case as it would necessitate additional post-processing steps, causing friction to  the non-technical educator's workflow in a real-world setting.

\begin{invsummary}
Given this detailed report: \\

\textcolor{brandred}{\textit{\{explanation\_selection\_response\}}}\\

And this summary: \\ \\ 
\textcolor{brandred}{\textit{\{explanation\_presentation\_response\}}}\\

Create a plot (i.e. line graph or heatmap) to help the student understand this explanation.
\end{invsummary}
\vspace{-5mm}

\subsubsection*{\il Visualization Examples}
\begin{figure}[h!]
    \centering
    \begin{minipage}{0.47\textwidth}
        \centering
        \includegraphics[width=\textwidth]{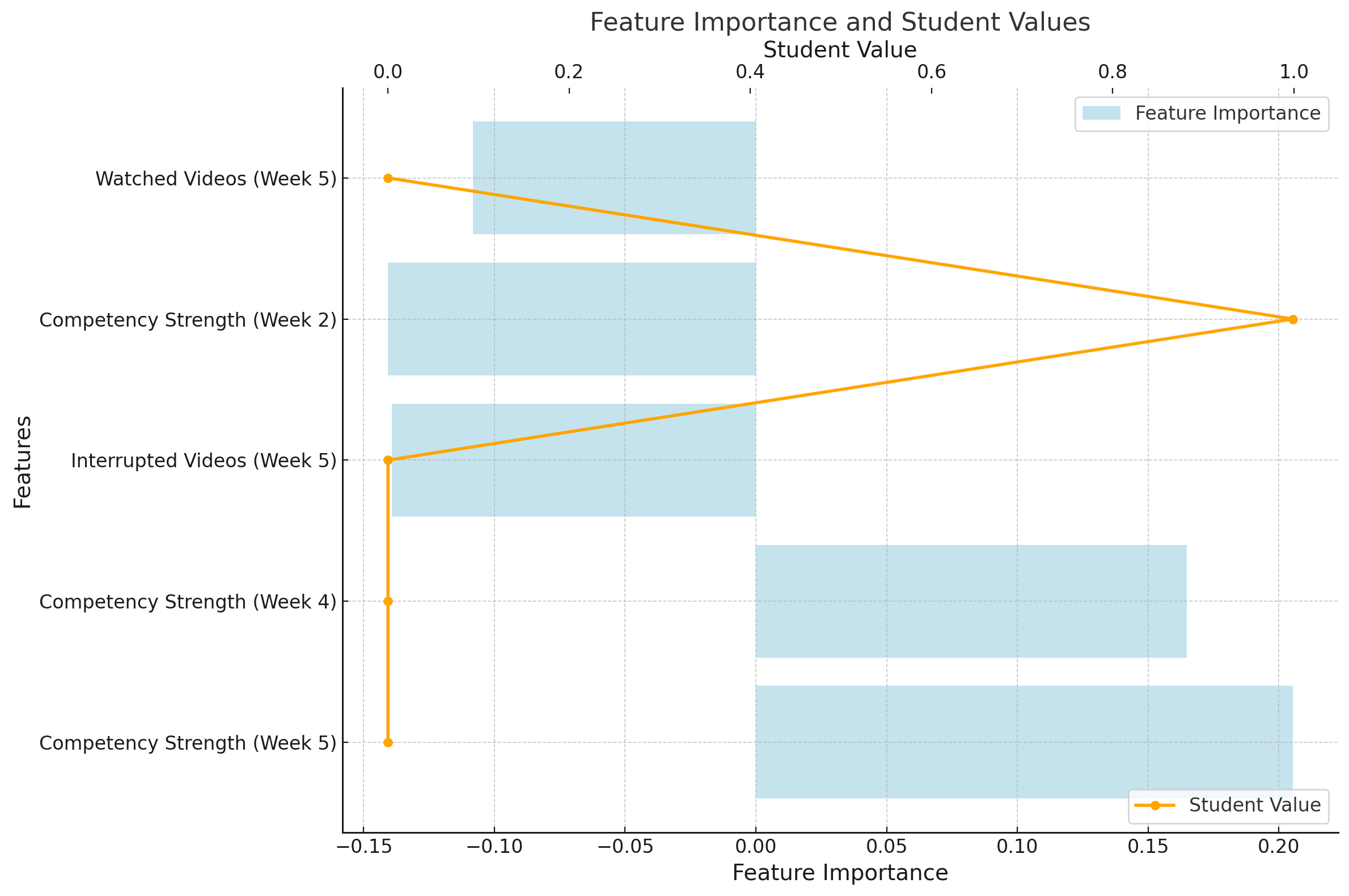} 
        \caption{A \il GPT-4o visual explanation for a sample student with the LIME explainer and the \textit{abnormal conditions} instruction.}
        \label{fig:image1}
    \end{minipage}
    \hfill
    \begin{minipage}{0.47\textwidth}
        \centering
        \includegraphics[width=\textwidth]{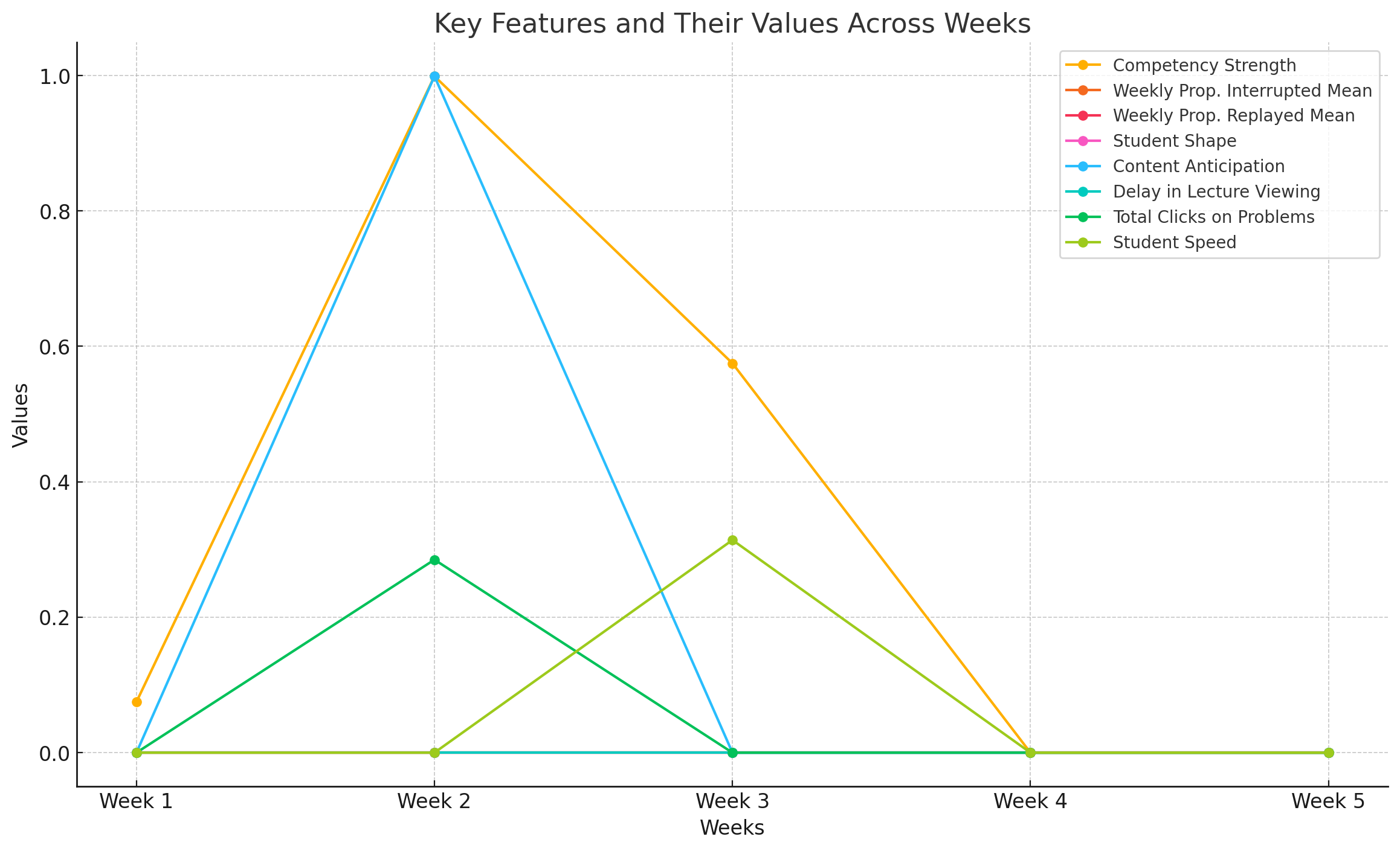} 
        \caption{A \il GPT-4o visual explanation for a sample student with the LIME explainer and the\textit{ chain-of-thought }instruction.}        \label{fig:image2}
    \end{minipage}
\end{figure}

\begin{figure}[h!]
    \centering
    \begin{minipage}{0.47\textwidth}
        \centering
        \includegraphics[width=\textwidth]{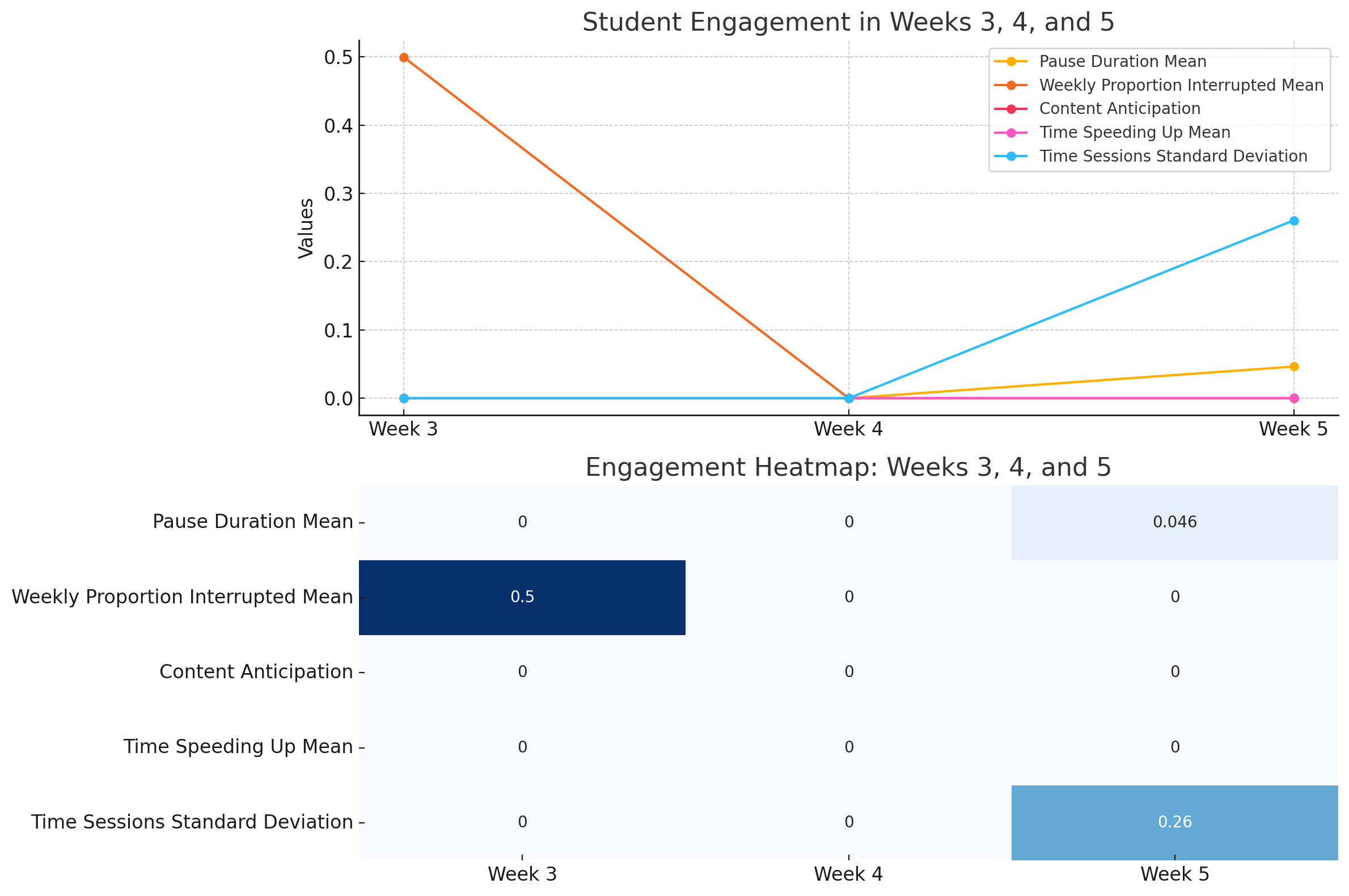} 
        \caption{A \il GPT-4o visual explanation for a sample student with the CEM explainer and the \textit{contrastive explanation} instruction.}
        \label{fig:image3}
    \end{minipage}
    \hfill
    \begin{minipage}{0.47\textwidth}
        \centering
        \includegraphics[width=\textwidth]{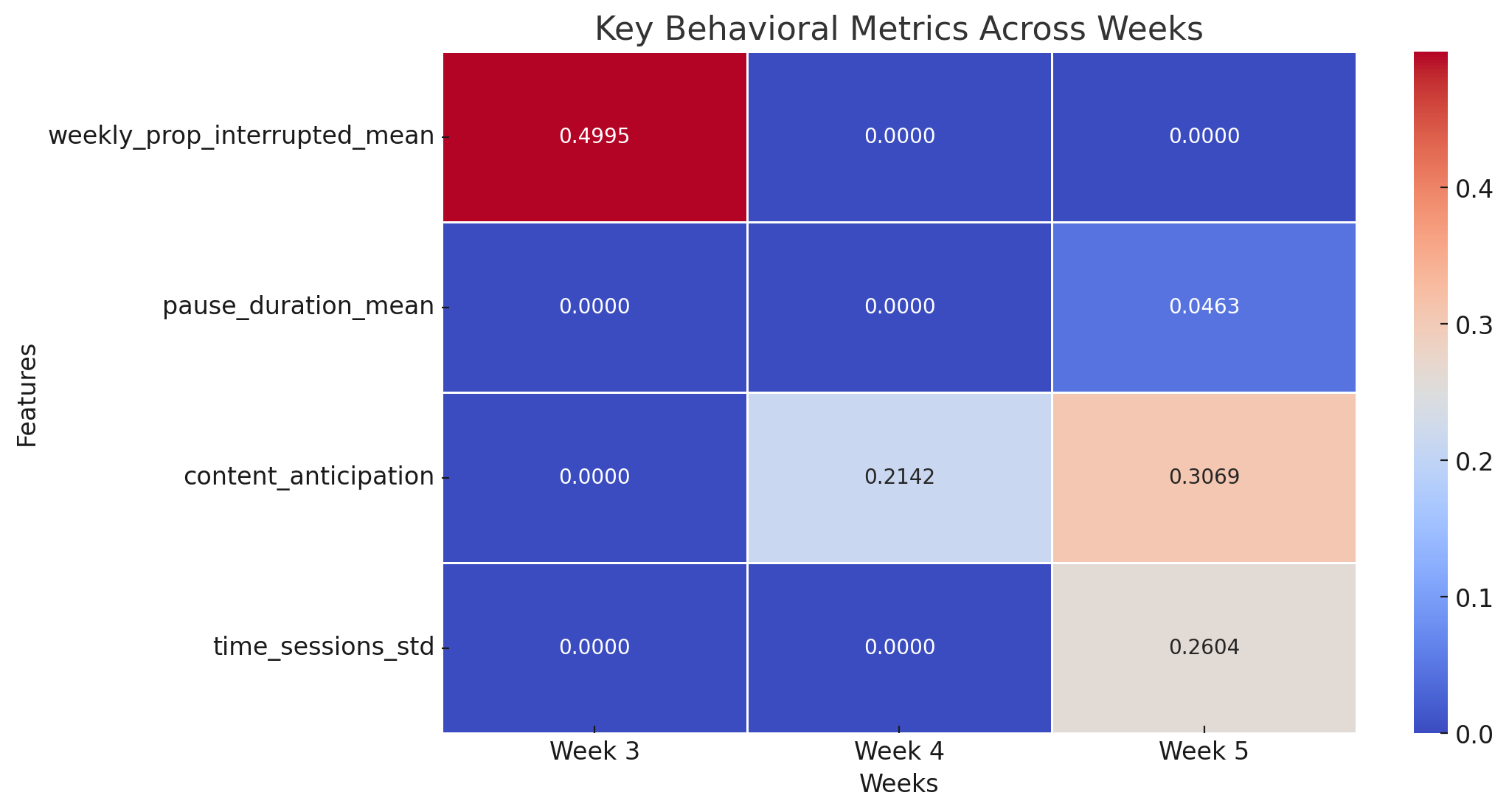} 
        \caption{An \il example GPT-4o visual explanation with the CEM explainer and the \textit{Pearl's explanation} instruction.}        \label{fig:image4}
    \end{minipage}
\end{figure}

\begin{figure}[h!]
    \centering
    \begin{minipage}{0.47\textwidth}
        \centering
        \includegraphics[width=\textwidth]{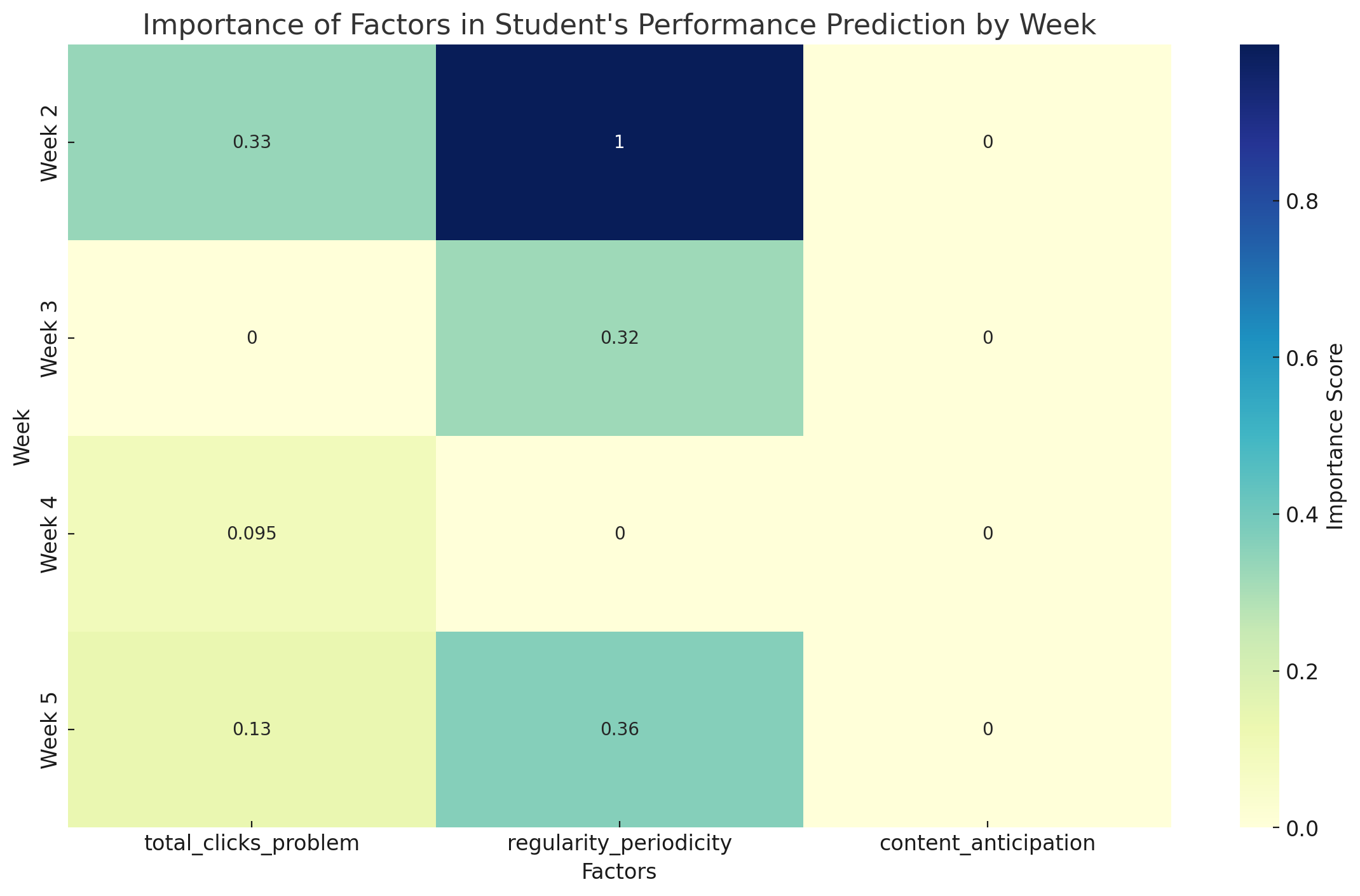} 
        \caption{A \il GPT-4o visual explanation for a sample student with the MC-LIME explainer and the \textit{Necessity Robustness} instruction.}
        \label{fig:image5}
    \end{minipage}
    \hfill
    \begin{minipage}{0.47\textwidth}
        \centering
        \includegraphics[width=\textwidth]{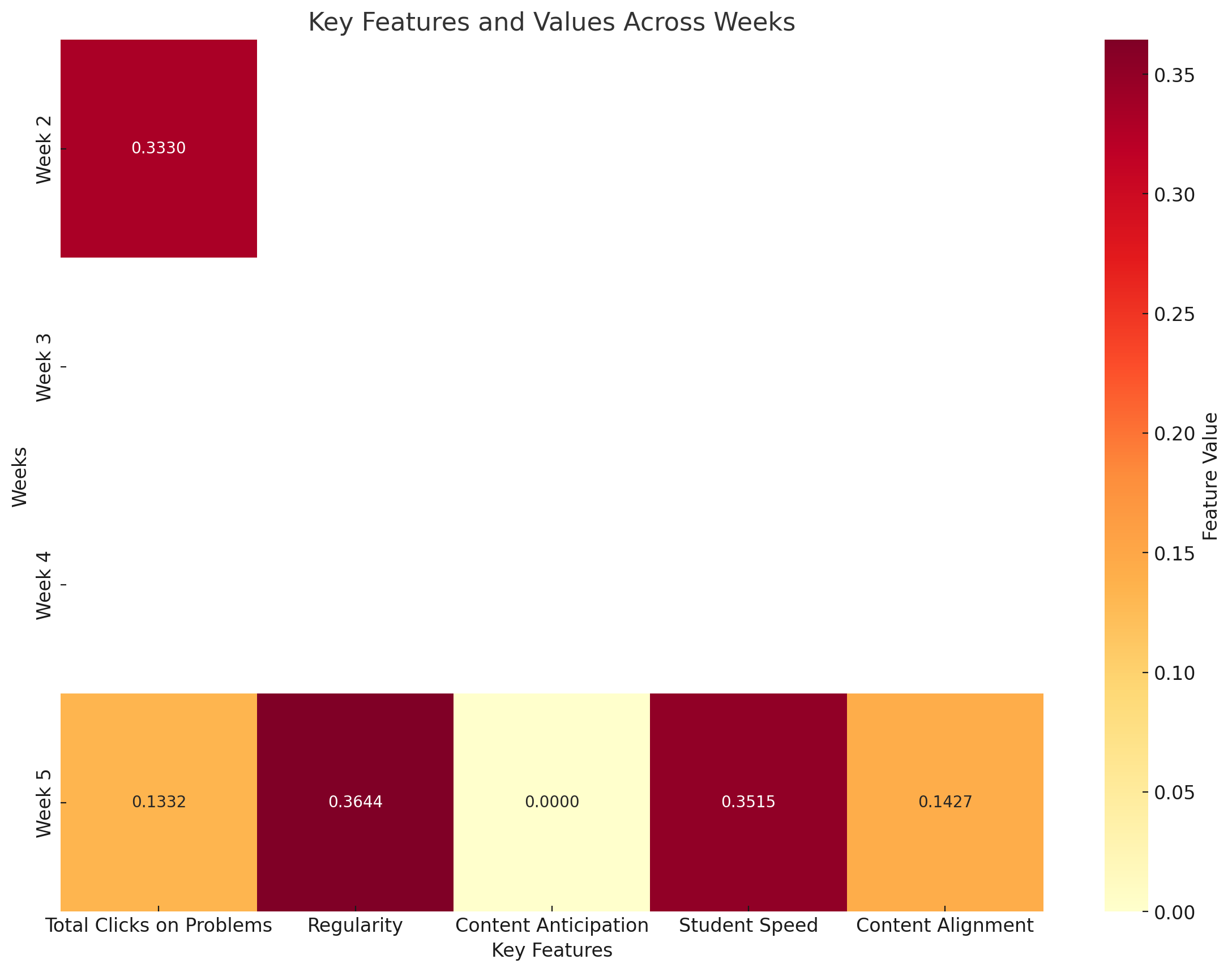} 
        \caption{An example \il GPT-4o visual explanation with the MC-LIME explainer and the \textit{Relevant Selection} instruction.}        \label{fig:image6}
    \end{minipage}
\end{figure}

Figures \ref{fig:image1}, \ref{fig:image2}, \ref{fig:image3}, \ref{fig:image4}, \ref{fig:image5}, and \ref{fig:image6} represent samples of GPT-4o generated \il visuals that were used in the student preference user study. Figures from all three source explainers and each of the six theories used in the study are included.

\subsection{Evaluation Prompts}
\label{appendix:exp-eval-prompt}

\begin{invsummary}
\subsubsection*{LLM-as-a-judge prompt}
    Based on the provided Input (if any) and Generated Text, answer the ensuing Questions with either a YES or NO choice. 

Your selection should be based on your judgment as well as the following rules: \begin{itemize}
    \item \textbf{YES}: Select YES if the generated text entirely fulfills the condition specified in the question. However, note that even minor inaccuracies exclude the text from receiving a YES rating. As an illustration, consider a question that asks, “Does each sentence in the generated text use a second person?” If even one sentence does not use the second person, the answer should NOT be ‘YES’. To qualify for a YES rating, the generated text must be entirely accurate and relevant to the question. 
    \item \textbf{NO}: Opt for NO if the generated text fails to meet the question's requirements or provides no information that could be utilized to answer the question. For instance, if the question asks, “Is the second sentence in the generated text a compound sentence?” and the generated text only has one sentence, it offers no relevant information to answer the question. Consequently, the answer should be NO.
\end{itemize}
\vspace{1em}

QUESTIONS:
\textcolor{brandred}{\textit{\{question\}}}

\vspace{1em}

FORMAT: 
A list of YES/NO answers separated by commas in a list format. Example: [answer1, answer2, ...]

\end{invsummary}

\section{Decomposed Question Evaluation}

\label{appendix:decomposed-qs}

\subsection{Human Annotation}
\label{appendix:human-annot}

Three experts annotated the 48 explanation selection responses for two students in their entirety, covering 3 explainers and 8 theories for each student. The interannotator agreement between the raters was high (0.71 Cohen's Kappa), as there was a 1 hour training workshop where the annotators discussed each of the criteria and a few sample entries together.

Then, the experts annotated the remaining 312 explanations separately (corresponding to 13 more students). Each annotation covers between five to nine individual questions responded to with a binary indicator as per Table \ref{appendix:tab-dqs}. In accordance with \citet{wang2023chatgpt,qin2024infobench}, we use GPT-4o as a annotator. Table \ref{tab:annot-agree-gpt} compares the human expert annotation with GPT-4o annotation, finding an average agreement of 94.68\%. The only type of explanation where the agreement is less that 90\% is for the Base Contrastive explanations (82.22\%, with a high standard deviation), where we found that GPT-4o is a lot more critical than human annotators. As the bias is on the side of GPT-4o grading more harshly (especially for the fourth question regarding context, which is the most difficult question to annotate) we found it appropriate to proceed.

\begin{table}[ht!]
\centering
  \resizebox{0.3\linewidth}{!}{%
  \begin{tabular}{@{}lc@{}}
  \toprule
  \multicolumn{1}{c}{\textbf{Theory}} & \textbf{Agreement} \\ \midrule
  {\color[HTML]{036400} Abnormal Conditions} & {\color[HTML]{036400} 98.89 ± 4.72} \\
  {\color[HTML]{9A0000} Base Contrastive} & {\color[HTML]{9A0000} 82.22 ± 9.01} \\
  {\color[HTML]{036400} Chain of Thought} & {\color[HTML]{036400} 95.11 ± 6.13} \\
  {\color[HTML]{036400} RaR + Contrastive} & {\color[HTML]{036400} 98.27 ± 3.39} \\
  {\color[HTML]{036400} Necessity Robustness} & {\color[HTML]{036400} 97.78 ± 2.62} \\
  Pearl Explanation & 95.83 ± 4.41 \\
  {\color[HTML]{036400} Relevance Selection} & {\color[HTML]{036400} 97.78 ± 3.58} \\
  Statistical Relevance & 91.56 ± 6.99 \\ \midrule
  \textbf{Overall} & \textbf{94.68 ± 5.20} \\ \bottomrule
  \end{tabular}%
  }
  \caption{\textbf{Agreement between Human and GPT-4o annotation.} Average percentage of answers with agreement between human and GPT-4o. Standard deviations were computed between the different explainers.}
  \label{tab:annot-agree-gpt}
\end{table}

\subsection{Explanation Selection Stage}

We detail the decomposed questions (DQs) used by GPT-4o and human experts to annotate the explanation responses.

\begin{longtable}{|>{\RaggedRight\arraybackslash}p{4cm}|>{\RaggedRight\arraybackslash}p{11cm}|}
\hline
\makecell{\textbf{Theory}} & \makecell{\textbf{Explanation Selection DQs}} \\ \hline
\endfirsthead
\hline
\makecell{\textbf{Theory}} & \makecell{\textbf{Explanation Selection DQs}} \\ \hline
\endhead
\hline
\endfoot

\begin{mdframed}
\textbf{Relevance Selection} 
\end{mdframed} & 
\begin{mdframed}
\begin{itemize}
    \item Is the generated text using the provided data extensively?
    \item Is the generated text analysis based largely on the explainer results provided?
    \item Is the generated text correctly using the model’s predicted outcome?
    \item Is the generated text considering the context (course structure)?
    \item Is the generated text selecting the causes that are most relevant to the explainer results?
    \item Is the generated text selecting the causes that are most relevant to the question?
    \item Is the generated text selecting the causes that are most relevant to the user?
    \item Is the generated text selecting some information that is not already shared with the student?
\end{itemize}
\end{mdframed} \\ \hline

\begin{mdframed}
\textbf{Abnormal Conditions} 
\end{mdframed} & 
\begin{mdframed}
\begin{itemize}
    \item Is the generated text using the provided data extensively?
    \item Is the generated text analysis based largely on the explainer results provided?
    \item Is the generated text correctly using the model’s predicted outcome?
    \item Is the generated text considering the context (course structure)?
    \item Is the generated text selecting the causes based on the abnormality of the causes?
    \item Is the generated text selecting the causes based on the temporality of the causes?
    \item Is the generated text selecting the causes based on the controllability of the causes?
    \item Is the selected explanation following the criteria of Abnormality, Temporality, Controllability?
\end{itemize}
\end{mdframed} \\ \hline

\begin{mdframed}
\textbf{Pearl Explanation} 
\end{mdframed} & 
\begin{mdframed}
\begin{itemize}
    \item Is the generated text using the provided data extensively?
    \item Is the generated text analysis based largely on the explainer results provided?
    \item Is the generated text correctly using the model’s predicted outcome?
    \item Is the generated text considering the context (course structure)?
    \item Is the generated text identifying variables that are determined by factors external to the model?
    \item Is the generated text identifying variables that are determined by relationships with other variables?
    \item Is the explanation selected containing only a set of factors that is minimal?
    \item Is the explanation considering variables that could potentially alter the model output?
\end{itemize}
\end{mdframed} \\ \hline

\begin{mdframed}
\textbf{Necessity Robustness Selection} 
\end{mdframed} & 
\begin{mdframed}
\begin{itemize}
    \item Is the generated text using the provided data extensively?
    \item Is the generated text analysis based largely on the explainer results provided?
    \item Is the generated text correctly using the model’s predicted outcome?
    \item Is the generated text considering the context (course structure)?
    \item Is the generated text selecting features that explain the model’s output?
    \item Is the generated text selecting a few of these sets of factors by considering the necessity criteria?
    \item Is the generated text selecting the most robust explanation?
\end{itemize}
\end{mdframed} \\ \hline

\begin{mdframed}
\textbf{RaR + Contrastive \\Explanation} 
\end{mdframed} & 
\begin{mdframed}
\begin{itemize}
    \item Is the generated text using the provided data extensively?
    \item Is the generated text analysis based largely on the explainer results provided?
    \item Is the generated text correctly using the model’s predicted outcome?
    \item Is the generated text considering the context (course structure)?
    \item Is the generated text reformulating the question in a “Why P rather than Q?” format?
    \item Is the generated text selecting the event P as the target event?
    \item Is the generated text selecting the event Q as the counterfactual event that did not occur?
    \item Is the generated text selecting the foil based on abnormality to the context, question, and user?
    \item Is the generated text selecting the foil based on relevance to the context, question, and user?
\end{itemize}
\end{mdframed} \\ \hline

\begin{mdframed}
\textbf{Base Contrastive \\Explanation} 
\end{mdframed} & 
\begin{mdframed}
\begin{itemize}
    \item Is the generated text using the provided data extensively?
    \item Is the generated text based only on the explainer results provided?
    \item Is the generated text correctly using the model’s predicted outcome?
    \item Is the generated text considering the context (course structure)?
    \item Is the generated text reformulating the question in a “Why P rather than Q?” format?
    \item Is the generated text determining the causes that explain the difference between P and not-Q?
    \item Is the generated text reasoning about which causes are a cause of P?
    \item Is the generated text reasoning about causes that aren’t present in the causal chain of Q?
\end{itemize}
\end{mdframed} \\ \hline

\begin{mdframed}
\textbf{Statistical Relevance} 
\end{mdframed} & 
\begin{mdframed}
\begin{itemize}
    \item Is the generated text using the provided data extensively?
    \item Is the generated text analysis based largely on the explainer results provided?
    \item Is the generated text correctly using the model’s predicted outcome?
    \item Is the generated text considering the context (course structure)?
    \item Is the generated text selecting the explanation based on likelihood?
\end{itemize}
\end{mdframed} \\ \hline

\begin{mdframed}
\textbf{Chain of Thought} 
\end{mdframed} & 
\begin{mdframed}
\begin{itemize}
    \item Is the generated text using the provided data extensively?
    \item Is the generated text analysis based largely on the explainer results provided?
    \item Is the generated text correctly using the model’s predicted outcome?
    \item Is the generated text considering the context (course structure)?
    \item Is the generated text selecting the explanation by step-by-step tasks?
\end{itemize}
\end{mdframed} 
\label{appendix:tab-dqs}
\end{longtable}

\subsection{Explanation Evaluation}
Across both explanation selection and presentation, the decomposed questions were generated and revised by two authors (who are well familiar with LLM prompting and explainability theories), then individually vetted by a computational social scientist. For explanation selection, the question ``Is the generated text considering the context (course structure)?" was iterated upon several times to find non-subjective wording. For explanation presentation, the question ``Is the generated text concise (readable within 5 minutes)?" was similarly iterated upon. Both of these questions are educational context dependent, and the phrasing inside the parentheses can be adapted to relate to tasks from different domains.

\begin{longtable}{|>{\RaggedRight\arraybackslash}p{4cm}|>{\RaggedRight\arraybackslash}p{11cm}|}
\hline
\makecell{\textbf{Theory}} & \makecell{\textbf{Explanation Presentation DQs}} \\ \hline
\endfirsthead
\hline
\makecell{\textbf{Theory}} & \makecell{\textbf{Explanation Presentation DQs}} \\ \hline
\endhead
\hline
\endfoot

\begin{mdframed}
\textbf{Relevance Selection}
\end{mdframed} & 
\begin{mdframed}
\begin{itemize}
    \item Is the generated text describing the student’s performance?
    \item Is the generated text considering a learning goal?
    \item Is the generated text describing the explanation findings?
    \item Is the generated text suggesting concrete actions that the student can take?
    \item Is the generated text written in the tone of a teacher talking to a student?
    \item Is the generated text concise (readable within 5 minutes)?
    \item Is the generated text describing relevant causes for the student?
    \item Is the generated text describing the information that a student knows already?
    \item Is the generated text describing new insights to the student?
\end{itemize}
\end{mdframed} \\ \hline

\begin{mdframed}
\textbf{Abnormal Conditions}
\end{mdframed} & 
\begin{mdframed}
\begin{itemize}
    \item Is the generated text describing the student’s performance?
    \item Is the generated text considering a learning goal?
    \item Is the generated text describing the explanation findings?
    \item Is the generated text suggesting concrete actions that the student can take?
    \item Is the generated text written in the tone of a teacher talking to a student?
    \item Is the generated text concise (readable within 5 minutes)?
    \item Is the generated text describing abnormal causes?
    \item Is the generated text describing recent events (temporality)?
    \item Is the generated text describing controllable factors?
\end{itemize}
\end{mdframed} \\ \hline

\begin{mdframed}
\textbf{Pearl Explanation}
\end{mdframed} & 
\begin{mdframed}
\begin{itemize}
    \item Is the generated text describing the student’s performance?
    \item Is the generated text considering a learning goal?
    \item Is the generated text describing the explanation findings?
    \item Is the generated text suggesting concrete actions that the student can take?
    \item Is the generated text written in the tone of a teacher talking to a student?
    \item Is the generated text concise (readable within 5 minutes)?
    \item Is the generated text describing factors that are external to the model?
    \item Is the generated text describing internal relationships with other variables?
    \item Is the generated text highlighting features that are actionable and focused on improvement?
\end{itemize}
\end{mdframed} \\ \hline

\begin{mdframed}
\textbf{Necessity Robustness \\Selection}
\end{mdframed} & 
\begin{mdframed}
\begin{itemize}
    \item Is the generated text describing the student’s performance?
    \item Is the generated text considering a learning goal?
    \item Is the generated text describing the explanation findings?
    \item Is the generated text suggesting concrete actions that the student can take?
    \item Is the generated text written in in the tone of a teacher talking to a student?
    \item Is the generated text concise (readable within 5 minutes)?
    \item Is the generated text describing the most necessary causes that explain the result?
    \item Is the generated text describing the robustness of the causes chosen?
\end{itemize}
\end{mdframed} \\ \hline

\begin{mdframed}
\textbf{RaR + Contrastive explanation}
\end{mdframed} & 
\begin{mdframed}
\begin{itemize}
    \item Is the generated text describing the student’s performance?
    \item Is the generated text considering a learning goal?
    \item Is the generated text describing the explanation findings?
    \item Is the generated text suggesting concrete actions that the student can take?
    \item Is the generated text written in in the tone of a teacher talking to a student?
    \item Is the generated text concise (readable within 5 minutes)?
    \item Is the generated text describing a target event that did occur (P)?
    \item Is the generated text describing a contrast event that did not occur (Q)?
    \item Is the generated text describing key differences between the target and the contrast event?
    \item Is the generated text explaining why the key differences are important for understanding the result?
\end{itemize}
\end{mdframed} \\ \hline

\begin{mdframed}
\textbf{Base Contrastive explanation}
\end{mdframed} & 
\begin{mdframed}
\begin{itemize}
    \item Is the generated text describing the student’s performance?
    \item Is the generated text considering a learning goal?
    \item Is the generated text describing the explanation findings?
    \item Is the generated text suggesting concrete actions that the student can take?
    \item Is the generated text written in the tone of a teacher talking to a student?
    \item Is the generated text concise (readable within 5 minutes)?
    \item Is the generated text describing a target event that did occur (P)?
    \item Is the generated text describing a contrast event that did not occur (Q)?
    \item Is the generated text describing key causes that explain the difference between the target and the contrast event?
    \item Is the generated text explaining why the key causes are important for understanding the result?
\end{itemize}
\end{mdframed} \\ \hline

\begin{mdframed}
\textbf{Statistical Relevance}
\end{mdframed} & 
\begin{mdframed}
\begin{itemize}
    \item Is the generated text describing the student’s performance?
    \item Is the generated text considering a learning goal?
    \item Is the generated text describing the explanation findings?
    \item Is the generated text suggesting concrete actions that the student can take?
    \item Is the generated text written in the tone of a teacher talking to a student?
    \item Is the generated text concise (readable within 5 minutes)?
    \item Is the generated text describing the likely causes that most contribute to the result?
    \item Is the generated text describing why the likely causes are important?
    \item Is the generated text concisely explaining why the likely causes lead to the result?
\end{itemize}
\end{mdframed} \\ \hline

\begin{mdframed}
\textbf{Chain of Thought}
\end{mdframed} & 
\begin{mdframed}
\begin{itemize}
    \item Is the generated text describing the student’s performance?
    \item Is the generated text considering a learning goal?
    \item Is the generated text describing the explanation findings?
    \item Is the generated text suggesting concrete actions that the student can take?
    \item Is the generated text written in in the tone of a teacher talking to a student?
    \item Is the generated text concise (readable within 5 minutes)?
    \item Is the generated text describing the explanation in a sequential (step-by-step) manner?
\end{itemize}
\end{mdframed}
\label{appendix:pres-tab-dqs}
\end{longtable}

\section{Generalization}
\label{appendix:generalization}

We conducted an additional experiment to test the generalizability of \il to a different educational context. Specifically, we employed the same BiLSTM model used in \cite{swamy2023trusting} and the LIME explainer on a flipped classroom course on Linear Algebra, where students alternated between online and in-person learning components. This new experiment introduced several key variations: a different student population (Bachelor’s students in STEM, taught in French), a novel educational setting (blended learning as opposed to fully online), and a smaller dataset (150 students compared to the original 17,000). As showcased in \cite{swamy2023trusting}, over two years, the flipped classroom course had almost twice as many videos as quizzes, with an average pass rate of 59\%. The exact prompts were used, changing only the details regarding the name of the course and the syllabus.

In these more challenging settings, \il demonstrated robust instruction-following capabilities, with only a minor decline in performance. The model's average faithfulness to the instructions dropped from 97\% to 95\% in Step 1 and from 98\% to 92\% in Step 2. These results highlight the system's capacity to generalize across diverse learning environments and populations, especially to smaller datasets that are harder to predict upon \cite{swamy2023trusting}. Our analysis provides evidence supporting \il's applicability beyond its original MOOC context, underscoring its potential for deployment in varied educational settings, and ease of extension to other domains.

\section{Student Preference Study}
\label{appendix:study}
\subsection{Background of Participants}
\label{appendix:study-background}

\begin{figure}[]
    \centering
    \includegraphics[width=0.95\textwidth]{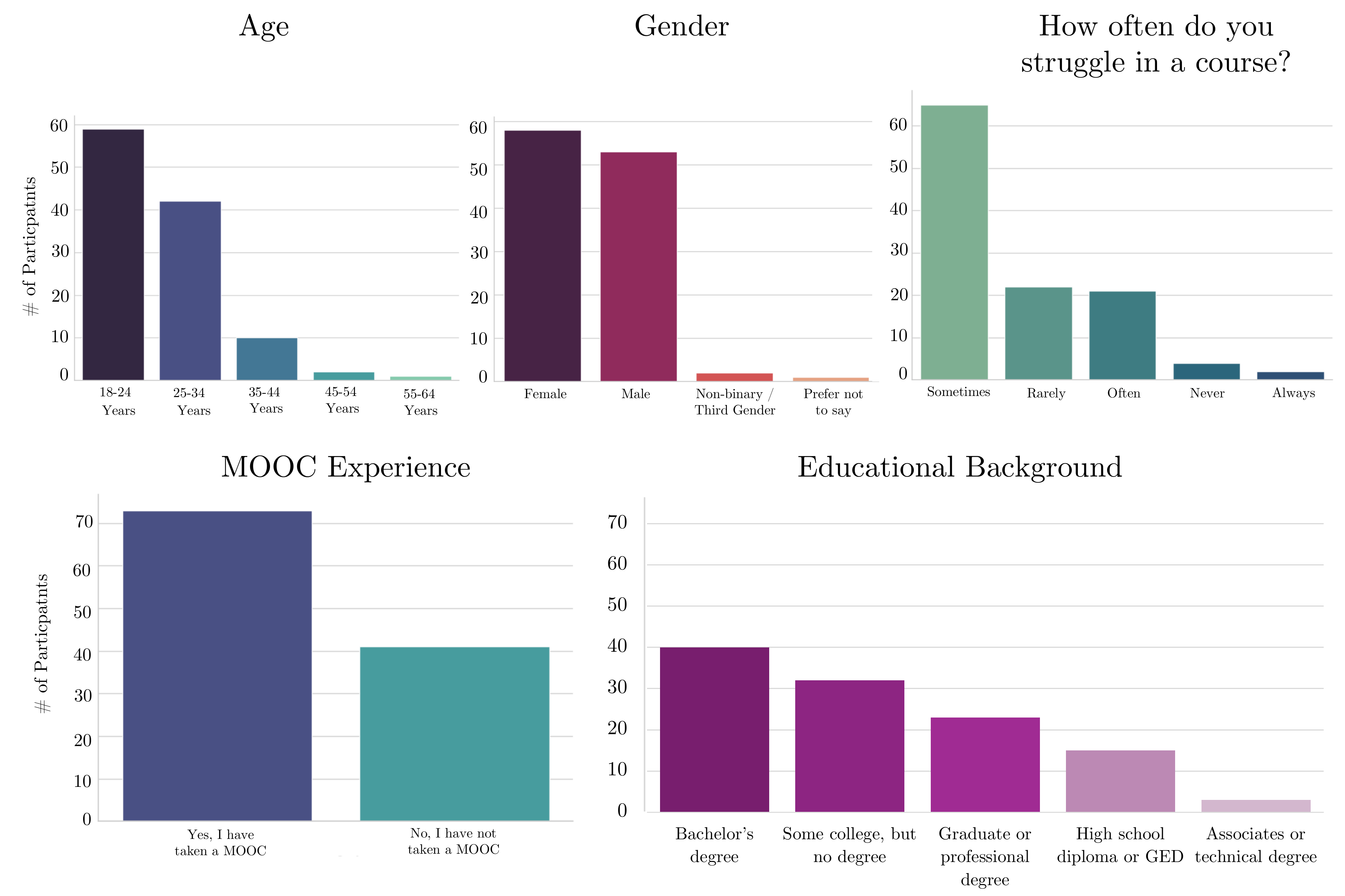}
    \caption{Demographics of students (age, gender, self-identified struggling in courses, MOOC Experience, and educational background) that participated in the user study.}
    \label{fig:userstudy_demo}
\end{figure}

We recruit 114 participants using \texttt{Prolific},\footnote{\url{www.prolific.com/}}, selecting the ones who identified their current status as a student and are over 18 years old. As our target participants have expertise with being a student after high school, they would be well-suited to understand both the educational context of the study and individually evaluate what explanations are most useful for them. During the study, we ask the participants whether they have ever taken an online course (MOOC), their level of education, and their level of ease with coursework. Detailed demographics distribution can be found in Figure \ref{fig:userstudy_demo}. The sample of participants is gender-balanced, and about half of them have taken or participated in creating a MOOC. The median completion time is 41 minutes, and the average reward per hour is £9.00. 

\subsection{Study Materials}
\label{appendix:study-materials}

In this section, we provide a detailed overview of the user study introduced in Section \ref{sec:student-preferences}, covering its design, content, and further analyses of the results. The study was developed over two pilot rounds, involving 6 participants from diverse backgrounds, with continuous and iterative revisions made based on their feedback. The survey received approval from the Human Research Ethics Committee (HREC) under application number HREC 065-2022/27.09.2022.

At the beginning of the study, the participants are shown the following welcome message and disclaimer:\\

\begin{mdframed}[
    backgroundcolor=green!10,
    linecolor=black,
    linewidth=1pt,
    innertopmargin=10pt,
    innerbottommargin=10pt,
    innerrightmargin=10pt,
    innerleftmargin=10pt,
    roundcorner=10pt,
    frametitlebackgroundcolor=gray!30,
    frametitlealignment=\raggedright 
]
Dear participant,\\

Thank you for participating in our study on model explanations. We are very grateful for your participation and your invaluable insight. Please read this Explanatory Statement in full before moving forward. If you would like further information regarding any aspect of this project, please contact us via the email address provided below.\\

We are a group of researchers from the ML4ED Laboratory at EPFL dedicated to improving education through technology. \textbf{The goal of this study is to evaluate different explanations to help a student improve their performance in an online course.}\\

- This survey has been approved by the EPFL Human Research Ethics Committee (HREC) under application number HREC 065-2022/27.09.2022. HREC reviews research proposals involving human participants to ensure that they are ethically acceptable.\\

 - All the personal information will be kept confidential and anonymized. Only demographic information is being recorded and will only be reported as aggregate in a way that prevents identification of any individual participant. You can freely withdraw at any time and any collected data you provided so far will be destroyed.\\

 - All data will be collected and stored safely and reported in an anonymous form, in accordance with the Swiss Federal law on data protection (“Loi fédérale sur la protection des données” – RS 235.1).\\

- Only anonymized or aggregated data may be used in follow-up research (subject to ethics approval), and made available to other researchers for further analysis and for verification of the conclusions reached by the research team.\\

- Only the principal investigator and the aforementioned researchers have access to the original data under strict confidentiality. Results from the project may be published in conference papers and/or journal articles. In any case, no personal data will be published (only aggregated, anonymous and/or anonymized data will be published).\\

- Personal data of participants will be stored for 5 years from the date of collection. During this time, participants have the right to access their data and request information about the processing of their personal data. In order to exercise this right, you need to contact the Principal Investigator.\\

By participating in this survey, you agree that your data can be used for scientific purposes.\\

\textbf{In the following study, you will be asked to compare explanations for approximately 20 minutes.} Please ensure that you have enough time to finish the study correctly. Unfinished or only partially answered studies will not be considered as taken part.\\

We ask you to approach the questions and exercises with seriousness and to complete them to the best of your ability. We will subsequently check questionnaires for seriousness and will have to discard questionnaires that do not meet this requirement.\\

Thank you for your help. If you encounter any problem with the survey, or if you want to give extra feedback, or receive additional information, feel free to contact us (vinitra.swamy@epfl.ch).\\
\end{mdframed}

\subsection{Content of the study}

First, we explain the setting of the study to the participants with the following introductory message:\\

\begin{mdframed}[backgroundcolor=orange!10, linecolor=black, linewidth=1pt, innertopmargin=10pt, innerbottommargin=10pt, innerrightmargin=10pt, innerleftmargin=10pt, roundcorner=10pt]
You are a student taking three online courses (MOOCs): Digital Signal Processing, African Cities, and Elements of Geometry. Since the courses are difficult, often with low passing rates, the teaching team wants to help students who are not doing well to perform better in the course by giving them personalized assistance, and encourage students who are already performing well to continue.\\

To do this, we have a very good model (over 90\% accurate) to predict students' success or failure using various weekly behavior features (such as number of video clicks or how accurately questions are answered on the weekly quizzes). We predict student performance early in the course (before the half-way point) as passing or failing behavior. We use the explanation of the prediction to give students additional, personalized feedback to help pass the course. \\

We want to compare these personalized feedback explanations according to several criteria:\\

\begin{itemize}
    \item Usefulness: This explanation is useful to understand the prediction based on my learning behavior.
    \item Trustworthiness: This explanation lets me judge if I should trust the suggestions.
    \item Actionability: This explanation helps me make a decision on how to improve my learning behavior.
    \item Completeness: This explanation has sufficient detail to understand why the prediction was made based on my learning behavior.
    \item Conciseness: Every detail of this explanation is necessary.
\end{itemize}
\end{mdframed}

We randomly sample 6 students' explanation responses, 2 from each course, and one passing and failing student for each course. For each student, we predict their success or failure with each model and generate an explanation with LIME, MC-LIME, and CEM, then feed it through the \il pipeline for explanation selection, presentation and visualization. We provide them to the participants. The ground truth (student's performance) and the models' exact performance are not provided to the participants so that we do not bias their assessment. 

\begin{figure}[ht!]
    \centering
    \begin{minipage}{0.49\linewidth}
        \centering
        \includegraphics[width=\linewidth]{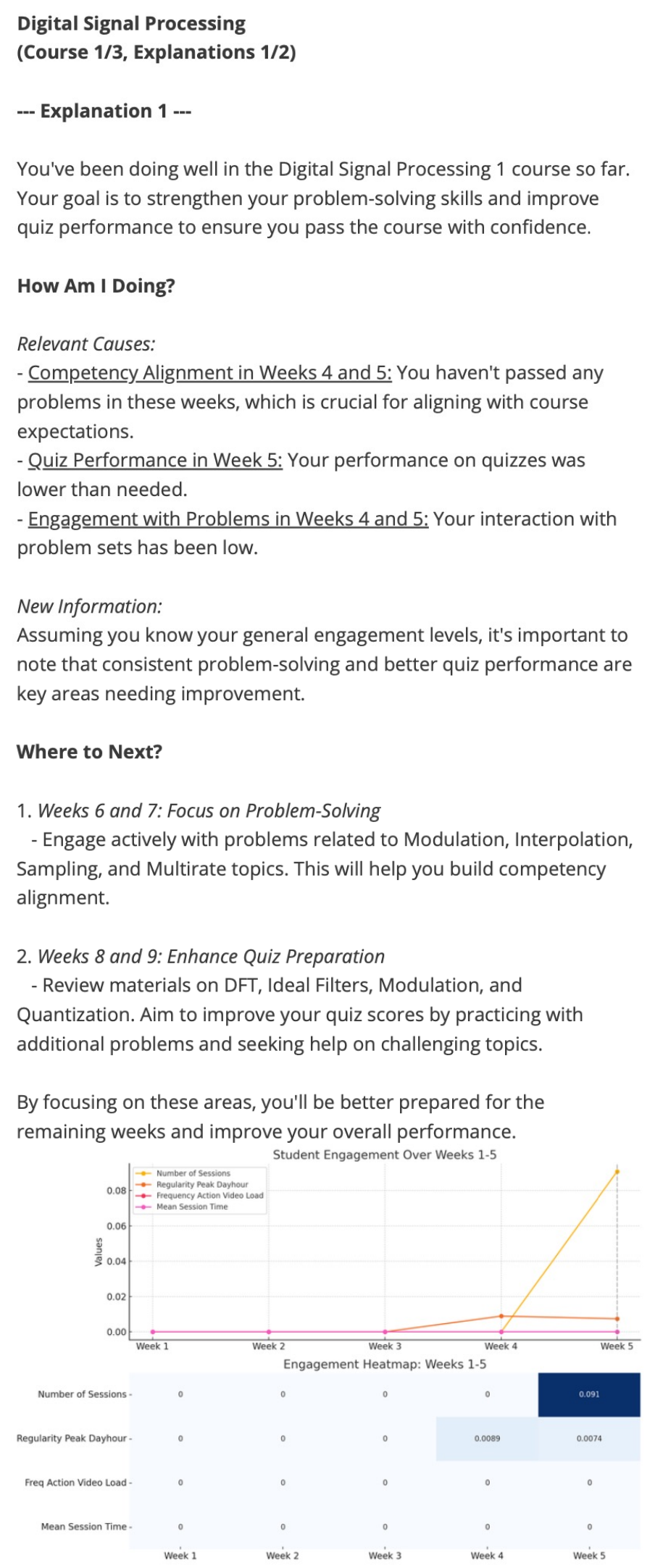}
        \caption{Example CEM \il explanation for Relevance Selection.}
        \label{fig:relevance}
    \end{minipage}
    \hfill
    \begin{minipage}{0.49\linewidth}
        \centering
        \includegraphics[width=\linewidth]{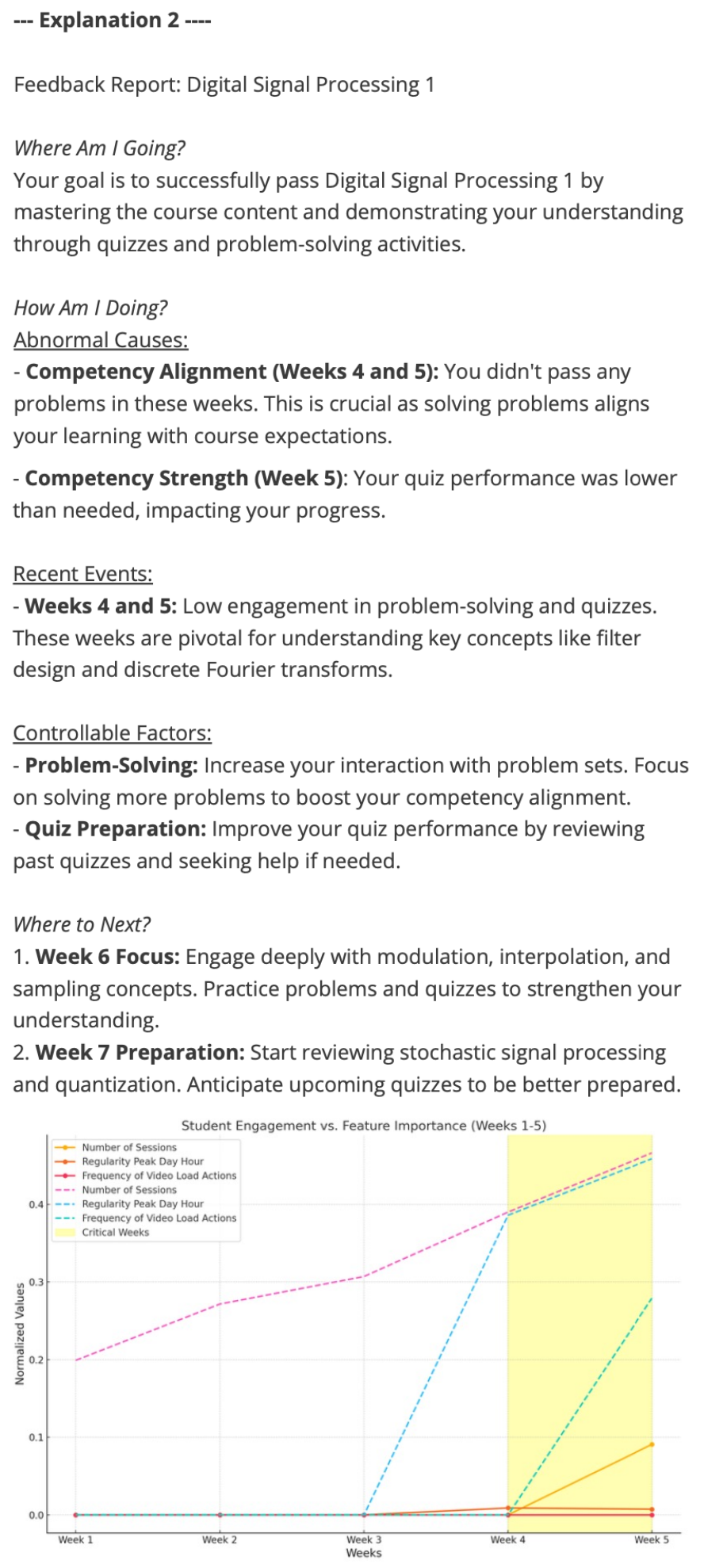}
        \caption{Example CEM \il explanation for Abnormal Conditions.\pagebreak}
        \label{fig:abnormal}
    \end{minipage}
\end{figure}

\begin{figure}[ht!]
    \centering
    \begin{minipage}{0.49\linewidth}
        \centering
        \includegraphics[width=\linewidth]{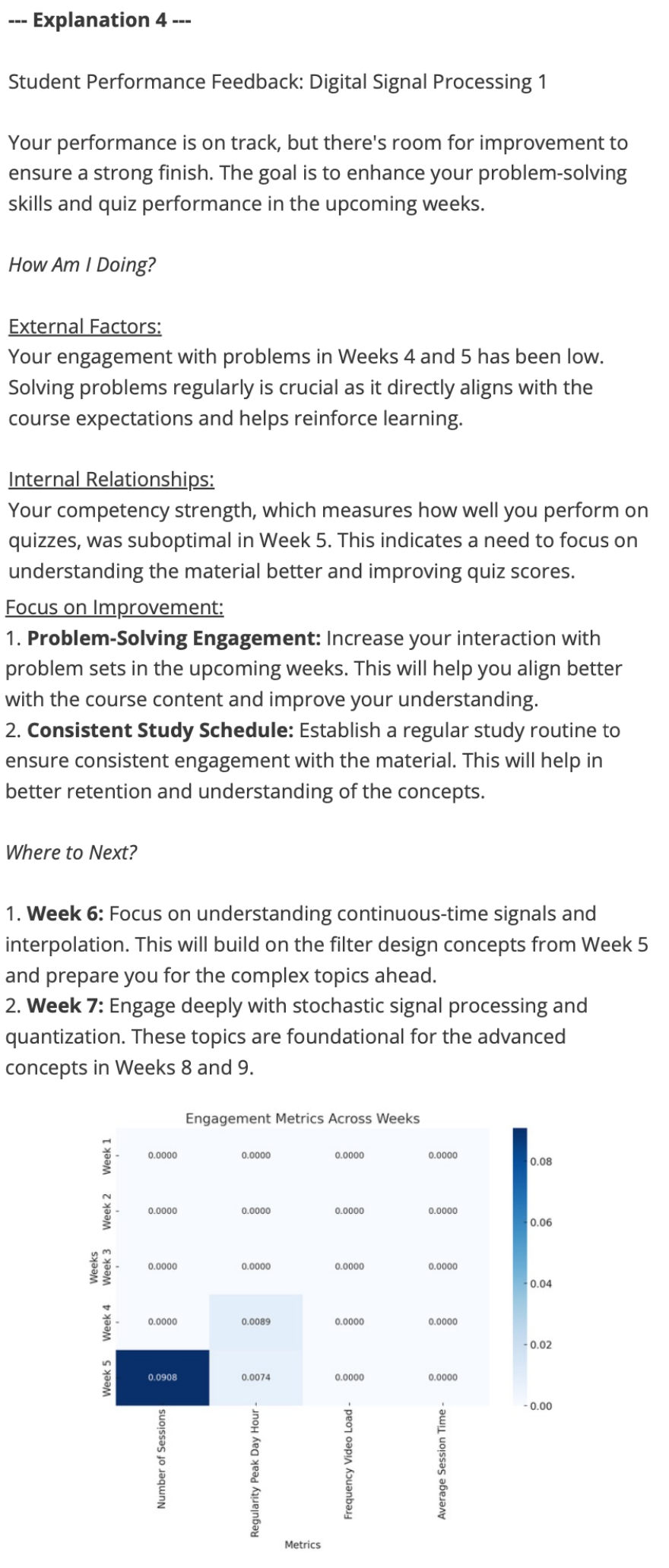}
        \caption{Example CEM \il explanation for Pearl Explanation.}
        \label{fig:pearl}
    \end{minipage}
    \hfill
    \begin{minipage}{0.49\linewidth}
        \centering
        \includegraphics[width=\linewidth]{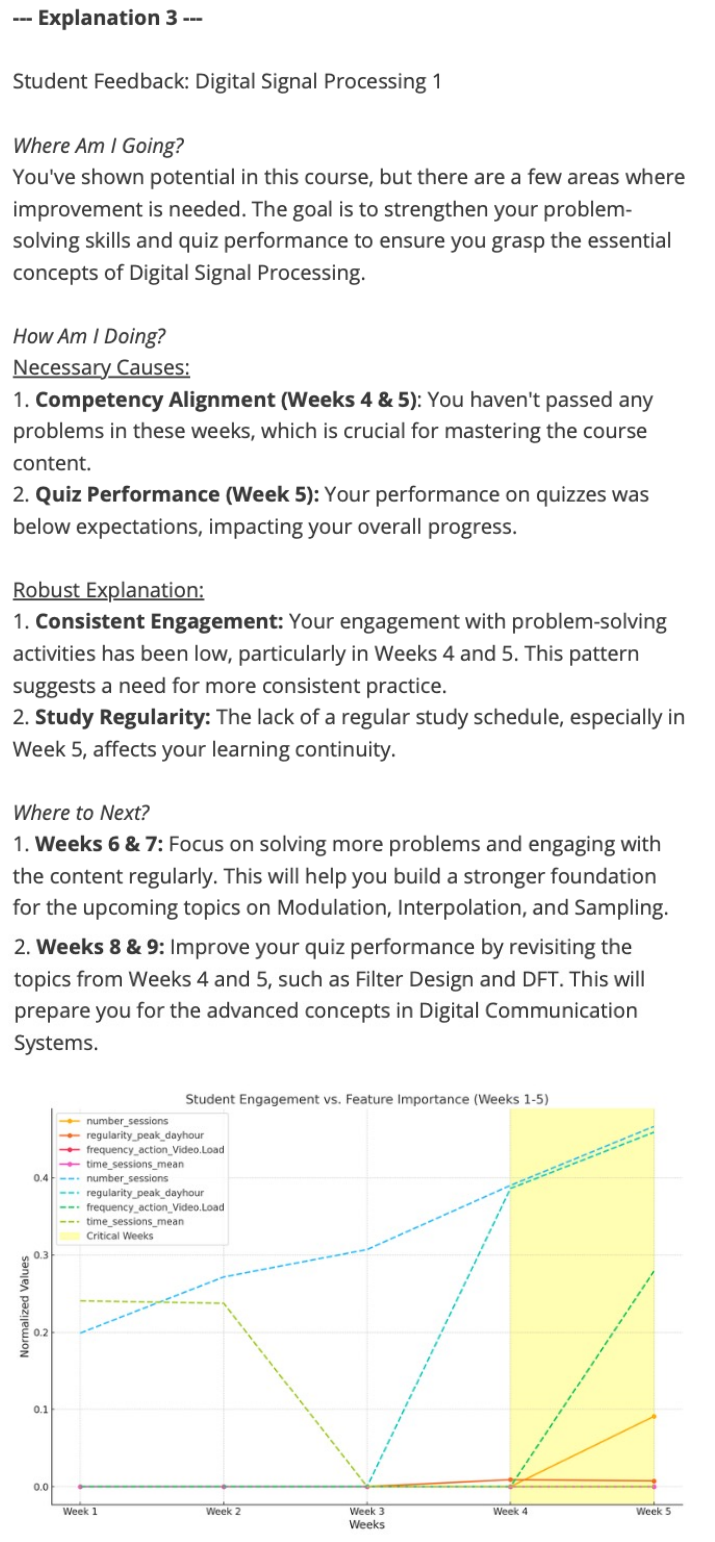}
        \caption{Example CEM \il explanation for Necessity Robustness.}
        \label{fig:nr}
    \end{minipage}
\end{figure}

\begin{figure}[ht!]
    \centering
    \begin{minipage}{0.49\linewidth}
        \centering
        \includegraphics[width=\linewidth]{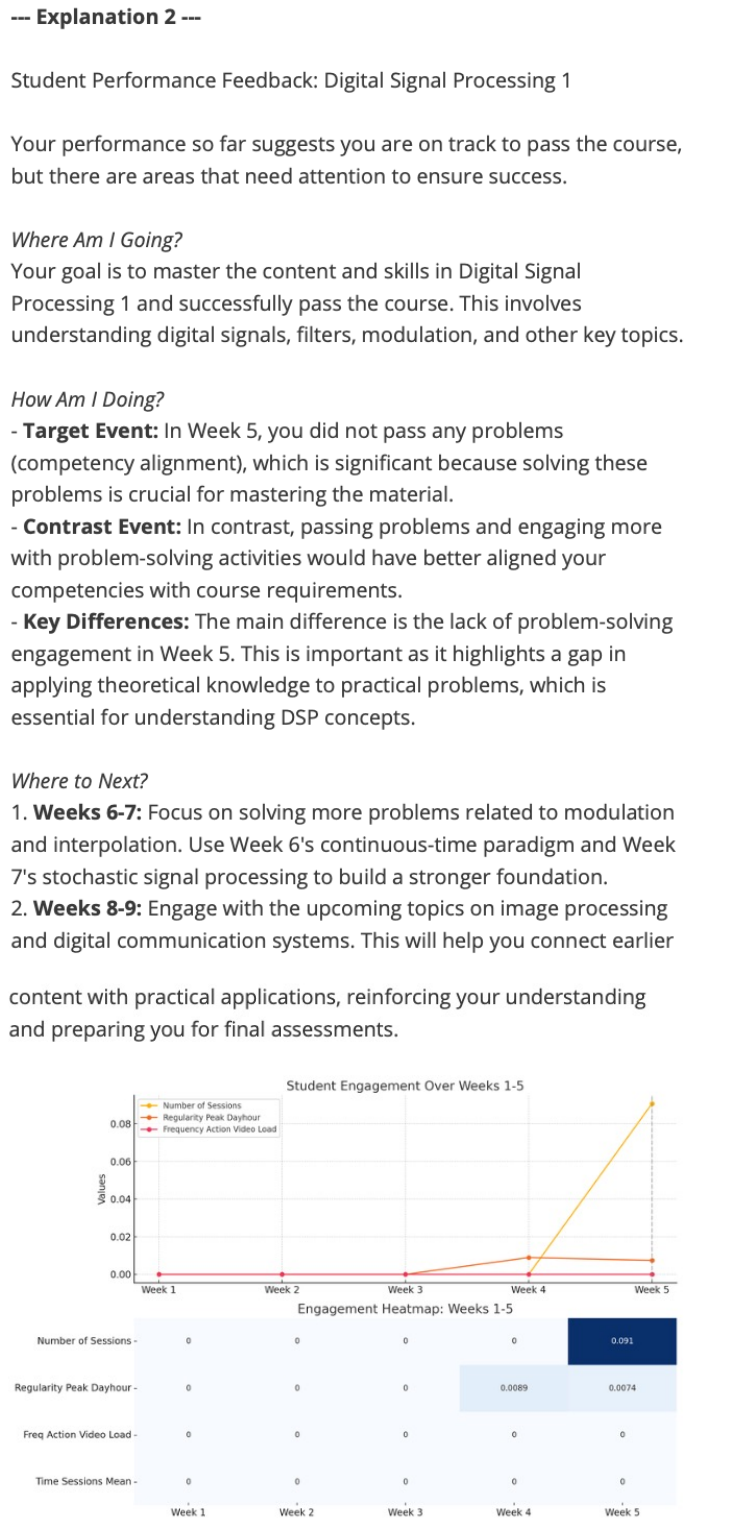}
        \caption{Example CEM \il explanation for Contrastive Explanation.}
        \label{fig:contrastive}
    \end{minipage}
    \hfill
    \begin{minipage}{0.49\linewidth}
        \centering
        \includegraphics[width=\linewidth]{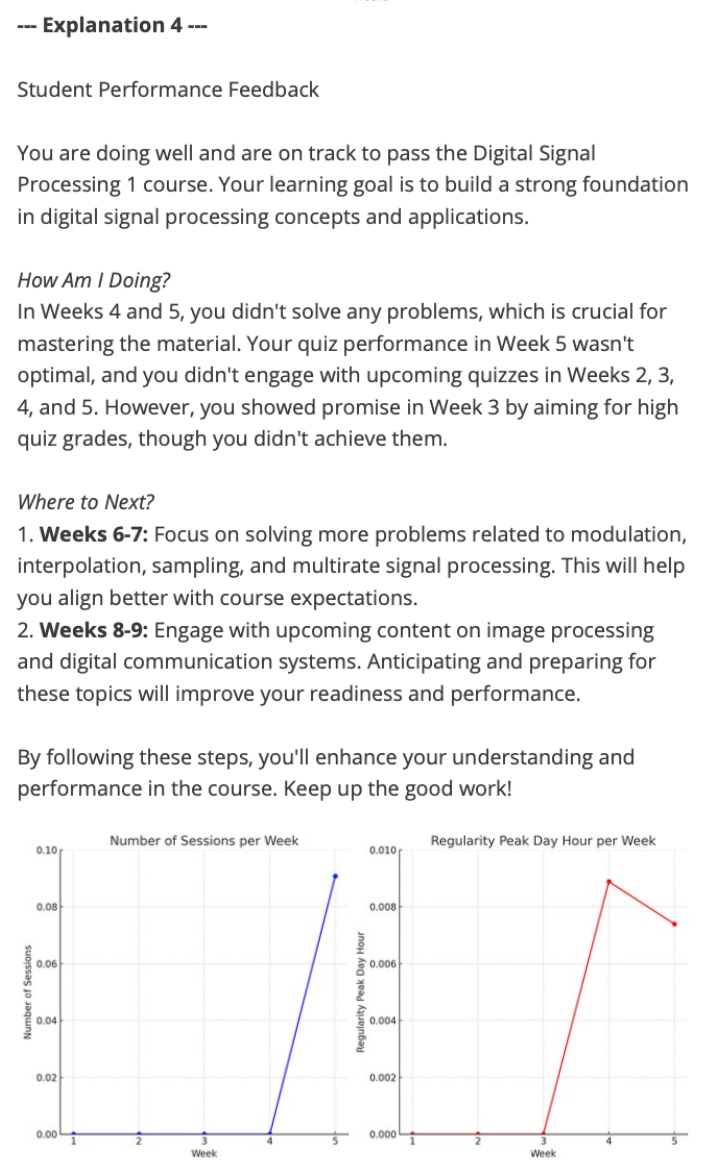}
        \caption{Example chain of thought (CoT) \il explanation.\pagebreak}
        \label{fig:cot}
    \end{minipage}
\end{figure}

\begin{figure}[ht!]
    \centering
    \begin{minipage}{0.3\linewidth}
        \centering
        \includegraphics[width=\linewidth]{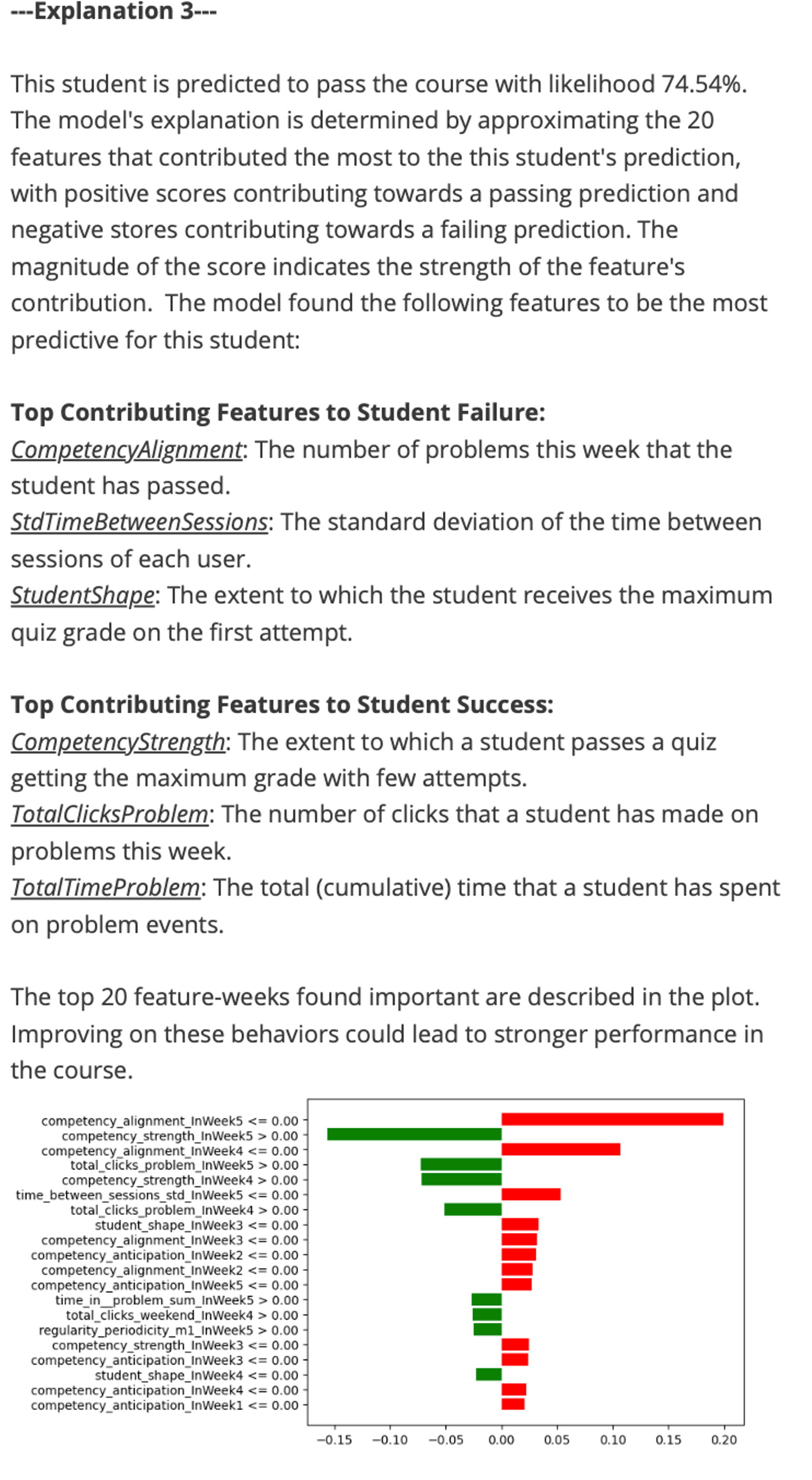}
        \caption{Example LIME (BASE) explanation.}
        \label{fig:LIME}
    \end{minipage}
        \hfill
    \begin{minipage}{0.3\linewidth}
        \centering
        \includegraphics[width=\linewidth]{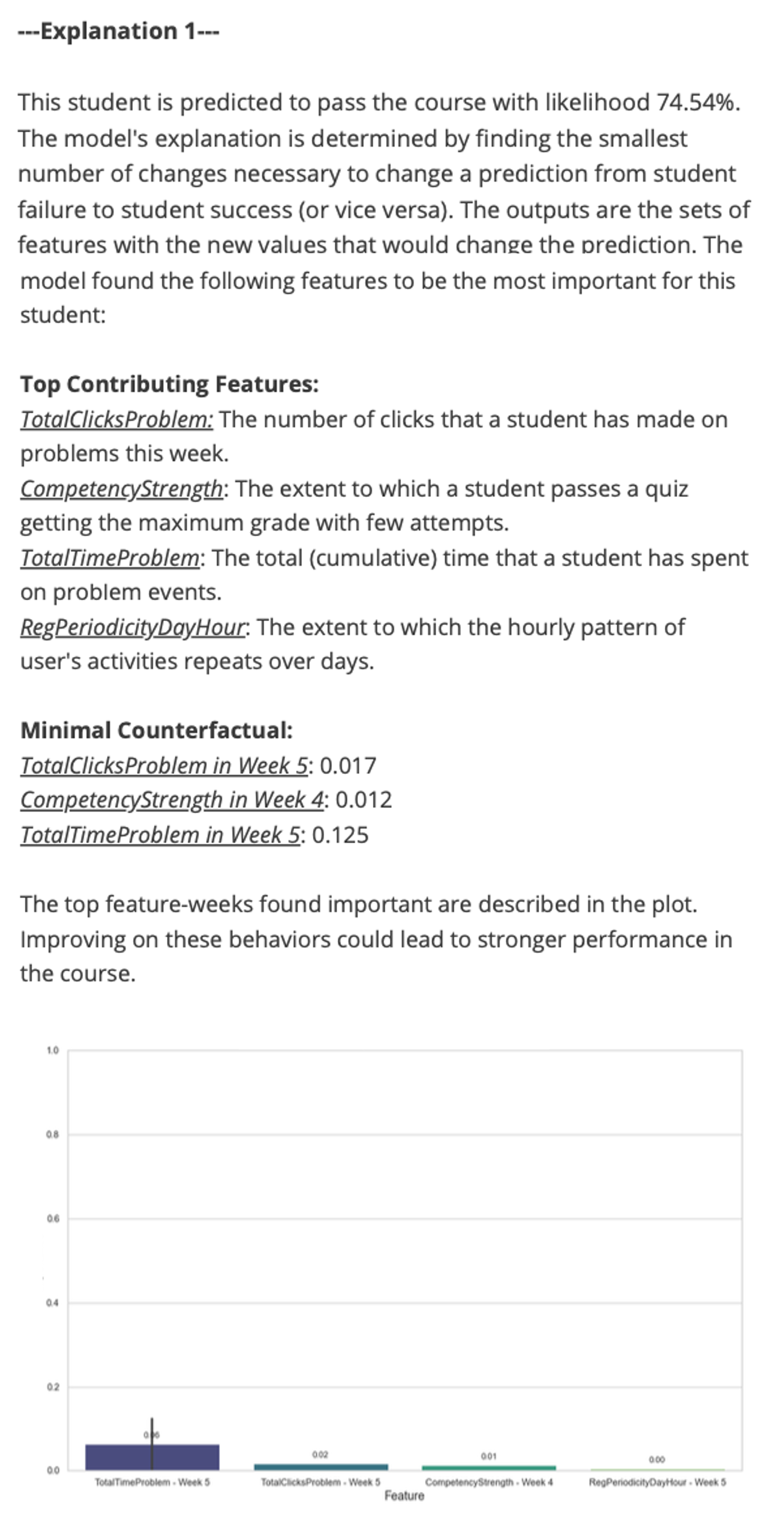}
        \caption{Example MC-LIME (BASE) explanation.}
        \label{fig:MCLIME}
    \end{minipage}
        \hfill
        \begin{minipage}{0.3\linewidth}
        \centering
        \includegraphics[width=\linewidth]{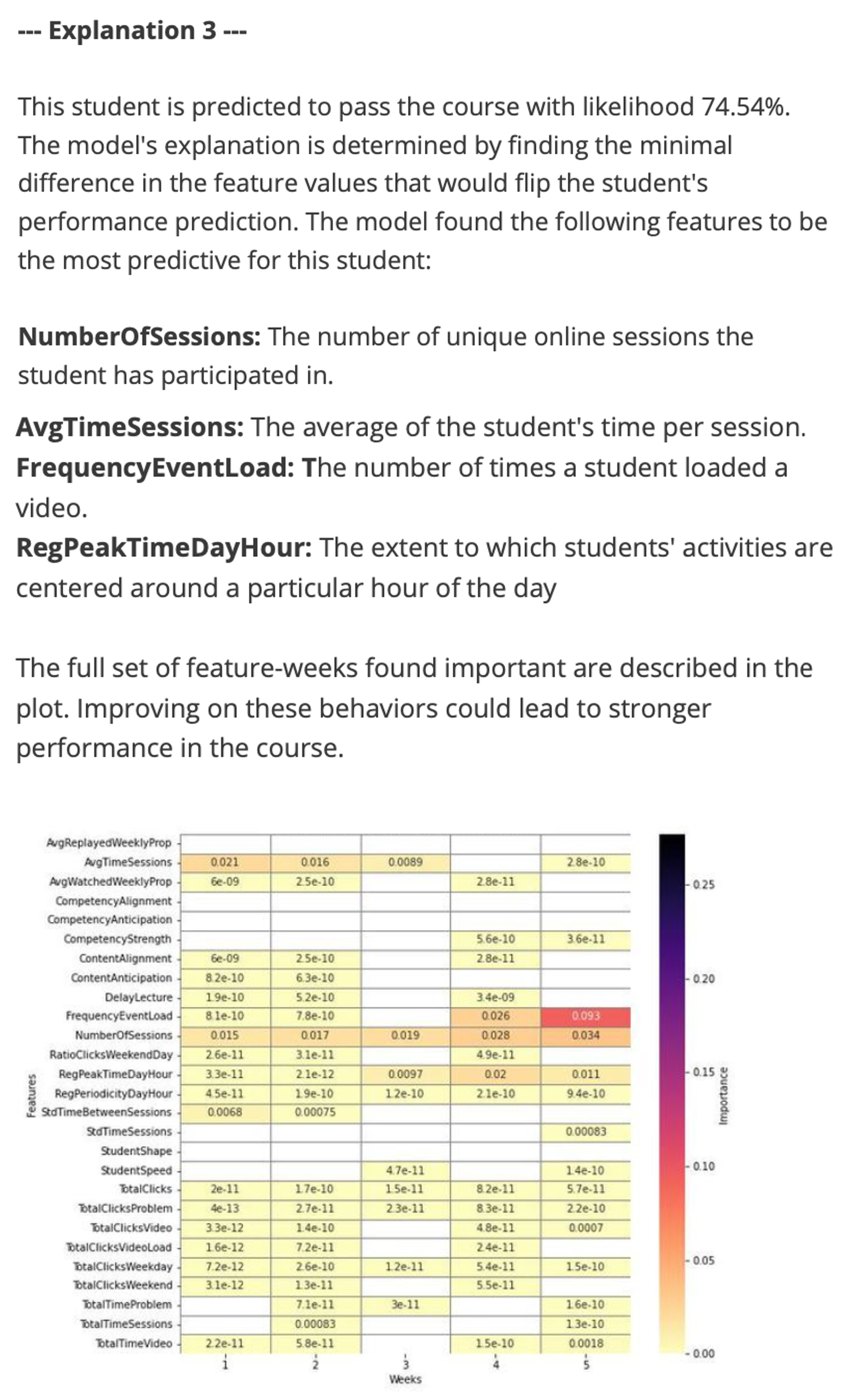}
        \caption{Example CEM (BASE) explanation.}
        \label{fig:CEM}
    \end{minipage}
\end{figure}

The content of the explanation obtained by each method differs greatly. We simplify the explanations and render them in textual and graph format to make them as easy to understand as possible to a human.  For baseline explanations (LIME, CEM, MC-LIME), we craft a text template in line with \citet{swamy2024interpretcc} and iterate on the wording with 2 learning scientists. We provide the description and list of features found important by the model and suggest that improving on these attributes can improve the student performance. For LIME, we show the default graphic that comes with the explanation (Fig. \ref{fig:LIME}). For CEM, we visualize all features found important across the 5 weeks of student progress in a heatmap (Fig. \ref{fig:CEM}), and describe the five most important features in the textual explanation. For MC-LIME, we present only the most concise counterfactual set found important based on the LIME scores and show how much those features need to change to impact the prediction (Fig. \ref{fig:MCLIME}). The remaining figures for all \il explanations are generated directly by GPT-4o, often choosing a line graph, a heatmap, or a bar plot that is directly related to the importance scores or feature values offered in the explanation. Several examples of \il plots are included in Fig. \ref{fig:relevance}, \ref{fig:abnormal}, \ref{fig:pearl}, \ref{fig:nr}, \ref{fig:contrastive}, \ref{fig:cot}.

Consider that the way we chose to present the explanations may influence participants' perceptions. Ideally, we would offer a comprehensive breakdown of how each model utilizes features and how the explanations are generated, allowing participants to fairly evaluate the explanation's quality. However, in a practical setting, the user encountering the explanation may lack the time or background knowledge to fully grasp these details. To address the trade-off between thoroughness and accessibility, we decided on a balanced approach, presenting each explanation with a simple graph and a concise explanatory text.

\begin{figure}[ht!]
    \centering
    \begin{minipage}{0.42\linewidth}
        \centering
        \includegraphics[width=\linewidth]{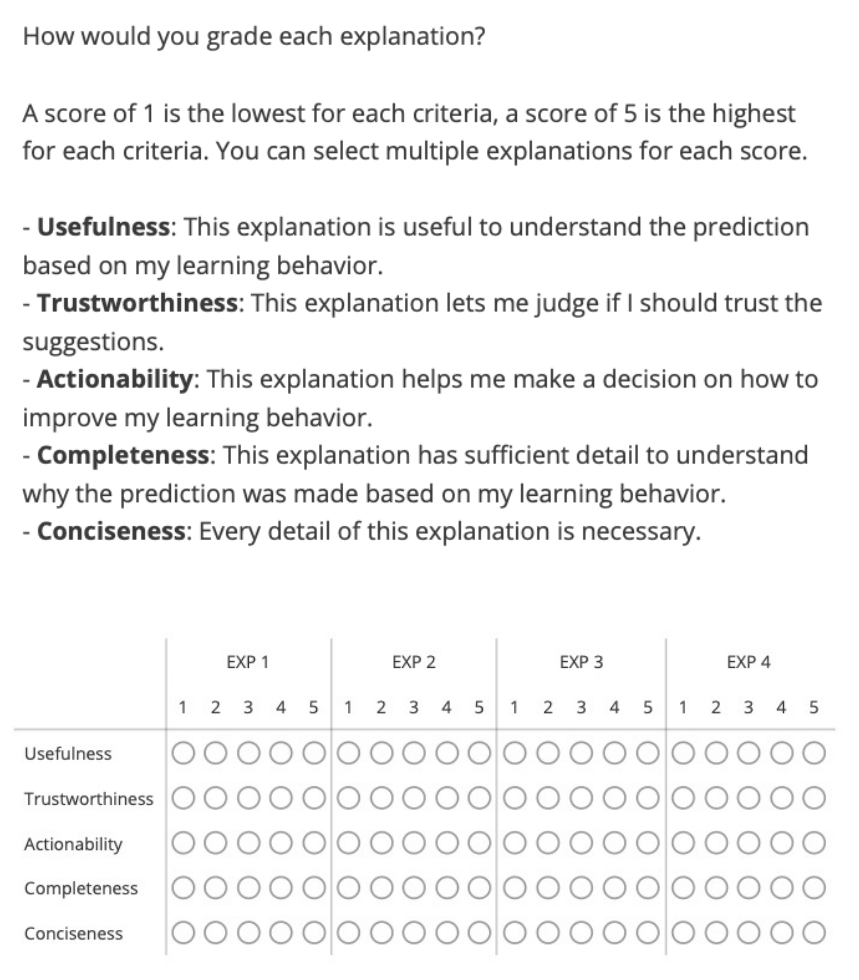}
        \caption{Grading criteria across four explanations presented at a time (one base explanation and three \il explanations).}
        \label{fig:answer-section}
    \end{minipage}
    \hfill
    \begin{minipage}{0.55\linewidth}
        \centering
        \includegraphics[width=\linewidth]{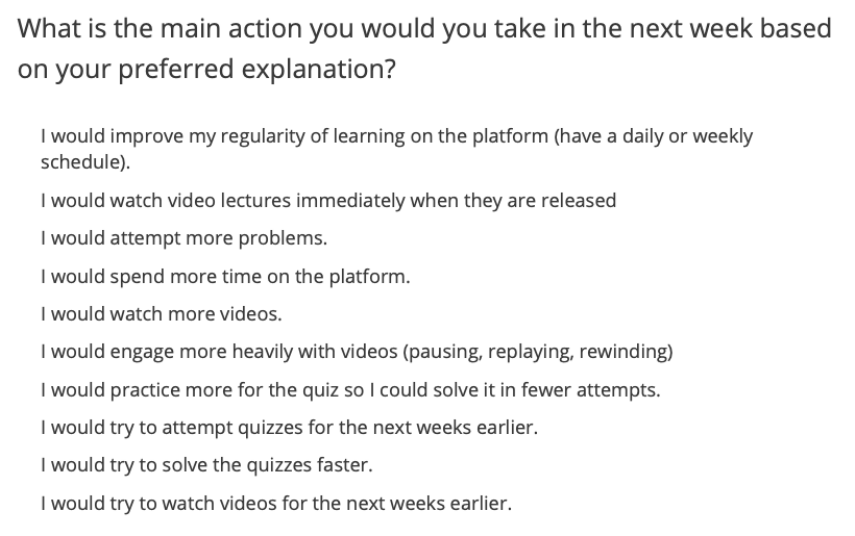}
        \caption{Actionability decision made by the students.}
        \label{fig:actionability}
    \end{minipage}
\end{figure}

For each student sample, participants were presented with four model explanations in a random order. They were then required to evaluate these explanations using Likert scales based on five predefined criteria. To familiarize participants with the evaluation process and to screen for inattentive responses, a practice question was included. Data from three participants who did not correctly answer the practice question were excluded from the final analysis. A screenshot of the answer section can be seen in Figure \ref{fig:answer-section}.

All study materials can be found in the \il repository: \texttt{https://github.com/epfl-ml4ed/iLLuMinaTE/} \texttt{tree/main/study-materials}, with the exact versions of the survey shown to students. There are six versions of the survey: two variations with 3 randomly chosen students each (one from each course) and three different source explainer versions (LIME, MC-LIME, CEM) for each student variation.

\pagebreak
\pagebreak
\subsection{Extended Results}
\label{appendix:study-results}

\begin{figure}[ht!]
    \centering
    \includegraphics[width=\linewidth]{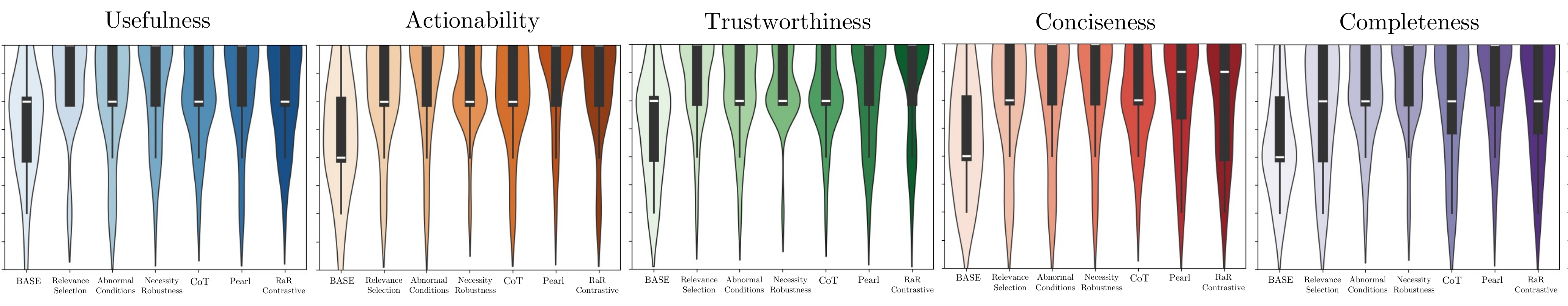}
    \caption{Results of student evaluation of different theories of explanations (Base, Relevance Selection, Abnormal Conditions, Necessity Robustness, CoT, Pearl, RaR + Contrastive) on all five axes of explanation.}
    \label{fig:violins}
\end{figure}

\begin{figure}[ht!]
    \centering
    \includegraphics[width=0.6\linewidth]{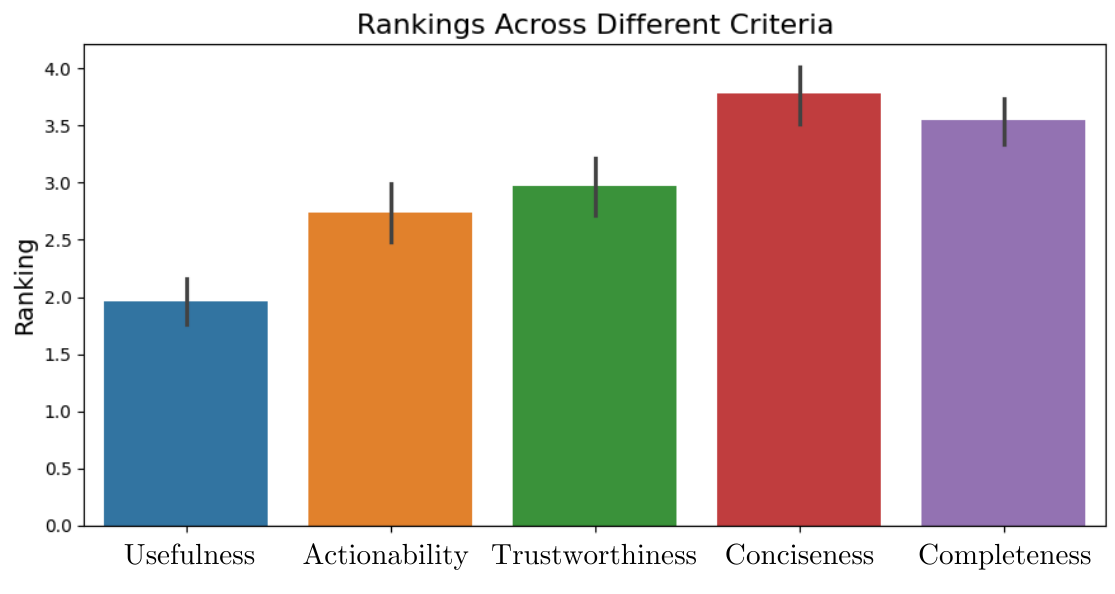}
    \caption{Results of student ranking of different explanation criteria. Lower ranking is better (more important).}
    \label{fig:ranking}
\end{figure}

The extended results of student preferences over five criteria (two of which are included in Fig. \ref{fig:exp3-criteria}) is explored in Fig. \ref{fig:violins}. Violin plots are chosen to reflect the distributions of student rankings of the criteria. Students also ranked the criteria by what they found most important (lower ranks are more important), as reflected in Fig. \ref{fig:ranking}. We note that there is a clear ordering from the student perspective over explainability criteria: usefulness, actionability, trustworthiness and then at a similar level completeness and conciseness.

\section{Actionability Experiments}
\label{appendix:actionability}

The following experiments indicate the results of the actionability simulation (with standard deviations) over explainers and theories. We note that MC-LIME is the most performant explainer, and contrastive explanations, closely followed by necessity-robustness leads to the highest quality (simulated) interventions. Note that these experiments were run over 114 study participants with six different base students that were simulated, but a larger scale study would be needed to provide more generalizable conclusions.

\begin{figure}[ht!]
    \centering
    \begin{minipage}[b]{0.41\linewidth}
        \centering
        \includegraphics[width=\linewidth]{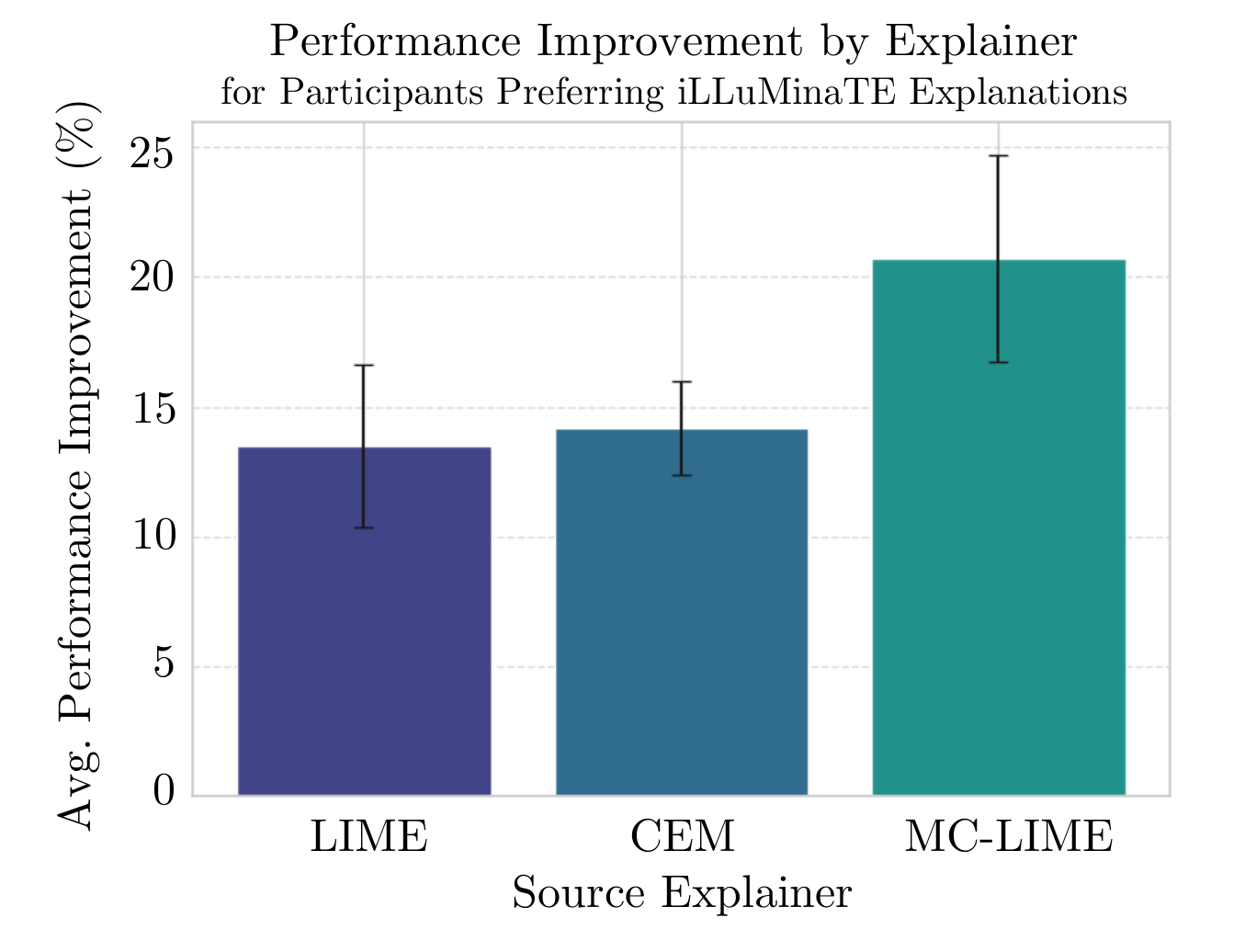}
        \label{fig:action1}
    \end{minipage}
    \hfill
    \begin{minipage}[b]{0.58\linewidth}
        \centering
        \includegraphics[width=\linewidth]{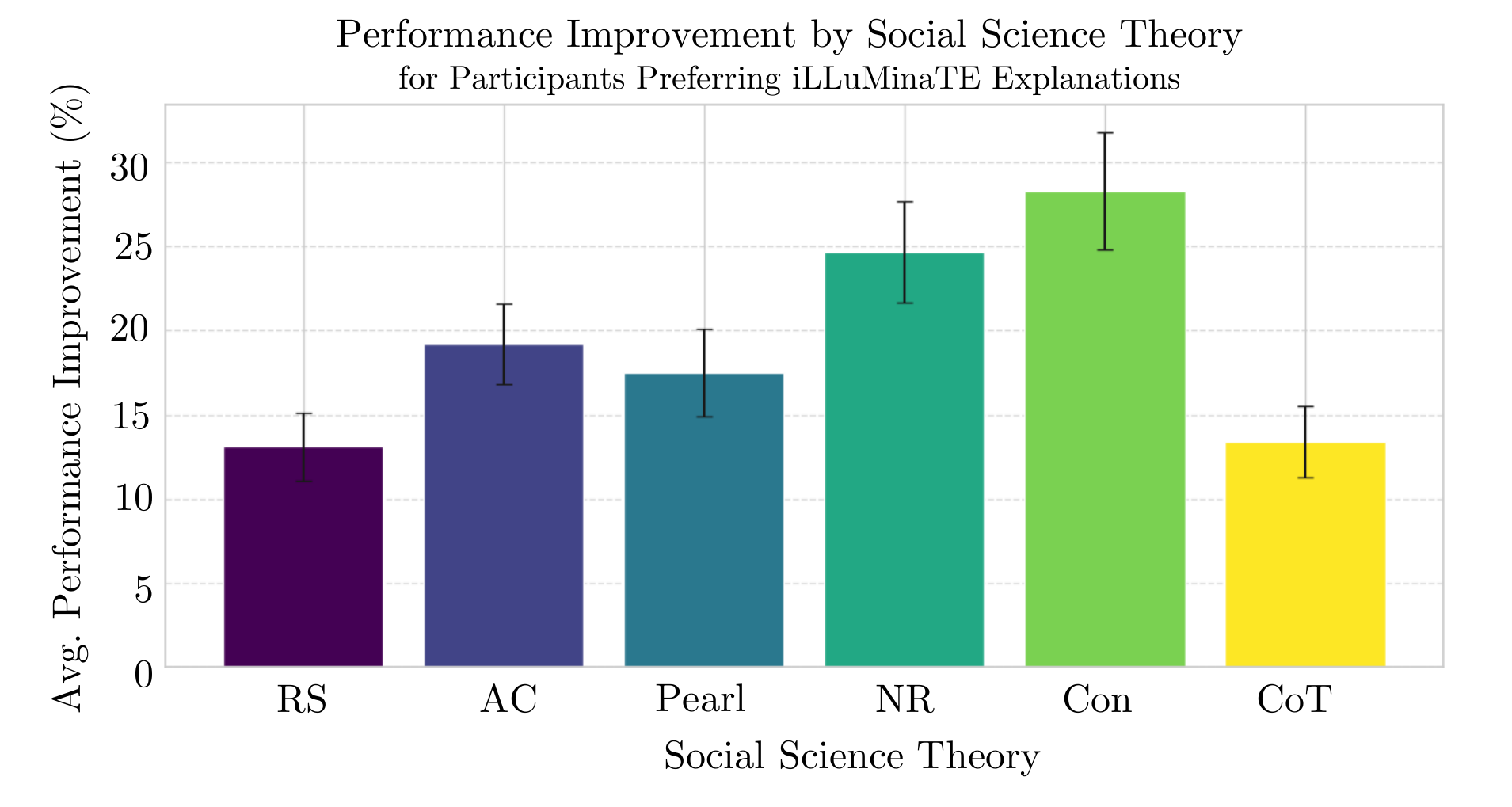}
        \label{fig:action2}
    \end{minipage}
    \caption{(Left) Actionability simulation across source explainers (LIME, CEM, MC-LIME), averaged over courses and theories. (Right) Actionability simulation across six  theories (Relevance Selection, Abnormal Conditions, Pearl's Model, Necessity and Robustness, Contrastive explanations, and Chain of Thought), averaged over courses and source explainer.}
\end{figure}

\section{Extended Background}

\subsection{What is a good explanation?}
To address this question, it is important to discuss what an explanation is, generally. Explanation can be seen through different lenses: it can be seen as causal or non-causal, as a process or a product of it. In this section, we will first look at different ideas about what makes a good explanation. Then, we will explore how humans understand and create explanations themselves. Finally, we will discuss how explanations are best communicated.

\subsection{Realist and Pragmatic Theories}

Largely, explanation theories can be divided into realist and pragmatic theories \cite{danks2022governance}. Realist theories emerge from the realms of the philosophy of science and logic, while pragmatic theories of explanation are rooted in studies of psychology and social sciences. 

The difference stands mostly on the consideration of the context as important factor in the definition of an explanation. Realist theories define an explanation a relation between theory and fact, while pragmatic theories a relation between theory, fact, and context.
This difference can be boiled down to \textit{whether radically false (i.e., not even approximately true) statements can be part of an explanation: realists say “no” while pragmatists say “yes"}. In this work, we focus mostly on pragmatic theories of explanation.

\subsection{Explanation as a process or as a product?}

Explanation can be seen as one or the other, or both \cite{lombrozo2006structure}. The theories that define explanation as a process could be categorized in three main groups \cite{miller2019explanation}:
\begin{itemize}
    \item Explanation as a causal process - Explanation is the process of inferring the real causes of an event.
    \item Explanation as a cognitive process - Explanation is a process that takes place in our cognition and it is closely related to abductive reasoning.
    \item Explanation as a social process - Explanation is a process of transferring knowledge between \textit{explainer} (the agent who explains a decision or event) and \textit{explainee} (the agent receiving the explanation).
\end{itemize}

Theories that see explanation as a product define it as an answer to a ``why-question'' \cite{miller2019explanation, molnar2020interpretable}.

\subsection{Explanation as a causal process}

For most of the philosophical theories, explanation is rooted in causality, starting from Hume's regularity theory \cite{hume2016enquiry} of causation that argues \textit{“there is a cause between two types of events if events of the first type are always followed by events of the second”}.

In broad terms, theories defined upon causality have important assumptions in common:
\begin{itemize}
    \item ``Everyday'' explanations don't differ from scientific explanations
    \item Explanation is a causal relationship
    \item Radically false statements can't be part of an explanation
\end{itemize}

Various philosophical theories emerged from causality such as the Deductive-Nomological Model (DN Model)\cite{sep-scientific-explanation} which can be summarized as \textit{``Whenever phenomenon X is observed to occur in the setting of conditions C, Y will be observed''} \cite{cambria2023survey}. An extension of these theories were the probabilistic theories, such as the Statistical Relevance model (SR model) that in general state that\textit{``event C is a cause of event E if and only if the occurrence of C increases the probability of E occurring''} \cite{miller2019explanation}.

Other models have been defined but all the theories in this category have strong limitations when applied to social or human realms. The main problem is that \textit{``the explanations that successfully follow these criteria seem to vary greatly in the extent to which they are explanatory deep or illuminating in the eye of the person receiving the explanation''} \cite{miller2019explanation}.
They miss relevance and bring little impact as in most cases there are no possible explanations that can enter in the realm of certainty and causality. Thus, from this group of theories we will keep only the SR model, as it mainly relies on likelihood and probability.

\subsection{Explanation as a cognitive process}
In general the cognitive process that we follow when we search for an explanation follows a general structure:
\textit{\begin{quote}
When we encounter something confusing, we use abductive reasoning to make sense of it and come up with an explanation. This process includes the mental simulation of hypothetical scenarios where the event didn't happen (counterfactuals) that lead us to the identification of potential causes. From these possibilities, we infer and select an explanation.
\end{quote}}

\subsubsection{Abductive Reasoning}
Abductive reasoning, also known as abduction, is a logical process in which an individual derives a hypothesis to explain an observed phenomenon \cite{peirce1997pragmatism}. It is often used to generate the best possible explanation for surprising or unexpected facts in a context of uncertainty. A common form of abductive inference follows this structure:
\begin{quote}
    \textit{The surprising fact, C, is observed. But if A were true, C would be a matter of course. Hence, there is reason to suspect that A is true.} \cite{peirce1997pragmatism}
\end{quote}

Alternatively, it can be expressed as:
\begin{quote}
    \textit{D is a collection of data (facts, observations, givens). H explains D (would, if true, explain D). No other hypothesis can explain D as well as H does. Therefore, H is probably true.} \cite{josephson1996abductive}
\end{quote}

The process could be summarized in these four steps:

\begin{enumerate}
    \item \textbf{Observation:} Observe some (presumably unexpected or surprising) events.
    \item \textbf{Hypothesis Generation:} Generate one or more hypotheses about these events.
    \item \textbf{Plausibility Judgment:} Judge the plausibility of the hypotheses.
    \item \textbf{Hypothesis Selection:} Select the ‘best’ hypothesis as the explanation.
\end{enumerate}

\subsubsection{Counterfactuals and Contrastive explanations}
Counterfactuals are very relevant to abductive reasoning as studies suggest that people don't simply explain why things happen, but rather why something happened one way (event P) instead of another way (event Q) \cite{molnar2020interpretable}. 
They were firstly described in Hume's work \cite{hume2016enquiry} as the \textit{counterfactual case}:

\begin{quote}
    \textit{“the cause should be understood relative to an imagined, counterfactual case: event C is said to have caused event E if, under some hypothetical counterfactual case the event C did not occur, E would not have occurred.”}
\end{quote}

In other words, explanations are always \textbf{contrastive}. We explain why P occurred by contrasting it with a situation where P didn't happen (Q), even if that situation is purely counterfactual.

\noindent \textbf{The Problem of Implicit Foils.} \citet{lipton1990contrastive} acknowledges a challenge. While explanations often highlight the target event (P), they rarely explicitly mention the foil (Q). For example, the question "Why did Elizabeth open the door?" could have countless foils: leaving it closed, opening the window, or someone else opening it. The unspecified foil, "not(Elizabeth open the door)," creates a vast pool of possibilities.

This theory \shortcite{van2002remote} further proposes a classification system for explanations based on the type of contrast:

\begin{itemize}
    \item Plain-Fact: Answers basic curiosity about an object's property (e.g., "Why does a magnet attract metal?"). These questions lack a foil.
    \item P-Contrast: Compares properties within an object (e.g., "Why is the sky blue rather than red?").
    \item O-Contrast: Compares properties between objects (e.g., "Why is a car red, while a bus is yellow?").
    \item T-Contrast: Compares properties of the same object across time (e.g., "Why was the water liquid yesterday but solid today?").
\end{itemize}

\citet{hesslow1988problem} argues that the explanandum, the event to be explained, should be viewed not as a simple property, but as a difference between the target object and a reference class (foils) with respect to a particular property. He emphasizes that the selection and weighting of causes depend on their \textit{explanatory relevance}. \citet{lipton1990contrastive}'s difference condition further clarifies that explaining "why P rather than Q" requires identifying a causal difference. This difference consists of a cause for P that is absent in the counterfactual scenario (not-Q).

Contrastive theory suggests that when selecting an explanation from multiple possible causes, people focus on the differences between the target event (P) and the counterfactual foil (Q).

\subsection{Explanation as a social process}
Furthermore, explanation can also be considered finally as a social process that is inherently part of a conversation between an \textit{explainer} and the \textit{explainee}.
Usually the explainer has enough knowledge or understanding of the causes of an event to be able to explain it to others.
The most relevant work in this category is Hilton's conversational model of explanation \cite{hilton1990conversational}.

\subsection{Why people ask for an explanation?}
The primary function of explanation goes beyond simply acquiring new information, it is about \textit{learning} and refining our understanding of the world.

Explanations can ultimately shape our causal beliefs and our prior knowledge. Malle argues that explanation helps us resolve contradictions or inconsistencies in our existing understanding and create a sense of shared meaning within a group of individuals \cite{malle2006mind}. Finally, explanations can serve as a tool for persuasion, influencing how others perceive the world.

Therefore, an explanation can have different goals in the setting of explainable AI. it is up to those designing the AI system to decide what's most important to explain in each situation. For example, an AI could be designed to build trust in the user, thus persuasion and other forms of manipulation could be more relevant than the actual truthfulness of the answer.

\subsection{Structure of explanation}

\subsubsection{Overton's scientific structure of explanation}

Overton’s model \cite{overton2012explanation} of scientific explanation defines five categories of properties or objects that are explained in science. These are:

\begin{itemize}
    \item Theories: These are sets of principles that form the building blocks for models.
    \item Models: An abstraction of a theory that represents the relationships between kinds and their attributes.
    \item Kinds: An abstract universal class that supports counterfactual reasoning. For example, an “arthropod” is a kind.
    \item Entities: An instantiation of a kind. For example, a specific spider is an entity.
    \item Data: Statements about activities, such as measurements and observations. For example, the fact that a particular spider has eight legs is data.
\end{itemize}

Overton argues that explanations of phenomena at one level must be relative to and refer to at least one other level, and that explanations between two such levels must refer to all intermediate levels.
\\
\subsubsection{Malle's theory}

\begin{figure}[t]
    \centering
    \includegraphics[width=0.35\textwidth]{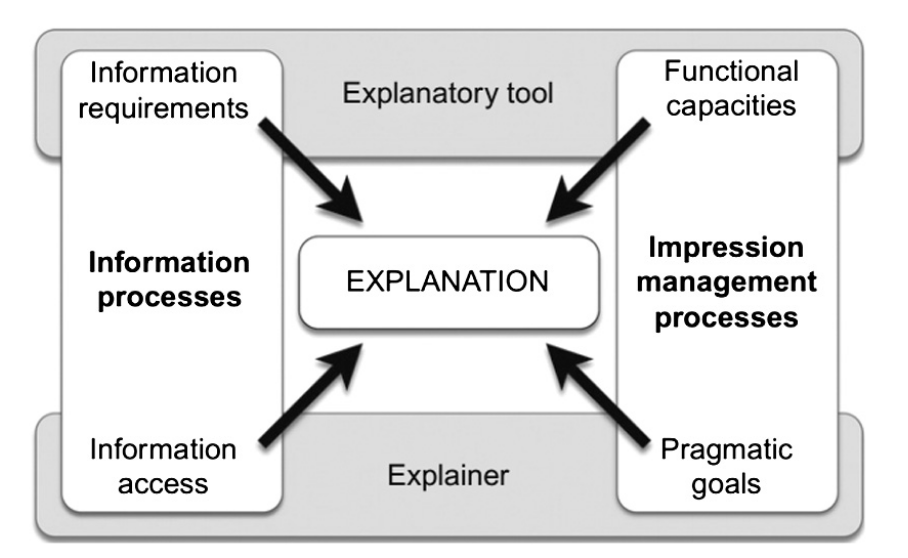}
    \caption{Malle's process model for behavior explanation from \citet{malle2006mind}}
    \label{fig:malle}
\end{figure}

\citet{malle2006mind} proposes a theory that breaks down the cognitive processes of explanation into two main groups:

\begin{enumerate}
    \item \textbf{Information Processes} — These involve the creation and assembly of explanations. 
    \item \textbf{Impression Management Processes} — These concern the social dynamics of providing explanations.
\end{enumerate}

Malle further divides these dimensions into two additional aspects (Figure \ref{fig:malle}), the tools and methods in the hands of the explainer for constructing and delivering explanations and the explainer’s perspective or prior knowledge. By considering these two dimensions, we have four items:

\begin{enumerate}
    \item Information Requirements — These are the necessary elements needed for a proper explanation. 
    \item Information Access — This refers to the information available to the explainer. Sometimes, critical information may be missing.
    \item Pragmatic Goals — This involves the purpose of the explanation, such as imparting knowledge to the listener, portraying someone in a certain light, or building trust.
    \item Functional Capacities — These are the inherent limitations and capabilities of the explanatory tools, which determine what goals can be achieved with them.
\end{enumerate}

In more simple terms these four elements could be described as \textbf{what information is needed}, \textbf{what is available}, \textbf{the intentional goals of the explainer} and the \textbf{constraints of the context and tools}. Malle's framework is fundamental to determine the Context Definition (Sec. \ref{appendix:exp-sel-prompt}) where we define these four elements for our LLMs to give an effective explanation.

\subsubsection{Hilton's conversational model of explanation}

Hilton's conversational model \cite{hilton1990conversational} emphasizes the social nature of explanation, contrasting it with simple causal attribution. He argues that explanations are conversations aimed at resolving a ``puzzle'' in the explainee's mind. This model consist of a two-stage process, which involves:

\begin{enumerate}
    \item Diagnosis: exploration of the possible causes of an event
    \item Explanation presentation: presenting the selected explanation in a socially relevant way
\end{enumerate}

Crucially, good explanations must be relevant to the explainee's specific question, not just provide any identified cause.  Hilton emphasizes that explanations follow the principles of cooperative conversation, including Grice's maxims \cite{grice1975logic} (see Appendix \ref{appendix:exp-pres-prompt}) for quality, quantity, relation, and manner. In essence, explanations should be truthful, provide the right amount of detail, stay focused on the question, and be clear and respectful. It highlights also that explanation should be tailored based on the explainee's existing knowledge, focusing on causes they may not already understand.

\subsection{\il's relationship to post-hoc explainers}

In our study we decided to use post-hoc explainers for the causal connection step. We have an instruction rule in the \il prompt framework to explicitly use the explainer's results. This decision was made due to the popularity of these explainers in the domain of AI for education \cite{hasib2022lime, 2022.EDM-short-papers.33}. However, as we seen from \cite{swamy2023trusting}, explainers in educational settings can disagree with each other. Thus, our work comparing LIME, CEM and MC-LIME begins to answer the question ``is the LLM-based communication of explanations different depending on which post hoc explainer is used?" -- we mostly identify that the choice of explainer does not lead to an overall difference in which theories students prefer (Sec. \ref{sec:student-preferences}). However, we do identify a significant increase of actionability for the MC-LIME source explainer over the other approaches (Sec. \ref{sec:actionability} and Appendix \ref{appendix:actionability}). A longer exploration with more source explainers is required to make generalizable statements (beyond the education use case) regarding how choice of source explainer impacts LLM communication of the explanation. \il is compatible with using interpretable-by-design models (i.e. Concept Bottleneck Models \cite{koh2020concept}, InterpretCC \cite{swamy2024interpretcc}) or in-hoc explainability methods (i.e. concept activation vectors for education \cite{asadi2023ripple}) as a causal connection step instead of post-hoc explainers.

\end{document}